\newif\ifconfver
\confvertrue        

\newif\ifplainver  
\plainvertrue

\newif\ifhide  
\hidetrue

\ifplainver
    \confverfalse   
\fi

\ifconfver
     \documentclass[10pt,twocolumn,twoside]{IEEEtran}
\else
    \ifplainver
        \documentclass[11pt]{article}
        \usepackage{fullpage}
    \else
        \documentclass[12pt,draftcls,onecolumn]{IEEEtran}
    \fi
\fi

\usepackage{calc,amsfonts,amssymb,amsmath,bm,url,color,theorem,graphicx,cite}
\usepackage{psfrag,float}
\usepackage{algorithm}
\usepackage{algorithmic}
\usepackage{soul}
\usepackage{enumerate}
\usepackage{bbm}
\usepackage{multirow}
\usepackage{subcaption}
\usepackage{array}
\usepackage{epstopdf}
\usepackage{diagbox}
\usepackage{titling}

\usepackage{etoolbox}%
\usepackage{xpatch}
\usepackage{blindtext}
\usepackage{tocloft}%
\usepackage{xr}
\usepackage{bibspacing}

\newlength{\articlesectionshift}%
\let\LaTeXStandardSection\section
\let\LaTeXStandardTheSection\thesection
\let\LaTeXStandardTheSubSection\thesubsection
\let\LaTeXStandardTheSubSubSection\thesubsubsection
\let\LaTeXStandardTheParagraph\theparagraph

\makeatletter
\newcounter{titlecounter}

\xpretocmd{\maketitle}{\ifnumgreater{\value{titlecounter}}{1}}{\clearpage}{}{} 
\xpatchcmd{\maketitle}{\let\maketitle\relax\let\@maketitle\relax}{\refstepcounter{titlecounter}\begingroup

\makeatother \addtocontents{toc}{\begingroup\addtolength{\cftsecindent}{-\articlesectionshift}}%
	\addcontentsline{toc}{section}{\protect{\numberline{\thetitlecounter}{\@title~ \@author}}}%
	\addtocontents{toc}{\endgroup}
}{%
	\typeout{Patching was successful}
}{%
	\typeout{patching failed}
}%

\def\@IEEEdestroythesectionargument#1{\LaTeXStandardSection{#1}}%

\xapptocmd{\maketitle}{%
	\renewcommand{\thesection}{\LaTeXStandardTheSection}%
	\renewcommand{\thesubsection}{\LaTeXStandardTheSubSection}%
	\renewcommand{\thesubsubsection}{\LaTeXStandardTheSubSubSection}%
	\renewcommand{\theparagraph}{\LaTeXStandardTheParagraph}%
}{}{}%

\@addtoreset{section}{titlecounter}

\newlength{\customfigwidth}
\newcolumntype{M}[1]{>{\centering\arraybackslash}m{#1}}

\definecolor{orange}{RGB}{255,107,0}

\definecolor{cpink}{rgb}{0.7, 0.11, 0.11}

\newtheorem{Fact}{Fact}
\newtheorem{Lemma}{Lemma}
\newtheorem{Prop}{Proposition}
\newtheorem{Theorem}{Theorem}

\newtheorem{Corollary}{Corollary}

\newtheorem{Asm}{Assumption}
\theorembodyfont{\rmfamily}
\newtheorem{Exa}{Example}

\newcommand\bw{\ensuremath{{\bm w}}}

\newcommand\bx{\ensuremath{{\bm x}}}
\newcommand\by{\ensuremath{{\bm y}}}

\newcommand\bh{\ensuremath{{\bm h}}}
\newcommand\bH{\ensuremath{{\bm H}}}

\newcommand\bz{\ensuremath{{\bm z}}}
\newcommand\bp{\ensuremath{{\bm p}}}

\newcommand\br{\ensuremath{{\bm r}}}
\newcommand\bR{\ensuremath{{\bm R}}}

\newcommand\bX{\ensuremath{{\bm X}}}
\newcommand\bZ{\ensuremath{{\bm Z}}}

\newcommand\bc{\ensuremath{{\bm c}}}
\newcommand\ba{\ensuremath{{\bm a}}}

\newcommand\bA{\ensuremath{{\bm A}}}

\newcommand\bb{\ensuremath{{\bm b}}}

\newcommand\bB{\ensuremath{{\bm B}}}

\newcommand\bd{\ensuremath{{\bm d}}}

\newcommand\bu{\ensuremath{{\bm u}}}
\newcommand\bv{\ensuremath{{\bm v}}}

\newcommand\bV{\ensuremath{{\bm V}}}
\newcommand\bU{\ensuremath{{\bm U}}}
\newcommand\bs{\ensuremath{{\bm s}}}

\newcommand{\Rbb}{\mathbb{R}}
\newcommand{\Cbb}{\mathbb{C}}

\newcommand{\setX}{\mathcal{X}}
\newcommand{\setR}{\mathcal{R}}
\newcommand{\setU}{\mathcal{U}}

\newcommand{\setW}{\mathcal{W}}
\newcommand{\setS}{\mathcal{S}}

\newcommand{\setD}{\mathcal{D}}

\newcommand{\setI}{\mathcal{I}}

\newcommand{\CN}{\mathcal{CN}}
\newcommand{\Exp}{\mathbb{E}}
\newcommand{\bvarphi}{{\boldsymbol{\varphi}}}
\newcommand{\balpha}{{\boldsymbol{\alpha}}}

\newcommand\jj{\ensuremath{{\frak j}}}
\newcommand{\eps}{\varepsilon}
\newcommand{\bzero}{{\bm 0}}
\newcommand{\bone}{{\bm 1}}
\newcommand{\bI}{{\bm I}}

\newcommand{\dec}{\mathrm{dec}}
\newcommand{\Diag}{\mathrm{Diag}}

\newcommand{\beq}{\begin{equation}}
\newcommand{\eeq}{\end{equation}}

\begin{document}

\bibliographystyle{IEEEtran}

\newcommand{\papertitle}{
Symbol-Level Precoding Through the Lens of \\ Zero Forcing and Vector Perturbation

}

\newcommand{\paperabstract}{
Symbol-level precoding (SLP) has recently emerged as a new paradigm for physical-layer transmit precoding in multiuser multi-input-multi-output (MIMO) channels.
It exploits the underlying symbol constellation structure, which the  conventional paradigm of linear precoding does not, to enhance symbol-level performance such as symbol error probability (SEP).
It also allows the precoder to take a more general form than linear precoding.
This paper aims to better understand the relationships between SLP and linear precoding,  subsequent design implications, and further connections beyond the existing SLP scope.
Focused on the quadrature amplitude modulation (QAM) constellations,
our study is built on a basic signal observation, namely, that SLP can be equivalently represented by a zero-forcing (ZF) linear precoding scheme augmented with some appropriately chosen symbol-dependent perturbation terms,
and that some extended form of SLP is equivalent to a vector perturbation (VP) nonlinear precoding scheme augmented with the above-noted perturbation terms.
We examine how insights arising from this perturbed ZF and VP interpretations can be leveraged to i) substantially simplify the optimization of certain SLP design criteria, namely, total or peak power minimization subject to SEP quality guarantees;
and ii) draw connections with some existing SLP designs.
We also touch on the analysis side by showing that, under the total power minimization criterion, the basic ZF scheme is a near-optimal SLP scheme when the QAM order is very high---which gives a vital implication that SLP is more useful for lower-order QAM cases.
Numerical results further indicate the merits and limitations of the different SLP designs derived from the perturbed ZF and VP interpretations.

}


\ifplainver


    \title{\papertitle}

    \author{Yatao Liu$^\dag$,    Mingjie Shao$^\dag$, Wing-Kin Ma$^\dag$ and Qiang Li$^\ddag$\\
    $^\dag$Department of Electronic Engineering, The Chinese University of Hong Kong, \\
    Hong Kong SAR of China\\
    $^\ddag$School of Information and Communication Engineering,\\
University of Electronic Science and Technology of China, Chengdu,
China
    }

    \maketitle

    \begin{abstract}
    \paperabstract
    \end{abstract}

\else
    \title{\papertitle}

    \ifconfver \else {\linespread{1.1} \rm \fi

    \author{Yatao Liu,  Mingjie Shao, Wing-Kin Ma and Qiang Li
    }

    \maketitle

    \ifconfver \else
        \begin{center} \vspace*{-2\baselineskip}
        \end{center}
    \fi

    \begin{abstract}
    \paperabstract
    \end{abstract}


    \begin{IEEEkeywords}\vspace{-0.0cm}
       Symbol-level precoding, symbol-error probability, zero-forcing, vector perturbation
    \end{IEEEkeywords}

    \ifconfver \else \IEEEpeerreviewmaketitle} \fi

 \fi

\ifconfver \else
    \ifplainver \else
        \newpage
\fi \fi

\section{Introduction}

%
%
%
    
Transmit precoding is a subject that has been studied for decades.
It plays a central role in the multiuser multi-input-multi-output (MIMO) scenarios, offering effective transmit signaling schemes to enable spatial multiplexing
and to enhance system throughputs.
Linear precoding is, by far, the most popular approach:
it is easy to realize at the symbol or signal level;
it has good design flexibility to cater for various design needs, such as those from cognitive radio, multi-cell coordination, cell-free MIMO and physical-layer security;
and there is a rich line of research concerning how we can design linear precoding for utilitarian throughput maximization, fair throughput allocation, etc.; see, e.g., \cite{Bengtsson2001,schubert2004solution,wiesel2005linear,Yu2007Transmitter,liu2013max,bjornson2014optimal,shi2011iteratively,shi2016sinr}.
The decades of transmit precoding research also led to beautiful ideas with nonlinear precoding, such as Tomlinson-Harashima precoding~\cite{windpassinger2004precoding} and vector perturbation (VP) precoding \cite{hochwald2005vector,Maurer2011Vector};
they take certain specific modulo-type nonlinear forms and may not be as flexible as linear precoding, but they can greatly improve performance compared to some simple linear precoding schemes such as the zero-forcing (ZF) scheme.
In linear precoding we often treat multiuser interference (MUI) as noise, or something to alleviate.
However, some recent research argues that MUI is not necessarily adversarial.
We can manipulate MUI at the symbol level to help us improve performance.
This idea is generally called {\em symbol-level precoding} (SLP) in the literature.

The currently popular way to define SLP is that we can choose any multi-antenna transmitted signals (absolute freedom rather than a linear form), and the aim is to enhance performance in a symbol-aware fashion, e.g., symbol error probability.
For the past decade researchers have been invoking various ideas that gradually evolved to the SLP defined above, and it is worthwhile to briefly recognize such original endeavors.
SLP is also known as
directional modulation~\cite{baghdady1990directional,daly2009directional,kalantari2016directional} and
constructive interference (CI) \cite{masouros2009dynamic,masouros2011correlation,masouros2015exploiting,haqiqatnejad2018constructive,li2018interference1,li2017exploiting},
depending on the context.
In the early 2010, Masouros {\it et al.} took the intuition that under phase shift keying (PSK) constellations, MUI can be characterized as constructive and destructive~\cite{masouros2009dynamic,masouros2011correlation}.
There,  linear precoders are designed such that, at the user side, the CI pushes the received signals deeper into the decision region.
Soon, this CI idea was exploited extensively for PSK constellations and in a more general nonlinear form~\cite{alodeh2015constructive,masouros2015exploiting,alodeh2016energy,krivochiza2017low,li2018interference1}.
Later, Alodeh {\it et al.} extended this interference manipulating concept to quadrature amplitude modulation (QAM) constellations~\cite{alodeh2017symbol},
and many subsequent works followed this adaptation~\cite{li2017exploiting,kalantari2017mimo,haqiqatnejad2018symbol,alodeh2020joint}. Lately, SLP has been applied to a number of scenarios, such as MIMO orthogonal frequency division multiplexing~\cite{studer2013aware,Yao2019}, physical-layer security~\cite{Liu2020,khandaker2018constructive} and intelligent reflecting surface~\cite{shao2020minimum};
see the overview papers \cite{alodeh2018symbol,li2020tutorial} for a more comprehensive introduction.

In addition to improved symbol-level performance over linear precoding, SLP allows us to have a better control with the transmitted signal amplitudes.
By comparison, linear precoding typically controls the average squared amplitudes, or powers.
The better amplitude control of SLP is particularly beneficial to the recent developments of large-scale or massive MIMO systems.
In such systems it is desirable that each antenna is employed with a low-cost radio frequency chain, wherein the power amplifiers trade a smaller linear amplification range for a higher power efficiency.
This necessitates the transmitted signals at each antenna to have low amplitude fluctuations at every time instant.
SLP has been adopted to deal with more amplitude stringent designs, such as peak-to-average-power ratio minimization \cite{spano2017spatial,Spano2018Symbol},
constant-envelope precoding \cite{mohammed2013per,pan2014constant,jedda2018quantized} and one-bit precoding~\cite{jacobsson2017quantized,swindlehurst2018reduced,sohrabi2018one,shao2019framework,shao2019one}.

SLP has been extensively employed in a variety of scenarios, as noted above,
and the flexibility of SLP as a precoding design framework has been the key factor with its recent prominence.
But we see fewer studies that work toward understanding the basic nature of SLP.
In particular, the connections between SLP and the existing precoding schemes were relatively under-explored in the prior literature.
Researchers realized that there are connections between SLP and ZF precoding; see, e.g., \cite{krivochiza2017low,li2018interference1,Li2021Interference}.
But the existing literature does not provide  a thorough enough investigation on such connections and the subsequent implications on precoding designs.

\subsection{This Work and The Contributions}

In this paper we study SLP through the lens of ZF and VP precoding,
with a focus on the QAM constellations.
We are interested in drawing connections between SLP and the existing precoding schemes, thereby revealing new insight.
We take the classic  single-cell multiuser multi-input-single-output (MISO) downlink as the scenario to study the problem;
and we consider a class of precoding designs that seek to minimize the transmission power, either as total power or as peak per-antenna power,
under the constraints that some symbol error probability (SEP) requirements are met.
Also, we study a general SLP structure wherein the received constellation ranges and phases, which are typically prefixed in the existing SLP designs, are part of our design variables.
Under the above problem setup, we raise the argument that {\em SLP can be regarded as a perturbed ZF scheme};
perturbations are injected into the symbols and on the channel nullspace, they are symbol-dependent, and they are designed for enhancing power efficiency.
As an extension not seen in the existing designs, we also argue that
SLP, with a suitable modification, can be regarded as a perturbed VP scheme.
Our study will revolve around how the SLP-ZF and SLP-VP relationships can be exploited to engage with the SEP-constrained designs more efficiently; we also seek to better understand the basic problem nature.
In addition, our study will lead us to draw connections with some existing SLP designs, which gives rise to an alternative explanation of the existing designs.

Some key contributions of this study should be highlighted.
On the theoretical side, we use the SLP-ZF relationship to show that, for the  SEP-constrained total power minimization design, the ZF scheme (without perturbations) is a near-optimal SLP scheme for very high-order QAM constellations.
This result is fundamentally intriguing, and it explains why we have never seen a numerical result that shows significant performance gains with SLP for very high QAM orders (see, e.g., \cite{alodeh2017symbol}).
It further leads to the vital implication that SLP is more useful for lower-order QAM constellations.

Another set of key contributions lies in design optimization.
We deal with SLP designs that jointly optimize SLP and the constellation ranges and phases; this is done over a block of symbols (typically a few hundreds in length, in practice).
The motivation is to work on a general design in an effort to enhance performance; as mentioned, the existing SLP solutions typically prefix the constellation ranges and phases.
The challenge arising is that the design problems are large-scale optimization problems.
We tackle the challenge by introducing an algorithmic method that exploits the problem structure provided by the SLP-ZF and SLP-VP relationships; it is a combination of the alternating minimization and proximal gradient methods.
A main issue there lies in finding a way to efficiently handle the large-scale problem nature, specifically, in the form of coupled constraints with a large number of optimization variables.
We deal with it by custom-building a proximal gradient method that exploits the coupled constraint structures in a very specific way (cf. Algorithm~\ref{AL_pro}).

Our numerical results also reveal useful insights as design guidelines.
They will be discussed in the conclusion section after we describe the different SLP designs derived from the SLP-ZF and SLP-VP relationships in the ensuing sections.

\subsection{Related Works}
	
	Let us give a further discussion with the relevant state of the art.
	This study focuses on the QAM constellations,
	and in this regard it is worthwhile to discuss the existing QAM-based SLP designs~\cite{alodeh2017symbol,li2017exploiting,kalantari2017mimo,haqiqatnejad2018symbol,alodeh2020joint,Li2021Interference}.
	The vast majority of the existing designs considered signal-to-noise ratio (SNR) or signal-to-interference-and-noise ratio (SINR) as the quality-of-service (QoS) metric,
	and they applied the CI notion, i.e., pushing symbols deeper into the correct decision regions, to enhance performance at the symbol level.
	Such designs will improve the SEP performance.
	They, however, do not work on SEP directly.
	Some recent studies directly use the SEP as the QoS metric~\cite{sohrabi2018one,shao2019framework,shao2019one};
	they appear in the context of one-bit and constant-envelope precoding, and they considered SEP performance maximization under power constraints.
	This study focuses on the classic multiuser downlink scenario and considers power minimization under SEP constraints.
	In fact, the reader will find that the SLP-ZF and SLP-VP relationships are particularly suitable tools for studying SEP-constrained designs.
	
	Exploring and exploiting the SLP-ZF relationship is the central theme of this study.
	As mentioned ealier, some prior studies already noticed and/or used the SLP-ZF relationship~\cite{krivochiza2017low,Li2021Interference,liu2018symbol}.
	The studies in~\cite{krivochiza2017low} and~\cite{Li2021Interference} considered the symbol-perturbed ZF structure and the per-symbol scaled ZF structure, respectively.
	These structures were proposed as specific forms of SLP, but their connections with SLP in its most general form were not studied.
	The conference version of this paper~\cite{liu2018symbol} showed the direct connection between SLP and perturbed ZF;
	the main results there will appear as Fact 2 and Proposition 1 in this paper.
	This present study takes the insight of our previous finding and sets its sight on a wider range of aspects, such as the SLP-VP relationship, the peak per-antenna power minimization design, and joint design with the SLP and the constellation ranges and phases.

\begin{table}
	\centering
	\caption{A summary of notations.} \label{table_notation}
	\renewcommand{\arraystretch}{1.1}
	\begin{tabular}{ c|c }
		\hline
		Notation & Definition\\
		\hline\hline
		$\circ$ & Hadamard product, $\bx \circ \by = [x_i y_i]_i$  \\
		\hline
		$\diamond$ & $\bx \diamond \by = [\Re(x_i) \Re(y_i) + \Im(x_i) \Im(y_i)]_i$ \\
		\hline
		$\ge$ & $\bx \ge \by$ means $x_i \ge y_i$ for all $i$ \\
		\hline
		$\ge_c$ & $\bx \ge_c \by$ means $\Re(\bx) \ge \Re(\by)$, $\Im(\bx) \ge \Im(\by)$ \\
		\hline
		$\|\cdot\|_\bR$ & Mahalanobis norm, $\| \bx \|_\bR = \sqrt{\bx^H \bR \bx}$, where $\bR$ is positive definite \\
		
		\hline \hline
		$\balpha_c$ & $\balpha_c=\balpha+\jj \balpha$, $\alpha_i=\frac{\sigma_v}{\sqrt{2}} Q^{-1}( \frac{1-\sqrt{1-\eps_i}}{2} )$ \\
		\hline
		$\setS$ & QAM constellation $\setS = \{ s_R + \jj s_I  \mid  s_R, s_I \in \{ \pm 1, \pm 3, \ldots, \pm (2L - 1) \} \}$ \\
		\hline
		$\rho$ & average symbol power, $\rho=\mathbb{E}_{s_t}[|s_t|^2]$, $s_t \in \setS$  \\
		\hline
		$\bs_t$ & symbol vector, $\bs_t \in \setS^K$  \\
		\hline
		$\bx_t$ & transmitted signal  \\
		\hline
		$\bu_t$ & symbol perturbation vector, cf. \eqref{rep_x}   \\
		\hline
		$\bB\bz_t$ & channel nullspace perturbation vector, cf. \eqref{rep_x} \\
		\hline
		$\bvarphi$ & constellation phase, with $|\varphi_i|=1$ \\
		\hline
		$\bd$ & constellation range, with $d_i=d_i^R+\jj d_i^I$, $d_i^R, d_i^I \ge 0$ \\
		\hline\hline
		$\bH$ & channel matrix  \\
		\hline
		$\bB$ & basis matrix of the nullspace of $\bH$ \\
		\hline
		$\bR$ & $\bR = (\bH \bH^H)^{-1}$ \\
		\hline
		$\bR_\bvarphi$ & $\bR_\bvarphi = {\rm Diag}(\bvarphi)^H\bR{\rm Diag}(\bvarphi) $ \\
		\hline
	\end{tabular}
\end{table}

\subsection{Notations}

We use $x$, $\bx$, $\bX$ and $\setX$ to denote a scalar, a vector, a matrix and a set, respectively;
$\Rbb$, $\Cbb$ and $\mathbb{Z}$ denote the set of all real numbers, complex numbers and integers, respectively;
$\bX^T$, $\bX^H$, $\bX^{-1}$ and ${\rm Tr}(\bX)$ are the transpose, Hermitian transpose, inverse and trace of $\bX$, respectively;
$\bx^*$ stands for the element-wise complex conjugate;
$\Re(\bx)$ and $\Im(\bx)$ are the real and imaginary components of $\bx$, respectively;
$|\bx|$ denotes the element-wise modulus of $\bx$;
$\langle \bx, \by \rangle \triangleq \Re( \bx^H \by)$ is the inner product of two vector $\bx, \by$;
${\rm card}(\setX)$ denotes the cardinality of the set $\setX$;
${\cal N}(\mu, {\sigma}^2)$ and ${\cal CN}(\mu, {\sigma}^2)$ denote the real and complex circularly symmetric Gaussian distribution with
mean $\mu$ and variance ${\sigma}^2$, respectively;
$\mathbb{E}_x[\cdot]$ denotes expectation of a random variable $x$.
Some  specialized notations will be defined later, and Table~\ref{table_notation} gives a summary of those notations and some commonly used symbols in the sequel.

\section{System Model}
\label{sec:model}

\subsection{Basics}

Consider a classic single-cell multiuser MISO downlink scenario, where a base station (BS) with $N$ transmit antennas simultaneously serves $K$ single-antenna users over a frequency-flat block faded channel.
The received signal $y_{i,t}$ of the $i$th user at symbol time $t$ can be modeled by
\begin{equation} \label{eq:sig_mod}
{y}_{i,t} = {\bh}_i^H {\bx}_{t} + {v}_{i,t}, \quad i=1,\ldots,{K},\ t=1,\ldots,T,
\end{equation}
where
${\bh}_i \in \Cbb^{{N}}$ represents the downlink channel from the BS to the $i$th user;
${\bx}_{t} \in \Cbb^{{N}}$ is the transmitted signal at symbol time $t$;
${v}_{i,t} \sim {\cal CN}(0, {\sigma}_v^2)$ is noise;
 $T$ is the transmission block length.

\begin{figure}[htb!]
	\centering
    \includegraphics[width=0.6\linewidth]{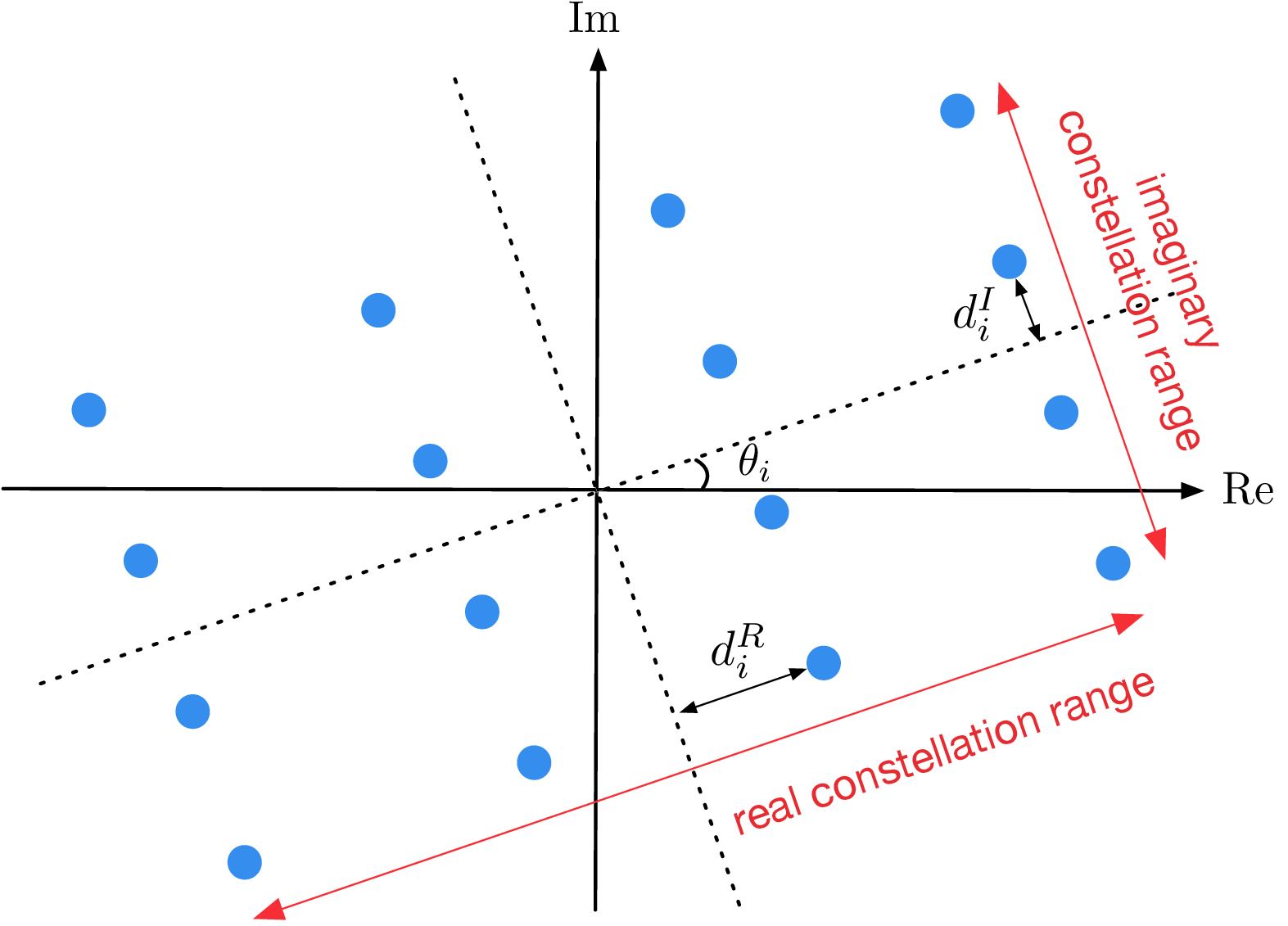}
	\caption{Illustration of the constellation experienced at the user's side. We assume 16-QAM.}
	\label{figd}
\end{figure}

Under the above scenario, the goal of precoding  is to simultaneously transmit data streams to multiple users, one for each user.
To describe, let $ {s}_{i,t} $ be the desired symbol of the $i$th user at symbol time $t$.
The symbols are assumed to be drawn from a quadrature amplitude modulation (QAM) constellation
\begin{equation}\label{eq:QAM_cons}
\setS = \{ s_R + \jj s_I  \mid  s_R, s_I \in \{ \pm 1, \pm 3, \ldots, \pm (2L - 1) \} \},
\end{equation}
where $L$ is a positive integer (the QAM size is $4L^2$); $\jj =\sqrt{-1}$.
Assuming perfect channel state information at the BS, we aim to design the transmitted signals $\bx_1,\dots,\bx_T$ such that the users will receive their desired symbols.
To be precise,
we want the noise-free part of $y_{i,t}$ in~\eqref{eq:sig_mod} to take the form
\begin{equation}\label{eq:approx}
\begin{split}
{\bh}_i^H {\bx}_{t} & \approx \varphi_i  ( d_i^R \Re(s_{i,t}) + \jj d_i^I \Im(s_{i,t}) ),
\end{split}
\end{equation}
where
$\varphi_i = e^{\jj \theta_i}$, $\theta_i \in [ 0, 2\pi ]$, is
the constellation phase rotation experienced by the $i$th user;
$d_i^R \ge 0$ and $d_i^I \ge 0$ describe the constellation range;\footnote{It is more accurate to say that $2(2L-1)d_i^R$ and $2(2L-1)d_i^I$ describe the constellation range, as seen in Figure~\ref{figd}, but we will call $d_i^R$ and $d_i^I$ the constellation range for the sake of convenience.} see Figure~\ref{figd}.
For notational conciseness, let us rewrite \eqref{eq:approx} as
\begin{equation}\label{eq:goal_vec}
\bH \bx_t \approx \bvarphi \circ (\bd \diamond \bs_t), \quad t=1,\dots,T,
\end{equation}
where
$\bH \!=\! [\bh_1,\dots,\bh_K]^H$ is the channel matrix;
$\bvarphi \!=\! [\varphi_1,\dots,\varphi_K]^T$ is the constellation phase rotation vector;
$\circ$ denotes the Hadamard product;
$\bd \!=\! [d_1,\dots,d_K]^T$, with $d_i = d_i^R + \jj d_i^I$, represents the constellation range vector;
$\bd \diamond \bs$ means that $[\bd \diamond \bs]_i = d_i^R \Re(s_i) + \jj d_i^I \Im(s_i)$ for all $i$;
$\bs_t = [s_{1,t},\dots,s_{K,t}]^T$ is the symbol vector at time $t$.
Our aim is to design $\bx_1,\dots,\bx_T$, as well as the constellation phase $\bvarphi$ and range $\bd$, such that a good approximation of \eqref{eq:goal_vec}, as indicated by some metric, will be yielded.
We will call such attempt symbol shaping in the sequel.

\subsection{Linear Precoding}

To provide intuition,  we first review how linear precoding performs symbol shaping.
In linear precoding, the transmitted signal $\bx_t$ takes the form
\begin{equation}\label{eq:LB}
\bx_t =  \sum_{i=1}^K \bw_i s_{i,t},
\end{equation}
where $\bw_i\in \Cbb^{N}$ is the precoding or beamforming vector of the $i$th user.
The noise-free part of the received signal $y_{i,t}$ is then given by
\[
\bh_i^H \bx_t = \bh_i^H \bw_i s_{i,t} + \sum_{j\neq i} \bh_i^H \bw_j s_{i,t},
\]
where $\bh_i^H \bw_i s_{i,t}$ is the desired symbol scaled by $\bh_i^H \bw_i$, and $\sum_{j\neq i} \bh_i^H \bw_j s_{i,t}$ is the multiuser interference (MUI).
In linear precoding, the MUI is often treated as Gaussian noise.
Also, the beamforming vectors are typically designed to maximize some utility defined over a certain quality-of-service (QoS) metric, e.g., the signal-to-interference-and-noise ratio (SINR)
\[
{\sf SINR}_i = \frac{ \rho | \bh_i^H \bw_i |^2}{ \sum_{j \neq i} \rho | \bh_i^H \bw_j |^2 + \sigma_v^2},
\]
where $\rho = \Exp[|s_{i,t}|^2]$ is the average symbol power;
or, power minimization under some target QoS requirements is sought.
The reader is referred to the literature~\cite{Bengtsson2001,schubert2004solution,wiesel2005linear,Yu2007Transmitter,liu2013max,bjornson2014optimal,shi2011iteratively,shi2016sinr} and the references therein for details.
Such QoS metric often ignores the constellation structure; the SINR defined above is an example.
On the other hand, from the perspective of  symbol shaping, the MUI $\sum_{j\neq i} \bh_i^H \bw_j s_{i,t}$ is seen as the approximation error in~\eqref{eq:approx};
$\varphi_i= e^{\jj \angle (\bh_i^H \bw_i)}$ is seen as the constellation phase rotation;
$d_i = (1+ \jj) |\bh_i^H \bw_i|$ is seen as the constellation range.

\subsection{SLP and Symbol Error Probability Characterization}

In symbol-level precoding (SLP), we attain symbol shaping by allowing the transmitted signals $\bx_t$'s to take any form to optimize certain constellation-dependent QoS metrics.
To put into context, consider the symbol error probability (SEP) as our QoS metric.
Assume that the users detect the symbols by the standard decision rule
\begin{align}\label{dec}
\hat{s}_{i,t} = \dec \Big( \frac{ \Re(\varphi_i^*  y_{i,t}) }{d_i^R}  \Big) + \jj \cdot \dec \Big( \frac{\Im(\varphi_i^* y_{i,t})}{d_i^I} \Big),
\end{align}
where $\dec(\cdot)$ denotes the decision function corresponding to $\{ \pm 1, \pm 3, \ldots, \pm (2L - 1) \}$.
Here, we assume that each user knows its corresponding constellation phase rotation $\varphi_i$ and range $d_i$;
the users can acquire them during the training phase.
The SEPs are given by
\begin{equation}\label{eq:overline_sep}
\overline{\sf SEP}_i = \frac{1}{T}\sum_{t=1}^T \underbrace{ {\rm Pr}( \hat{s}_{i,t} \neq s_{i,t} \mid s_{i,t} )}_{\triangleq~ {\sf CSEP}_{i,t}},
\end{equation}
where $\overline{\sf SEP}_i$ is the SEP of the $i$th user;\footnote{ Note that, under the assumption of independent and identically distributed $s_{i,t}$'s, we have $\overline{\sf SEP}_i \to \mathbb{E}_{s_{i,t}}[{\rm Pr}( \hat{s}_{i,t} \neq s_{i,t} )]$ as $T \to \infty$.}
${\sf CSEP}_{i,t}$ is the SEP of $\hat s_{i,t}$ conditioned on $s_{i,t}$.
We are particularly interested in making sure that every $\overline{\sf SEP}_i$ will meet, or be better than, a given value $\eps_i > 0$; i.e.,
\[
\overline{\sf SEP}_i \le \eps_i, \quad i=1,\dots,K.
\]
Dealing with the above SEP quality constraints is difficult, and as a compromise we consider
\begin{equation}\label{eq:SEP_const}
{\sf CSEP}_{i,t}  \leq \eps_i, \quad i=1,\dots,K,\ t=1,\dots,T,
\end{equation}
which will guarantee $\overline{\sf SEP}_i \le \eps_i$.

The SEP quality guarantees in \eqref{eq:SEP_const} can be turned to some more convenient forms.
Before we present it, we want to provide the intuition.
Consider the following example.
\begin{Exa}\label{exa:SEP}
The intuition is best illustrated by reducing the problem to the real-valued case;
i.e., $\bh_i$, $\bs_t$, $\bd$ and $\bx_t$ are real-valued; $\bvarphi = \bone$; the constellation is $\{\pm 1,\pm 3,\dots,\pm (2L-1)\}$;
$\hat s_{i,t} = \dec (y_{i,t}/d_i)$;
$v_{i,t} \sim \mathcal{N}(0,\sigma_v^2)$.
It can be shown that
\begin{equation*}
\begin{aligned}
{\sf CSEP}_{i,t} &  \left\{
\begin{array}{ll}
\!\!\leq  2 Q \Big( \frac{ d_i - |\bh_i^T \bx_t - d_i s_{i,t}| }{\sigma_v} \Big), &\! | s_{i,t} | < 2L - 1  \\[2ex]
\!\!= Q \Big( \frac{ d_i + (\bh_i^T \bx_t - d_i s_{i,t}) }{\sigma_v} \Big), &\! s_{i,t} = 2L - 1 \\[2ex]
\!\!= Q \Big( \frac{ d_i - (\bh_i^T \bx_t - d_i s_{i,t}) }{\sigma_v} \Big), &\! s_{i,t} = -2L + 1
\end{array}
\right.,
\end{aligned}
\end{equation*}
where $Q(x) = \int_{x}^{\infty} \frac{1}{\sqrt{2\pi}} e^{-z^2 / 2} dz$;
see, e.g., \cite{liu2018symbol,shao2019framework}.
Figure~\ref{fig_sep_illu} shows an illustration of how ${\sf CSEP}_{i,t}$ is derived.
Applying the above expression to \eqref{eq:SEP_const}, the SEP quality guarantees in~\eqref{eq:SEP_const} are satisfied if
\begin{equation*}
\begin{aligned}
\left\{
\begin{array}{ll}
\!\! |\bh_i^T \bx_t - d_i s_{i,t}| \le d_i-\textstyle{\sigma_v} Q^{-1}( \frac{\eps_i}{2} ), &\! | s_{i,t} | < 2L - 1  \\[2ex]
\!\!\bh_i^T \bx_t - d_i s_{i,t} \ge \textstyle{\sigma_v} Q^{-1}(\eps_i) - d_i, &\! s_{i,t} = 2L - 1 \\[2ex]
\!\!\bh_i^T \bx_t - d_i s_{i,t} \le d_i - \textstyle{\sigma_v} Q^{-1}(\eps_i), &\! s_{i,t} = -2L + 1
\end{array}
\right.
\end{aligned}
\end{equation*}
In particular, observe that the above inequalities are linear with respect to (w.r.t.) $\bx_t$ and $\bd$---what we meant by convenient.\hfill $\blacksquare$
\begin{figure}
	\centering
    \includegraphics[width=0.7\linewidth]{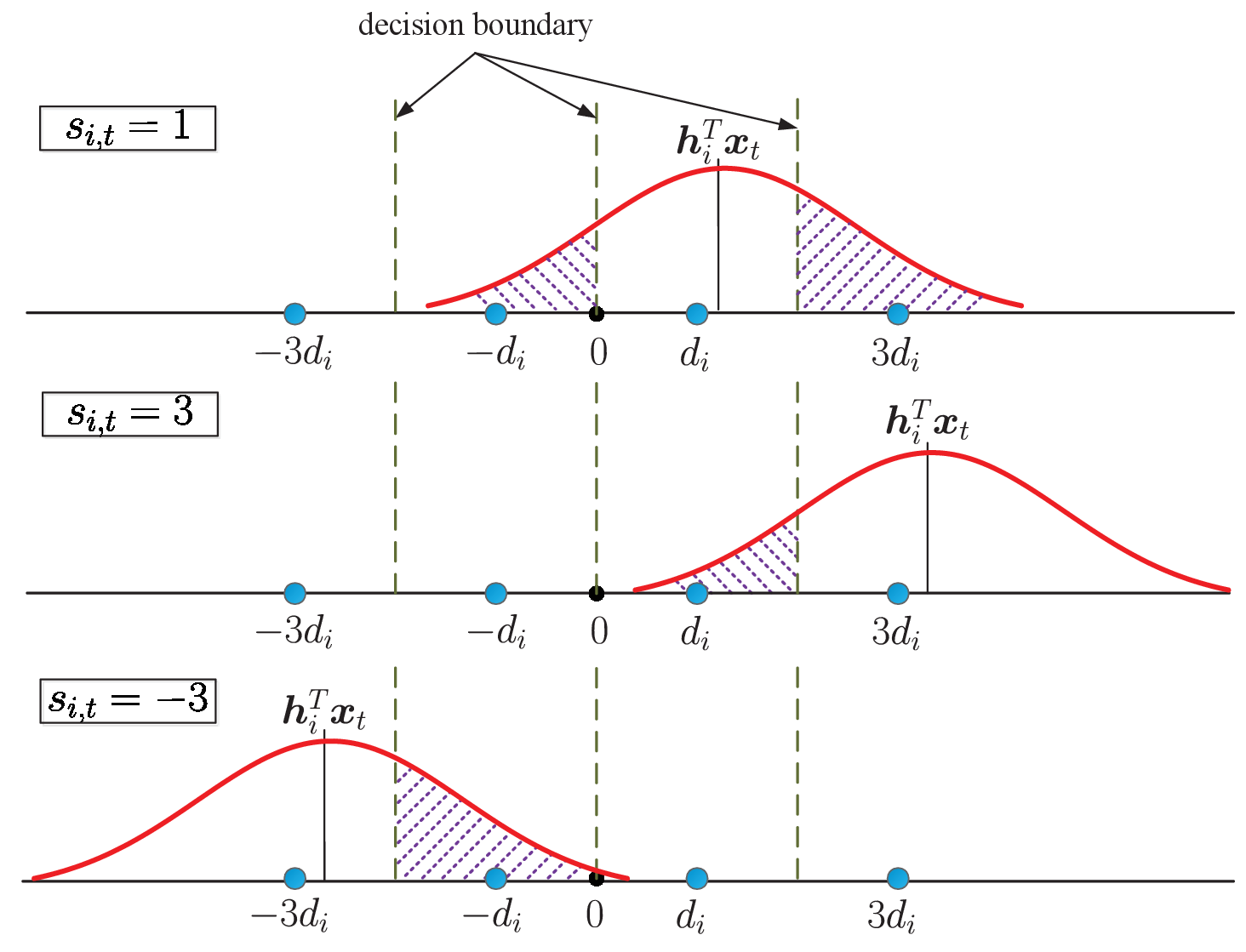}
	\caption{{Illustration of the conditional SEP for $\setS = \{\pm 1, \pm 3\}$. According to \eqref{eq:sig_mod}, $y_{i,t}$ is ${\cal N}(\bh_i^T \bx_t, \sigma_v^2)$ distributed. The shaded area corresponds to the conditional SEP in~\eqref{eq:overline_sep}.}}
	\label{fig_sep_illu}
\end{figure}
\end{Exa}
By taking the above idea in Example \ref{exa:SEP} to the complex-valued case, we get the following result.
\begin{Fact}\label{fac:SEP}
The SEP quality guarantees in \eqref{eq:SEP_const} hold for all $i$ if
\begin{equation}\label{eq:SEP_const1}
-\bd + \ba_{t} \leq_c  \bvarphi^* \circ (\bH\bx_t)- \bd \diamond \bs_t   \leq_c \bd - \bc_{t},
\end{equation}
where
$\bx \ge_c \by$ means that $\Re(\bx) \ge \Re(\by)$, $\Im(\bx) \ge \Im(\by)$;
$\ba_t \!\! =\!\! [ a_{1,t}, \ldots, a_{K,t} ]^T$, $a_{i,t} \!\!=\!\! a_{i,t}^R \!+\! \jj a_{i,t}^I$, $\bc_t \!\!=\!\! [ c_{1,t}, \ldots, c_{K,t} ]^T$, $c_{i,t} = c_{i,t}^R + \jj c_{i,t}^I$;
\begin{equation}\label{eq:ac}
\begin{aligned}
a_{i,t}^R & = \left\{
\begin{array}{ll}
\alpha_i, & | \Re(s_{i,t}) | < 2L - 1  \\
\beta_i, & \Re(s_{i,t}) = 2L - 1 \\
-\infty, & \Re(s_{i,t}) = -2L + 1
\end{array}
\right.  \\
c_{i,t}^R & = \left\{
\begin{array}{ll}
\alpha_i, & | \Re(s_{i,t}) | < 2L - 1  \\
-\infty, & \Re(s_{i,t}) = 2L - 1 \\
\beta_i, & \Re(s_{i,t}) = -2L + 1
\end{array}
\right.
\end{aligned}
\end{equation}
and
\begin{equation}\label{eq:alpha_beta}
\textstyle \alpha_i  =   \frac{\sigma_v}{\sqrt{2}} Q^{-1}\Big( \frac{1 - \sqrt{1- \eps_i}}{2}  \Big),~
\beta_i =  \frac{\sigma_v}{\sqrt{2}} Q^{-1} ( 1  -  \sqrt{1- \eps_i});
\end{equation}
$a_{i,t}^I$ and $c_{i,t}^I$ are defined by the way as $a_{i,t}^R$ and $c_{i,t}^R$ in \eqref{eq:ac}, specifically, by replacing ``$R$'' with ``$I$'' and ``$\Re$'' with ``$\Im$''.
\end{Fact}
We relegate the proof of Fact~\ref{fac:SEP} to Appendix~\ref{sec:proof_fact1}.

Intuitively, the constellation range $\bd$ should not be too small in order to achieve certain SEP guarantees.
To quantify that, consider the following assumption:
\begin{Asm}\label{Asm2}
The QAM order $L$ (cf. \eqref{eq:QAM_cons}) has $L\geq 2$; i.e., high-order and non-constant modulus QAM cases.
Each user's symbol stream $s_{i,1},\dots,s_{i,T}$ has at least one symbol $s_{i,t}$ such that $| \Re(s_{i,t}) | < 2L - 1$ and $| \Im(s_{i,t}) | < 2L - 1$; that is, $s_{i,t}$ is an inner constellation point (ICP) of the QAM constellation.
\end{Asm}
We have the following result.
\begin{Fact}\label{fac:d}
Suppose that Assumption~\ref{Asm2} holds.
Any constellation range $\bd$ satisfying the SEP quality guarantees~\eqref{eq:SEP_const1} must satisfy
\[
\bd \ge_c \balpha_c,
\]
where $\balpha_c = \balpha + \jj \balpha$; the $\alpha_i$'s were defined in~\eqref{eq:alpha_beta}.
\end{Fact}
\noindent{\em Proof:}
Suppose that $s_{i,t}$ is an ICP.
From~\eqref{eq:SEP_const1} we see that $-d_i + \alpha_{c,i} \le_c d_i - \alpha_{c,i}$, which reduces to $d_i \ge_c \alpha_{c,i}$.
\hfill $\blacksquare$


\section{A New Look at SLP}
\label{sec:SLP_TTP}

In this section, we introduce a new way to represent SLP, which will enable us to link SLP with linear precoding.

\subsection{Precoding via the Lens of Zero-Forcing}
\label{sec:slp_zf}

Let us make the following assumption.
\begin{Asm}\label{Asm1}
The channel matrix $\bH$ has full row rank.
\end{Asm}
The following result will be key to our developments.

\begin{Fact}\label{fac:rep}
Suppose that Assumption \ref{Asm1} holds.
Let $\bs_t \in \Cbb^K$, $\bd \in \Cbb^K$ and $\bvarphi \in \Cbb^K$, with $|\bvarphi| = \bone$, be given.
Any $\bx_t \in \Cbb^N$ can be represented by
	\begin{equation}\label{rep_x}
	\bx_t = \bH^\dag ( \bvarphi \circ ( \bd \diamond \bs_{t}  + \bu_t ) ) +  \bB \bz_t,
	\end{equation}
for some $\bu_t\in \Cbb^K$ and $\bz_t \in \Cbb^{N-K}$,
where $\bH^\dag \triangleq \bH^H (\bH \bH^H )^{-1}$ is the pseudo-inverse of $\bH$, and $\bB\in \Cbb^{N \times (N-K)}$ is a basis matrix for the nullspace of $\bH$.
Also, under the representation~\eqref{rep_x}, we can equivalently represent the SEP quality guarantee \eqref{eq:SEP_const1} in Fact~\ref{fac:SEP} by
\begin{equation}\label{eq:SEP_const2}
 -\bd + \ba_t \leq_c  \bu_t  \leq_c \bd - \bc_t.
\end{equation}
\end{Fact}

\noindent{\em Proof:} \
Let $\setR \subseteq \Cbb^N$ denote the range space of $\bH^H$.
Let $\setR^\perp$ be the orthogonal complement of $\setR$, which is also the nullspace of $\bH$.
Any $\bx_t \in \Cbb^N$ can be decomposed into $\bx_t = \bar{\bx}_t + \tilde{\bx}_t$, where $\bar{\bx}_t \in \setR$ and $\tilde{\bx}_t \in \setR^\perp$.
By letting $\bB$ be a basis matrix for $\setR^\perp$, we can represent $\tilde{\bx}_t$ by $\tilde{\bx}_t = \bB \bz_t$ for some $\bz_t$.
In the same vein, we can write $\bar{\bx}_t = \bH^H\br_t$ for some $\br_t$.
Let $\bs_t$, $\bd$, $\bvarphi$ ($|\bvarphi| = \bone$) be given.
Choose $\bu_t$ such that
\begin{equation}\label{eq:rt}
\br_t = (\bH \bH^H )^{-1}(\bm \varphi \circ (\bd \diamond\bs_t+ \bu_t)),
\end{equation}
or equivalently,
\begin{equation*}\label{eq:ut}
\bu_t = \bvarphi^{*} \circ ( \bH \bH^H \br_t ) -  \bd \diamond \bs_{t}.
\end{equation*}
Putting \eqref{eq:rt}, $\bar{\bx}_t = \bH^H\br_t$ and $\tilde{\bx}_t = \bB \bz_t$ into $\bx_t = \bar{\bx}_t + \tilde{\bx}_t$ gives the representation in \eqref{rep_x}.
Furthermore, putting \eqref{rep_x} into~\eqref{eq:SEP_const1} gives the result in \eqref{eq:SEP_const2}.
\hfill $\blacksquare$

\medskip

Fact~\ref{fac:rep} shows two key revelations.
Firstly, an SLP scheme is equivalent to a zero-forcing (ZF) scheme with a symbol perturbation $\bu_t$ and a nullspace perturbation $\bB \bz_t$.
From that point of view we can regard SLP as instances of ZF, with suitable perturbations.
The most obvious one is the traditional ZF scheme $\bx_t^{\sf ZF} = \bH^\dag ( \bd \diamond \bs_{t} ) $ itself, which is an instance of \eqref{rep_x} with $\bu_t = \bzero$, $\bz_t = \bzero$ and $\bvarphi = \bone$.
Secondly, we see from \eqref{eq:SEP_const2} that the SEP quality guarantee~\eqref{eq:SEP_const1} depends only on the symbol perturbation component $\bu_t$ and the constellation range $\bd$.
This result will substantially simplify our designs.

\subsection{SLP is Symbol-Perturbed ZF}

Let us further examine the implications of the SLP-ZF relationship in Fact~\ref{fac:rep} by considering SLP designs.
Consider an SLP design that minimizes the total transmission power (TTP) under the SEP quality guarantee in~\eqref{eq:SEP_const1}; i.e.,
\ifconfver
\begin{equation} \label{eq:TTP_final}
\begin{aligned}
\min_{  \bd, \bvarphi, \bX }
& ~  \frac{1}{T} \sum_{t=1}^T \| \bx_t \|_2^2 \\
{\rm s.t.} ~& \!\! - \!\bd \!+\! \ba_{t} \!\leq_c \! \bvarphi^* \!\!\circ \!(\bH\bx_t)\!-\! \bd \diamond \bs_t  \! \leq_c\! \bd \!- \!\bc_{t}, ~\! t =1,\dots,T,\\
& ~  \bd \ge_c \balpha_c, \quad |\bvarphi| = \bm 1,
\end{aligned}
\end{equation}
\else
\begin{equation} \label{eq:TTP_final}
\begin{aligned}
\min_{  \bd, \bvarphi, \bX }
& ~  \frac{1}{T} \sum_{t=1}^T \| \bx_t \|_2^2 \\
{\rm s.t.} ~& -\bd + \ba_{t} \leq_c \bvarphi^* \circ (\bH\bx_t)- \bd \diamond \bs_t  \leq_c \bd- \bc_{t}, ~ t =1,\dots,T,\\
& ~ \bd \ge_c \balpha_c, \quad |\bvarphi| = \bm 1,
\end{aligned}
\end{equation}
\fi
where $\bX = [ \bx_1,\dots,\bx_T ]$;
note that we jointly optimize the transmitted signal $\bX$ and the received constellation phase $\bvarphi$ and range $\bd$, and that the constraint $\bd \ge_c \balpha_c$ is due to Fact~\ref{fac:d}.
Using the alternative SLP representation in Fact~\ref{fac:rep}, we have the following result.

\begin{Prop} \label{prop:slp_zf}
	Suppose that Assumption \ref{Asm1} holds.
	Then an optimal $ \bX^\star = [ \bx_1^\star,\dots,\bx_T^\star ]$ to Problem~\eqref{eq:TTP_final} is given by
	\begin{equation*}\label{eq:SLP_pert_ZF}
	\bx_t^\star =  \bH^\dag ( \bvarphi^\star \circ ( \bd^\star \diamond \bs_{t}  + \bu_t^\star ) ), \quad t = 1,\dots,T,
	\end{equation*}
	where $( \bd^\star, \bvarphi^\star, \bU^\star)$, $\bU^\star= [\bu_1^\star, \dots, \bu_T^\star]$, is an optimal solution to
	\begin{equation} \label{eq:TTP_ref}
	\begin{aligned}
	\min_{ \bd, \bvarphi, \bU }
	& ~  \frac{1}{T} \sum_{t=1}^T \|  \bd \diamond \bs_{t}  + \bu_t  \|_{\bR_{\bvarphi}}^2 \\
	{\rm s.t.}
	& ~ -\bd + \ba_t \leq_c  \bu_t  \leq_c \bd - \bc_t, \quad t=1,\dots,T, \\
	& ~ \bd \ge_c \balpha_c,\quad  |\bvarphi| = \bm 1,
	\end{aligned}
	\end{equation}
	with $\bR_{\bvarphi} = {\rm Diag}(\bvarphi)^H \bR {\rm Diag}(\bvarphi)$; $\bR = (\bH \bH^H )^{-1}$;   $\| \bx \|_\bR \triangleq \sqrt{ \bx^H \bR \bx }$.
\end{Prop}

\noindent{\em Proof:}
Substituting~\eqref{rep_x} into
Problem~\eqref{eq:TTP_final} gives
\begin{equation*}
\begin{aligned}
\min_{ \bd, \bvarphi,\{\bu_t, \bz_t\}_{t=1}^T}
& ~   \frac{1}{T} \sum_{t=1}^T \big( \|  \bd \diamond \bs_{t}  + \bu_t  \|_{\bR_{\bvarphi}}^2 + \| \bB \bz_t \|_2^2 \big) \\
{\rm s.t.}
& ~ -\bd + \ba_t \leq_c  \bu_t  \leq_c \bd - \bc_t, \quad t=1,\dots,T, \\
& ~ \bd \ge_c \balpha_c ,\quad |\bvarphi| = \bm  1.
\end{aligned}
\end{equation*}
We see that any optimal solution to the above problem must have $\| \bB \bz_t \|_2^2 = 0$, or equivalently, $\bz_t = \bzero$, for all $t$.
The proof is complete.
\hfill $\blacksquare$

\medskip

Proposition~\ref{prop:slp_zf} indicates that the optimal SLP scheme under the TTP minimization design~\eqref{eq:TTP_final} is a symbol-perturbed ZF scheme, with the nullspace components being shut down.

\subsection{ZF is a Near-Optimal SLP for Very Large QAM Sizes}
\label{sec:ZF_SLP}

We showed in the preceding subsection that the optimal SLP scheme under the TTP minimization design~\eqref{eq:TTP_final} is a symbol-perturbed ZF scheme.
In fact, we can even show that the optimal SLP scheme reduces to the basic ZF scheme---without symbol perturbations---under certain assumptions.
Let us set the stage by assuming the following:
\begin{Asm}\label{Asm3}
The symbols $s_{i,t}$'s are independently and identically distributed (i.i.d.) and are uniformly distributed on the QAM constellation $\setS$.
\end{Asm}
\begin{Asm}\label{Asm4}
The transmission block length $T$ tends to infinity.
\end{Asm}

The SLP design problem~\eqref{eq:TTP_ref} under Assumptions~\ref{Asm2}--\ref{Asm4} can be written as
\begin{equation}\label{eq:TTP_thm}
\begin{aligned}
f_{\sf SLP} =
\min_{\bd \ge_c \balpha_c, |\bvarphi| = \bone }
& ~ g(\bd,\bvarphi),
\end{aligned}
\end{equation}
where
\begin{align*}
g(\bd,\bvarphi) &= \mathbb{E}_{\bs_t} \Big[ \min_{ \bu_t \in \setU(\bs_t,\bd) }\| \bd \diamond \bs_{t}  + \bu_t  \|_{\bR_\bvarphi}^2 \Big],\\
\setU(\bs_t,\bd) &= \setU(s_{1,t},d_1) \times \cdots \times \setU(s_{K,t},d_{K}), \\
\setU(s_{i,t},d_i)  &=  \{ u_i \in \Cbb | -d_i + a_{i,t} \leq_c  u_i  \leq_c d_i - c_{i,t} \}.
\end{align*}
Let
\begin{equation}\label{eq:ZF}
\bx^{\sf ZF}_t =\bH^\dag (  \balpha_c \diamond \bs_{t} )
\end{equation}
be our benchmark ZF scheme.
Note that the ZF scheme~\eqref{eq:ZF} is a feasible solution to Problem~\eqref{eq:TTP_thm}, with $\bd = \balpha_c$, $\bvarphi = \bone$ and $\bu_t = \bzero$.
Our result is as follows.

\begin{Theorem}\label{Them1}
    Suppose that Assumptions~\ref{Asm2}--\ref{Asm4} hold.
	Also, suppose that  $\eps_1 = \cdots = \eps_K = \eps$.
	Then the optimal value $f_{\sf SLP}$ of Problem~\eqref{eq:TTP_thm} satisfies
	\begin{equation*}
	\kappa f_{{\sf ZF}} \le f_{\sf SLP}
	\le  f_{{\sf ZF}} ,
	\end{equation*}
	where
    $f_{{\sf ZF}}  =  \mathbb{E}_{\bs_t}\big[\big\|\bx_t^{\sf ZF}\big\|_2^2 \big] $ is the  TTP of ZF, and
    \begin{equation*}
	\begin{aligned}
	\kappa &=\Big( 1 - \frac{1}{L} \Big)^{2K} \frac{2L-3}{2L+1}   \frac{(2L-1)(2L-3) -3}{(2L-1)(2L-3) -3 + \frac{3\lambda_{\rm max}(\bR)}{\lambda_{\rm min}(\bR)}}.
	\end{aligned}
	\end{equation*}
\end{Theorem}

The proof of Theorem~\ref{Them1} is shown in Appendix~\ref{sec:proof_thm1}.
Theorem~\ref{Them1} suggests that the TTP ratio between SLP and ZF is lower bounded by $\kappa~(\kappa < 1)$.
In particular, $\kappa$ increases as $L$ increases, and $\kappa \to 1$ as $L \to \infty$.
This leads to the following important conclusion:

\begin{Corollary}\label{Cor1}
Under Assumptions~\ref{Asm2}--\ref{Asm4},
the optimal SLP scheme under the TTP minimization design~\eqref{eq:TTP_ref} approaches the ZF scheme~\eqref{eq:ZF} as the QAM size tends to infinity.
\end{Corollary}

Corollary~\ref{Cor1} suggests that, for very high-order QAM, we may simply use ZF.
It explains why we have not seen a numerical result that shows significant gains with SLP for very high-order QAM; see, e.g.,~\cite{alodeh2017symbol}.
Our numerical results will illustrate that the ZF scheme is indeed near-optimal for very large $L$.
On the other hand, our numerical results will also indicate that, for smaller $L$, the optimal SLP scheme can have significant TTP reduction over the ZF scheme.

\section{SLP Schemes for TTP Minimization}
\label{sec:SLP_schemes}

We now turn to the aspect of tackling the TTP minimization SLP design~\eqref{eq:TTP_ref}.
Let us recapitulate Problem \eqref{eq:TTP_ref}:
    \begin{equation} \label{eq:TTP_recap}
    \begin{aligned}
    \min_{ \bd, \bU, \bvarphi }
    & ~f_{\sf TTP}(\bd,\bU,\bvarphi) \triangleq  \frac{1}{T} \sum_{t=1}^T \|  \bd \diamond \bs_{t}  + \bu_t  \|_{\bR_{\bvarphi}}^2 \\
    {\rm s.t.}
    & ~ -\bd + \ba_t \leq_c  \bu_t  \leq_c \bd - \bc_t, \quad t=1,\dots,T, \\
    & ~ \bd \ge_c \balpha_c,\quad  |\bvarphi| = \bm 1.
    \end{aligned}
    \end{equation}
We should briefly mention the problem nature.
Problem~\eqref{eq:TTP_recap} is  a large-scale problem since $T$ is large in practice, say, a few hundreds.
The objective function of~\eqref{eq:TTP_recap}
is convex w.r.t. either $\bvarphi$ or $(\bd,\bU)$, but not w.r.t. both.
Also, the unit-modulus constraint $|\bvarphi| = \bone$ is non-convex.

\subsection{Alternating Minimization over $(\bd,\bU)$ and $\bvarphi$}
\label{sec:TTP_AM}

We tackle Problem \eqref{eq:TTP_recap}  in an approximate fashion by alternating minimization (AM).
Specifically, we alternatingly minimize the objective function over $(\bd,\bU)$ and $\bvarphi$:
\begin{subequations}\label{eq:AM}
\begin{align}
    \bvarphi^{k+1} \in \arg\min_{\bvarphi \in {\cal P}} ~& f_{\sf TTP}(\bd^k,\bU^k,\bvarphi), \label{eq:AM1}\\
    (\bd^{k+1}, \bU^{k+1}) \in \arg\min_{(\bd, \bU) \in \setW } ~& f_{\sf TTP}(\bd,\bU,\bvarphi^{k+1}),\label{eq:AM2}
\end{align}
\end{subequations}
where
\[
{\cal P} = \{\bvarphi \in \Cbb^K | |\bvarphi| =\bm 1 \},
\]
\begin{align*}
\setW\! =\! \{(\bd, \bU) \in \Cbb^K \times \Cbb^{K \times T}| -\bd + \ba_t\! \leq_c \! \bu_t  \!\leq_c\! \bd - \bc_t,~ \forall t,  
~ \bd \ge_c\!  \balpha_c\}.
\end{align*}

Let us describe how the above minimizations are handled.
First, the problem in \eqref{eq:AM1} can be shown to be
\begin{equation}\label{eq:phi}
\min_{\bvarphi \in {\cal P}}~ \bvarphi^H \bar{\bR} \bvarphi,\quad
\end{equation}
where
$\bar{\bR} = \frac{1}{T} \sum_{t=1}^T \Diag( \bd^k \diamond \bs_{t}  +  \bu_t^k )^H \bR \Diag( \bd^k \diamond \bs_{t}  +  \bu_t^k )$.
Problem \eqref{eq:phi} is a unit-modulus quadratic program;
it is non-convex, but in practice it can be efficiently approximated by a variety of methods, such as
semidifinite relaxation~\cite{luo2010semidefinite} and the proximal gradient (PG) method~\cite{boumal2016nonconvex ,tranter2017fast,attouch2013convergence}.
We choose the PG method to approximate Problem~\eqref{eq:phi}, and the method is shown in Algorithm~\ref{AL_PG}.
Note that $\langle \cdot,\cdot \rangle = \Re(\bx^H \by)$ is the inner product;
$\nabla f$ is the gradient of $f$;\footnote{ Since $f$ deals with complex variables, we define  $\nabla f(\bx) = \nabla_{\Re(\bx)} f(\bx) + \jj \, \nabla_{\Im(\bx)} f(\bx)$ where $\nabla_{\Re(\bx)} f(\bx)$ and $\nabla_{\Im(\bx)} f(\bx)$ are the gradients w.r.t. the real and imaginary parts of $\bx$, respectively.}
$\Pi_{\setX}(\bx) \in \arg\min_{\by \in \setX}\|\bx-\by\|_2^2$ denotes a projection of $\bx$ onto $\setX$.
Also, we have
\begin{equation*}
\by =\Pi_{{\cal P}} (\bx) \Leftrightarrow y_i =
\begin{cases}
{x_i}/{|x_i|},\quad &  x_i\neq 0\\
\mbox{any $x$ with $|x|=1$}, \quad & x_i = 0
\end{cases} 
\end{equation*}
The PG method is guaranteed to converge to a critical point (under some assumptions)~\cite{attouch2013convergence};
we discuss the details in the supplemental material of this paper.
\begin{algorithm}
	\caption{PG method for handling  $\min f(\bvarphi)$ s.t. $\bvarphi \in {\cal P}$}
	\begin{algorithmic}[1]
\STATE given an initialization $\bvarphi^0$, $0 < \alpha < 1$
\STATE $\ell = 0$
\REPEAT
\STATE $\bm \varphi^{\ell+1} = \Pi_{{\cal P}} \left( \bm \varphi^{\ell} - \frac{\alpha}{L_\ell} \nabla f(\bm \varphi^{\ell}) \right)$;
$L_{\ell}$ is such that
\begin{align*}
\! \!\! \!\! \!\! \!\! \!\! \!
 f(\bm \varphi^{\ell+1}) &\le \! f(\bm \varphi^\ell )\! + \!\langle \nabla f(\bm \varphi^\ell ),\bm \varphi^{\ell+1} \! - \!\bm \varphi^\ell  \rangle \! + \! \frac{L_{\ell}}{2}\|\bm \varphi^{\ell+1}\! - \!\bm \varphi^\ell \|_2^2,
\end{align*}
which can be obtained by line search~\cite{beck2017first} or by setting $L_\ell$ as a Lipschitz constant of $\nabla f$ (assuming that it exists)
\STATE $\ell = \ell+1$
\UNTIL{some stopping rule holds}
	\end{algorithmic}\label{AL_PG}
\end{algorithm}

Second, the problem in~\eqref{eq:AM2} can be expressed as
    \begin{equation} \label{eq:TTP_nophase}
    \begin{aligned}
    \min_{ \bm \xi }
    & ~ \phi (\bm \xi)\triangleq\frac{1}{T} \sum_{t=1}^T \|  \bd \diamond \bs_{t}  + \bu_t  \|_{\bR_{\bvarphi}}^2, \quad
\mbox{s.t.}~ \bm \xi \in \setW,
    \end{aligned}
    \end{equation}
where $\bm \xi = (\bd, \bU)$; $\bvarphi=\bvarphi^{k+1}$.
Problem~\eqref{eq:TTP_nophase} is a convex quadratic program with linear constraints.
While we can call off-the-shelf convex optimization software, such as CVX~\cite{grant2008cvx}, to solve Problem~\eqref{eq:TTP_nophase}, it is computationally prohibitive to do so in practice---this is because Problem~\eqref{eq:TTP_nophase} is a large-scale problem.
Our solution is a custom-built one, leveraging on the structure of the constraints to improve the efficiency of solving Problem~\eqref{eq:TTP_nophase}.
We use the accelerated proximal gradient (APG) method for convex optimization~\cite{beck2017first}, shown in Algorithm~\ref{AL_APG}.
The APG method is known to converge to the optimal solution at a rate of $\mathcal{O}(1/\ell^2)$ (under some assumptions)~\cite{beck2017first}.

\begin{algorithm}[htb!]
	\caption{APG method for solving $\min \phi(\bm \xi)$ s.t. $\bm \xi \in \setW$, where $\phi$ and $\setW$ are convex.}
	\begin{algorithmic}[1]
		\STATE given an initialization $\bm \xi^0$
		\STATE $\ell = 0$, $\nu_{-1} = 0$, $\bm \xi^{-1} = \bm \xi^{0}$
        \REPEAT
\STATE $ \nu_{\ell} = \textstyle(1+\sqrt{1+4\nu_{\ell-1}^2})/{2}$
\STATE $\bp^\ell =  \bm \xi^{\ell} + \frac{\nu_{\ell-1}-1}{\nu_\ell} ( \bm \xi^{\ell} - \bm \xi^{\ell-1})$
\STATE $\bm \xi^{\ell+1} = \Pi_{\setW} (\bp^\ell - L_\ell^{-1} \nabla \phi(\bp^{\ell}))$; $L_{\ell}$ is such that
\begin{align*}
\!\!\!\!\!\!\!\! \phi(\bm \xi^{\ell+1}) \!\le \! \phi(\bp^\ell )\! + \! \langle \nabla \phi(\bp^\ell ),\bm \xi^{\ell+1}\!-\!\bp^\ell  \rangle \! + \! \frac{L_{\ell}}{2}\|\bm \xi^{\ell+1}\!-\!\bp^\ell \|_2^2,
\end{align*}
which can be obtained by line search~\cite{beck2017first} or by setting $L_\ell$ as a Lipschitz constant of $\nabla f$ (assuming that it exists)
\STATE $\ell = \ell+1$
\UNTIL{some stopping rule holds}
	\end{algorithmic}\label{AL_APG}
\end{algorithm}

The computational efficiency of APG hinges on whether the projection $\Pi_{\setW}$ can be computed easily.
Although the coupling of $\bd$ and $\bu_t$'s in the constraints makes the projection seemingly not too easy to compute, it turns out that $\Pi_{\setW}$ can be solved in a semi-closed form fashion.
Specifically, given a point $\tilde{\bm \xi} = (\tilde{\bd}, \tilde{\bU})$, the projection $\Pi_{\setW}(\tilde{\bm \xi})$ is to solve
\begin{equation} \label{eq:proj1}
\begin{aligned}
\min_{ \bd, \bU } & ~
\sum_{t=1}^T \|{\bu}_t - \tilde{\bu}_t\|_2^2 + \|{\bd} - \tilde{\bd}\|_2^2 \\
{\rm s.t.}
& ~ -\bd + \ba_t \leq_c  \bu_t  \leq_c \bd - \bc_t, \quad t=1,\dots,T, \\
& ~ \bd \ge_c \balpha_c.
\end{aligned}
\end{equation}
Observe that Problem~\eqref{eq:proj1} is separable w.r.t. each coordinate $i = 1,\dots,K$ and also w.r.t. the real and imaginary components.
Hence, solving Problem~\eqref{eq:proj1} amounts to solving $2K$ independent subproblems, and
all the subproblems share the same structure as follows
\begin{equation}\label{eq:proj2}
\begin{aligned}
\min_{ d, u_{1}, \ldots, u_{T} } & ~
\sum_{t=1}^T (u_{t} - \tilde{u}_{t})^2 + (d - \tilde{d})^2 \\
{\rm s.t.}
& ~ -d + a_{t} \leq  u_{t}   \leq d - c_{t}, \quad  t=1,\ldots,T, \\
& ~ d \geq \alpha.
\end{aligned}
\end{equation}
We outline how Problem \eqref{eq:proj2} is solved.
The idea is to first eliminate the variables $u_t$'s by plugging
the solutions of $u_t$'s given $d$ into \eqref{eq:proj2}.
The resulting problem for $d$ is to solve a series of one-dimensional quadratic programs over different intervals, which admit closed-form solutions.
By comparing all solutions of $d$ over all the intervals, the one that gives the smallest objective value is the projection solution.
We show the projection solution in Algorithm~\ref{AL_pro} and relegate the mathematical details to  Appendix \ref{app:alg1_proof}.

\begin{algorithm}[htb!]
	\caption{A fast solution to Problem~\eqref{eq:proj2}}
	\begin{algorithmic}[1]
		\STATE
		{\bf input:} $[~ \tilde{u}_1, \cdots, \tilde{u}_{T}, \tilde{d} ~]$.
		
		\STATE set $ \setD_1 \triangleq \{ {c}_{t} + \tilde{u}_{t} ~\big|~ {c}_{t} + \tilde{u}_{t} \ge {\alpha}, ~\forall t \}$.
		\STATE set $\setD_2 \triangleq \{ {a}_{t} - \tilde{u}_{t} ~\big|~ {a}_{t} - \tilde{u}_{t} \ge {\alpha}, ~\forall t \}$.
		\STATE set $\tilde{\setD} \triangleq \{{\alpha}\} \cup \setD_1 \cup \setD_2 \cup \{+\infty\}$.
		\STATE sort the elements of $\tilde{\setD}$ in an ascending order to obtain $\setD \triangleq\{\omega_1,\dots,\omega_{{\rm card}(\setD)}\}$.
		\STATE {\bf for} $p = 1,\cdots, {\rm card}(\setD)-1$
        \STATE \quad set ${\cal T}_p \triangleq \{ t ~\big|~ \omega_{p+1} \le {c}_{t} + \tilde{u}_{t} \} $.
		\STATE \quad set ${\cal L}_p \triangleq \{ t ~\big|~ \omega_{p+1} \le {a}_{t} - \tilde{u}_{t} \} $.
		\STATE \quad compute $ {d}^{p} = \max \{ \omega_{p}, \min\{ \omega_{p+1}, \hat{d}^{p}\} \}$, where
\begin{equation*}
\hat{d}^{p} = \frac{ \sum_{t\in {\cal T}_p} ({c}_{t} +\tilde{u}_{t}) + \sum_{t\in {\cal L}_p} ({a}_{t} - \tilde{u}_{t}) + \tilde{d}}{ 1 +  {\rm card}({\cal T}_p)  + {\rm card}({\cal L}_p) }.
\end{equation*}
		\STATE \quad compute
\begin{equation*}
\textstyle f^{p} \! = \!  \sum\limits_{t\in {\cal T}_p} ( {d}^{p}- {c}_{t} - \tilde{u}_{t})^2 + \! \sum\limits_{t\in {\cal L}_p} (- {d}^{p} + {a}_{t} - \tilde{u}_{t})^2 + ( {d}^{p}-\tilde{d})^2.
\end{equation*}
		\STATE {\bf end for}
		\STATE  compute $ d= {d}^{\tilde{p}}$, where $\tilde{p} = \arg \min_p f^{p}$.
		\STATE  compute $u_{t} = \max \{ - d + a_{t}, \min \{ \tilde{u}_{t},d - c_{t}  \}  \},~ \forall t$.
		\STATE  {\bf output:} $[~ {u}_1, \cdots, {u}_{T}, {d} ~]$.
	\end{algorithmic}\label{AL_pro}
\end{algorithm}

\begin{table}[htbp]
\centering
\captionsetup{justification=centering}
	\begin{tabular}{ c|c|c | c  }
\hline
  Subproblem  &    Gradient  &   Projection & \multirow{2}{*}{ Per-iteration Complexity}  \\
  and Method &   Calculation  &   Calculation &  \\
\hline\hline
PG for \eqref{eq:phi} & ${\cal O}(K^2)$  & ${\cal O}(K)$ & ${\cal O}(K^2)$ \\ \hline
APG for \eqref{eq:TTP_nophase} & ${\cal O}(K^2 T)$ & ${\cal O}(K T^2)$ & ${\cal O}(K^2 T + K T^2)$\\
\hline 
\end{tabular}
\caption{Computational complexity of AM.}\label{tb:computation}
\end{table}
Table~\ref{tb:computation} summarizes the per-iteration complexities of the PG method for Problem \eqref{eq:phi} and the APG method for Problem \eqref{eq:TTP_nophase}.
It is worth noting that the computations of the gradient and projection operations contribute to the main complexity.

\subsection{Does the Alternating Minimization Converge?}

A curious question is whether the AM method~\eqref{eq:AM} for Problem~\eqref{eq:TTP_recap} guarantees convergence to a critical point.
From a mathematical optimization viewpoint, this aspect is subtle.
AM is known to have provable critical-point convergence for a class of optimization problems that have convex constraints; see, e.g.,~\cite{razaviyayn2013unified}.
But our problem has unit modulus constraints $|\bvarphi|=\bone$, and this
makes the convergence analysis challenging.
It turns out that, by taking insight from the proximal AM framework in mathematical optimization~\cite{attouch2010proximal}, we can answer the question.
Simply speaking, by modifying the AM update~\eqref{eq:AM1} as
\[
\bvarphi^{k+1} \in \arg\min_{\bvarphi \in {\cal P}} ~ f_{\sf TTP}(\bd^k,\bU^k,\bvarphi) + \frac{\tau}{2}\|\bvarphi - \bvarphi^k\|_2^2,
\]
for some $\tau > 0$,
and by initializing the PG method for the above update with $\bvarphi^k$, we can show convergence to a critical point.
The result is quite technical, however, and we relegate it to the supplemental material of this paper.

On the other hand, we should note that the original AM method~\eqref{eq:AM} works well in our numerical study.

\subsection{A Suboptimal SLP Scheme}
\label{sec:semi_ZF}

We study a suboptimal, but computationally efficient, alternative of the above SLP design.
Specifically we follow the same AM method as in~\eqref{eq:AM}, but we prefix the constellation range as $\bd = \balpha_c$.
There are two reasons for this.
First, if we prefix the constellation range $\bd$, the TTP minimization problem in \eqref{eq:AM2}, or \eqref{eq:TTP_nophase}, will be decoupled into a multitude of per-symbol-time TTP minimization problems
\begin{equation}\label{eq:szf_tp_pro}
\begin{aligned}
\min_{\bu_t}
& ~ \| \bd \diamond \bs_t  + \bu_t \|_{\bR_{\bvarphi}}^2 \\
{\rm s.t.}
& ~ -\balpha_c + \ba_t \leq_c  \bu_t  \leq_c \balpha_c - \bc_t
\end{aligned}
,\mbox{ for } t=1,\dots,T,
\end{equation}
which are computationally much easier to solve than Problem~\eqref{eq:TTP_nophase}.
Second, by observing the objective function of \eqref{eq:szf_tp_pro}, it seems that reducing the constellation range $\bd$ should reduce the power.
This intuition drove us to choose the smallest, $\bd = \balpha_c$.
We support our intuition by the following result.
\begin{Fact}\label{fac:zf}
Consider the TTP minimization problem~\eqref{eq:TTP_recap} with the symbol perturbation $\bU$ prefixed as $\bU = \bzero$.
Suppose that Assumptions \ref{Asm2}--\ref{Asm4} hold.
Then an optimal solution to the aforementioned problem is $\bm d = \bm \alpha_c$, $\bvarphi = \bone$;
the corresponding SLP is the ZF scheme in~\eqref{eq:ZF}.
\end{Fact}
\noindent
The proof of Fact~\ref{fac:zf} is relegated to Appendix~\ref{app:factzf_proof}.
While we are unable to prove similar results when the symbol perturbations $\bU$ are present,
Fact~\ref{fac:zf} gives us the insight that $\bd = \balpha_c$ may be a reasonable choice.

Let us write down the above suboptimal SLP scheme.
\begin{subequations}\label{eq:AM_semizf}
\begin{align}
&\bvarphi^{k+1} \in \arg\min_{\bvarphi \in {\cal P}} ~ f_{\sf TTP}(\balpha_c,\bU^k,\bvarphi), \label{eq:AM_semizf1}\\
&\begin{aligned}
\bu_t^{k+1} = \arg\min_{\bu_t}
& ~ \| \balpha_c \diamond \bs_t  + \bu_t \|_{\bR_{\bvarphi^{k+1}}}^2 \\
{\rm s.t.}
& ~ -\balpha_c + \ba_t \leq_c  \bu_t  \leq_c \balpha_c - \bc_t,\label{eq:AM_semizf2}
\end{aligned}\\
 \notag&\mbox{ for } t=1,\dots,T.
\end{align}
\end{subequations}
We will call the above scheme the semi-ZF SLP scheme; the reason will be given later.
Every problem in \eqref{eq:AM_semizf2} is a convex quadratic program with simple bound constraints, and it can be efficiently solved in a variety of ways, e.g., by the active set method~\cite{stark1995bounded}, ADMM~\cite{boyd2011distributed}, and the APG method~\cite{beck2017first}.
We will use the APG method (c.f., Algorithm~\ref{AL_APG}) to solve \eqref{eq:AM_semizf2} when we implement the semi-ZF SLP scheme in the numerical simulation section.

\subsection{Relationship with the Existing SLP Solutions}
\label{sec:rel_exsit}

The semi-ZF SLP scheme in~\eqref{eq:AM_semizf} has strong connections with the existing SLP solutions.
We illustrate the connections by considering the real-valued case in Example~\ref{exa:SEP};
the complex-valued counterpart is just a notationally more complicated version, and we will omit it.
By examining the constraints of \eqref{eq:AM_semizf2}, we notice that \eqref{eq:AM_semizf2} can be written as
\begin{equation}\label{eq:szf_tp_pro1}
\begin{aligned}
\min_{\bu_t}
& ~ \| \balpha \circ \bs_t  + \bu_t \|_{\bR}^2 \\
{\rm s.t.}
& ~ u_{i,t}
\begin{cases}
=0,  &  |s_{i,t}|<2L-1\\
\ge \beta_i-\alpha_i,   & s_{i,t}=2L-1 \\
\le \alpha_i-\beta_i,   & s_{i,t}=-(2L-1)
\end{cases}, \ i=1,\dots,K;
\end{aligned}
\end{equation}
(as a minor note, $\alpha_i=\sigma_vQ^{-1}(\eps_i/2)$, $\beta_i=\sigma_vQ^{-1}(\eps_i)$).
Equation~\eqref{eq:szf_tp_pro1} gives the physical interpretation that, if $s_{i,t}$ is an ICP, we set the corresponding symbol perturbation $u_{i,t}$ as $0$; or, we perform ZF partially.
This is why we call the scheme semi-ZF SLP.
Problem~\eqref{eq:szf_tp_pro1} resembles the existing SLP solutions, which were derived from different formulations.

As a representative example, consider the constructive interference power minimization (CIPM) design~\cite{alodeh2017symbol} and the subsequent variant~\cite{krivochiza2017low}.
The idea there starts with achieving a set of signal-to-noise ratio (SNR) requirements
\[
\frac{ \mathbb{E}_{\bx_t}[|\bh_i^T \bx_t|^2] }{\sigma_v^2} \ge \zeta_i, \quad i=1,\dots,K,
\]
where $\zeta_i > 0$ is the SNR target of the $i$th user.
The idea is then turned to the symbol level, giving rise to the following design formulation
\begin{equation}\label{eq:exit_work1}
\begin{aligned}
\min_{\bx_t}
& ~ \| \bx_t \|_2^2 \\
{\rm s.t.}
& ~ \frac{\bh_i^T \bx_t}{\sigma_v}
\begin{cases}
=\sqrt{\frac{\zeta_i}{\rho}} s_{i,t},  &  |s_{i,t}|<2L-1\\
\ge \sqrt{\frac{\zeta_i}{\rho}} s_{i,t},  & s_{i,t}=2L-1 \\
\le \sqrt{\frac{\zeta_i}{\rho}} s_{i,t},   & s_{i,t}=-(2L-1)
\end{cases}, \  i=1,\dots,K.
\end{aligned}
\end{equation}
Here, recall that $\rho = \mathbb{E}[|s_{i,t}|^2]$ is the average symbol power.
In particular, the authors of CIPM applied the constructive interference (CI) notion, i.e., pushing symbols deeper into the correct decision regions, by applying it on outer constellation points (OCPs) only.

The subsequent variant of the CIPM design in~\cite{krivochiza2017low} plugs the symbol-perturbed ZF structure\footnote{As a minor note, the work~\cite{krivochiza2017low} applied the symbol-perturbed ZF structure as a specific form of SLP. It did not provide the reasoning; like the one in Fact~\ref{fac:rep} and Proposition~\ref{prop:slp_zf}. }
\[
\bx_t = \bH^\dag (\bd \circ \bs_t + \bu_t), ~\bd = [\sigma_v\sqrt{\zeta_1/\rho},\dots,\sigma_v\sqrt{\zeta_K/\rho}]^T
\]
into \eqref{eq:exit_work1} to get
\begin{equation}\label{eq:exit_work2}
\begin{aligned}
\min_{\bu_t}
& ~ \| \bd \circ \bs_t + \bu_t \|_\bR^2 \\
{\rm s.t.}
& ~ u_{i,t}
\begin{cases}
=0,\quad &  |s_{i,t}|<2L-1\\
\ge 0, \quad & s_{i,t}=2L-1 \\
\le 0, \quad & s_{i,t}=-(2L-1)
\end{cases} , \ i=1,\dots,K.
\end{aligned}
\end{equation}
Now, we see that the CIPM formulation in \eqref{eq:exit_work2} looks very similar to the semi-ZF SLP formulation in~\eqref{eq:szf_tp_pro1}.
However, it is worth noting that the vast majority of the existing SLP solutions were not derived from the SEP metric, while our design considers the SEP quality constraints and did not use the CI notion.
Hence our design provides an alternative path to explain the existing SLP solutions.

\section{SLP Schemes for PPAP Minimization}
\label{sec:PPAP}

In this section, we describe how our SLP designs can be modified to handle the peak per-antenna power (PPAP) minimization design.
The problem is formulated as follows:
\ifconfver
\begin{equation} \label{eq:PPAP}
\begin{aligned}
\min_{ \bX , \bd, \bvarphi }
& ~ f_{\sf PPAP}(\bX, \bd, \bvarphi) \triangleq \max_{t=1,\dots,T}\|{\bx}_t\|_{\infty}^2 \\
{\rm s.t.} ~& \!\! - \!\bd \!+\! \ba_{t} \!\leq_c \! \bvarphi^* \!\!\circ \!(\bH\bx_t)\!-\! \bd \diamond \bs_t  \! \leq_c\! \bd \!- \!\bc_{t}, ~\! t =1,\dots,T,\\
& ~ \bd \ge_c \balpha_c, \quad |\bvarphi| = \bm 1.
\end{aligned}
\end{equation}
\else
\begin{equation} \label{eq:PPAP}
\begin{aligned}
\min_{ \bX , \bd, \bvarphi}
& ~  \max_{t=1,\dots,T}\|{\bx}_t\|_{\infty}^2 \\
{\rm s.t.} & -\bd + \ba_{t} \leq_c \bvarphi^* \circ (\bH\bx_t)- \bd \diamond \bs_t  \leq_c \bd- \bc_{t}, \quad t =1,\dots,T, \\
& ~ \bd \ge_c \balpha_c, \quad |\bvarphi| = \bm 1.
\end{aligned}
\end{equation}
\fi
We should note that we minimize the PPAP at all the symbol times;
the existing linear precoding formulations typically deal with the peak average power $\max_{n=1,\dots,N} \mathbb{E}[|x_{n,t}|^2]$~\cite{Yu2007Transmitter}.
Substituting the representation \eqref{rep_x} to Problem~\eqref{eq:PPAP} gives
\begin{equation} \label{eq:PPAP_pro}
\begin{aligned}
\min_{ \bd, \bvarphi, \bU, \bZ }
& ~ f_{\sf PPAP}(\bd, \bvarphi, \bU, \bZ)  \\
{\rm s.t.}
& -\bd + \ba_{t} \leq_c \bu_t  \leq_c \bd- \bc_{t},  \quad t =1,\dots,T, \\
& ~ \bd \ge_c \balpha_c, \quad |\bvarphi| = \bm 1,
\end{aligned}
\end{equation}
where
\[
f_{\sf PPAP}(\bd, \bvarphi, \bU, \bZ) \triangleq  \max\limits_{t=1,\dots,T} \|\bH^\dag ( \bvarphi \circ ( \bd \diamond \bs_{t}  + \bu_t ) ) +  \bB \bz_t\|_{\infty}^2.
\]
Note that the nullspace components $\bz_t$'s, which are shut down in the TTP minimization (cf. Proposition \ref{prop:slp_zf}), are part of the design variables.

Our optimization strategy is identical to that for TTP minimization in the preceding section.
Specifically, we apply AM between $\bvarphi$ and $(\bd,\bU,\bZ)$.
The new challenge is that $f_{\sf PPAP}$ is non-smooth.
We circumvent this issue by log-sum-exponential (LSE) approximation
\begin{equation}\label{LSE}
\textstyle\max\{ x_1, \ldots, x_N \} \approx \delta \log \Big( \sum_{i=1}^{N} e^{x_i/\delta} \Big),
\end{equation}
for a given smoothing parameter $\delta>0$.
It is known that the right-hand side of \eqref{LSE} is smooth, and the approximation in \eqref{LSE} is tight when $\delta \to 0$.
Applying \eqref{LSE} to $f_{\sf PPAP}$ yields
\begin{align}
f_{\sf PPAP}(\bd,\bvarphi,\bU,\bZ) &\approx
\delta \log \!\Big( \! \sum_{n=1}^{N} \sum_{t=1}^T e^{ \frac{|\tilde{\bh}_n^H ( \bvarphi\circ  ( \bd \diamond \bs_{t}  + \bu_t ))  +  \tilde{\bb}_n^H \bz_t|^2}{\delta}} \Big)  \nonumber \\
&   \triangleq \hat{f}_{\sf PPAP}(\bd,\bvarphi,\bU,\bZ), \label{eq:hat_ppap}
\end{align}
where $\tilde{\bh}_n^H$ and $\tilde{\bb}_n^H$ denote the $n$th row of $\bH^\dag$ and $\bB$, respectively.
The rest of the operations are same as the AM in Section~\ref{sec:TTP_AM}:
we minimize $\hat{f}_{\sf PPAP}$ over $|\bvarphi|=\bone$ by the PG method in Algorithm~\ref{AL_PG}, and we minimize $\hat{f}_{\sf PPAP}$ over $(\bd,\bU,\bZ)$ by the APG method in Algorithm~\ref{AL_APG}.

Like the suboptimal semi-ZF scheme in Section~\ref{sec:semi_ZF}, we can pre-fix $\bd = \balpha_c$ to reduce the computational cost.
It is worthwhile to note that the resulting minimization of $\hat{f}_{\sf PPAP}$ over $(\bd,\bU,\bZ)$ with $\bd = \balpha_c$ is, in essence, solving
\begin{equation} \label{szf_ppap_pro}
\begin{aligned}
\min_{ \bu_t, \bz_t}
& ~\|\bH^\dag (\bvarphi \circ( \balpha_c \diamond \bs_t  + \bu_t )) + \bB \bz_t\|_{\infty}^2 \\
{\rm s.t.} ~
&  -\balpha_c + \ba_t \leq_c \bu_t \leq_c \balpha_c - \bc_t,
\end{aligned}
\end{equation}
for $t=1,\ldots, T$;
or, in words, we are minimizing the PPAP of all the symbol times.
Moreover, if we further pre-fix $\bvarphi=\bone$ (no phase optimization) and $\bU=\bzero$ (no symbol perturbations), then our design reduces to
\begin{equation} \label{eq:nullzf_pro}
\begin{aligned}
\min_{ \bz_t }
& ~  \|\bH^\dag (\balpha_c \diamond \bs_t)  + \bB \bz_t\|_{\infty}^2
\end{aligned}
\end{equation}
for $t=1,\dots,T$, which is a nullspace-assisted ZF scheme (more precisely, the design reduces to the LSE approximation of \eqref{eq:nullzf_pro}).
Our numerical results will show that even the nullspace-assisted ZF scheme provides significant PPAP reduction, compared to the state-of-the-art schemes such as the basic ZF scheme and the linear precoding design under peak per-antenna average power minimization~\cite{Yu2007Transmitter}.

\section{When SLP Meets Vector Perturbation}
\label{sec:SLP-VP}

The SLP designs in the previous sections can also be extended to cover vector perturbation (VP) precoding.

\subsection{A Review of VP}
Let us first review the working principle of VP~\cite{hochwald2005vector,Maurer2011Vector}.
To facilitate, consider the real-valued case in Example~\ref{exa:SEP}.
Also, assume $\bd = \bone$.
The transmitted signals in VP are
\begin{equation}\label{eq:VP1}
\bx_t
=  ~ \bH^\dag ( \bs_{t}  +4L \bm \gamma_t ),
\end{equation}
for some integer vector $\bm \gamma_t \in \mathbb{Z}^K$.
VP looks like yet another perturbed ZF scheme, but the key idea lies in the detection.
The users detect the symbols by a modulo-type detection
\[
\hat{s}_{i,t} \! = \! \dec (\! \mathcal{M}( y_{i,t} ) )\!,
\]
where
\[
\textstyle\mathcal{M}(y) = y - \Big\lfloor \frac{y + 2L}{4L} \Big\rfloor4L
\]
is the modulo operation, with the modulo constant given by $4L$; $\lfloor x \rfloor$ denotes the maximum integer that is less than or equal to $x$.
In the absence of noise, one can verify that $\mathcal{M}(y_{i,t})=s_{i,t}$.
This further translates into the fact that the VP term $\bm \gamma_t$ does not affect the decision accuracies or SEPs.
The role played by the VP term, however, is to improve power efficiency.
We can reduce the transmitted power by designing an appropriate $\bm \gamma_t$; e.g., for TTP minimization,
\[
\min_{ \bm \gamma_t \in \mathbb{Z}^K } \| \bH^\dag ( \bs_{t}  +4L \bm \gamma_t ) \|_2^2.
\]
The above problem is computationally hard, but in practice it can be solved by sphere decoding~\cite{damen2003maximum} if $K$ is not too large.

\subsection{Connecting VP and SLP}
Next, we show how VP and SLP are connected.
Consider SLP under the modulo-type detection:
\begin{align}\label{dec_slpvp}
\hat{s}_{i,t} \! = \! \dec \Big(\! \mathcal{M}\Big( \frac{ \Re(\varphi_i^*  y_{i,t}) }{d_i^R} \Big) \! \Big)\! +\! \jj \cdot \dec \Big(\! \mathcal{M}\Big(  \frac{\Im(\varphi_i^* y_{i,t})}{d_i^I} \Big)\!\Big).
\end{align}
Following the SEP result in~\cite{liu2018symbol,shao2019framework} or in Section~\ref{sec:model}, it is shown that the SEP quality guarantee ${\sf CSEP}_{i,t} \le \eps_i$ holds if
\begin{equation}\label{eq:SEP_cond_VP}
-d_i + \alpha_{c,i} \leq_c \varphi_i^* \bh_i^H\bx_t - d_i \diamond (s_{i,t}+4L \gamma_{i,t})  \leq_c d_i- \alpha_{c,i},
\end{equation}
for some complex integer $\gamma_{i,t}$, i.e., $ \gamma_{i,t} \in \mathbb{Z}_C \triangleq \{ a+\jj b~ |~ a,b \in \mathbb{Z} \}$, where $\alpha_{c,i} = \alpha_i +\jj \alpha_i$ and $\alpha_i$ is defined in~\eqref{eq:alpha_beta}.
Define
\begin{equation}\label{eq:mu}
 \bm \mu_{t} = \bm \varphi^*\circ( \bH\bx_t) - \bd \diamond (\bs_{t}+4L \bm\gamma_{t}),
\end{equation}
where $\bm \gamma_t =[\gamma_{1,t},\ldots,\gamma_{K,t} ]^T$, such that \eqref{eq:SEP_cond_VP} can be rewritten as
\[
-\bd + \balpha_c \leq_c  \bm \mu_t  \leq_c \bd - \balpha_c,~ \forall t.
\]
By substituting the representation of $\bx_t$ in \eqref{rep_x} into~\eqref{eq:mu}, the symbol perturbation takes the form
\begin{equation*}
  \bu_t  = 4L \bd \diamond \bm \gamma_t +\bm \mu_t.
\end{equation*}
We see that the symbol perturbation $\bu_t$ consists of two terms.
The first term $4L\bd \diamond \bm \gamma_t$, or simply $\bm \gamma_t$, is referred to as the vector perturbation;
the second term $\bm \mu_{t}$ plays a similar role as the symbol perturbation $\bu_t$ in the previous sections (recall $-\bd + \ba_{t} \leq_c \bu_t  \leq_c \bd- \bc_{t}$ in the previous SLP designs).
As a result, the transmitted signal $\bx_t$ can be expressed as
\begin{equation}\label{eq:VP_rep}
\begin{split}
\bx_t
= & ~ \bH^\dag ( \bvarphi \circ ( \bd \diamond( \bs_{t}  +4L \bm \gamma_t ) +\bm \mu_t )) +  \bB \bz_t.
\end{split}
\end{equation}
The expression \eqref{eq:VP_rep} suggests that
{\it  SLP under the modulo detection \eqref{dec_slpvp} takes a form that is the VP extension of the symbol-perturbed, nullspace-assisted, ZF scheme.}
In particular,
if we choose $\bd \! = \! \bm \alpha_c$, $\bvarphi \! = \! \bm 1$, $\bm\mu_t=\bzero$, $\bz_t=\bzero$ such that
\begin{equation*}
\bx_t = \bH^\dag (  \balpha_c \diamond (\bs_{t}  + 4L \bm \gamma_t ) ),
\end{equation*}
the resulting scheme is essentially the VP scheme in~\eqref{eq:VP1}.

Next, we specify  SLP designs under \eqref{dec_slpvp}-\eqref{eq:VP_rep}.
The VP-extended SLP designs for TTP minimization and PPAP minimization are, respectively, given by

\begin{equation} \label{eq:TTP_slpvp_ref}
\begin{aligned}
\min_{ \bd, \bvarphi, \bm \Xi, \bm \Gamma }
& ~  \frac{1}{T} \sum_{t=1}^T \|  \bvarphi \circ ( \bd \diamond( \bs_{t}  +4L \bm \gamma_t ) +\bm \mu_t ) \|_\bR^2 \\
{\rm s.t.}
& ~ -\bd + \balpha_c \leq_c  \bm \mu_t  \leq_c \bd - \balpha_c, \quad t=1,\dots,T, \\
& ~ \bm \gamma_t \in \mathbb{Z}_C^K, \ t =1,\dots,T,  ~ \bd \ge_c \balpha_c,\  |\bvarphi| = \bm 1,
\end{aligned}
\end{equation}
and
\ifconfver
\begin{equation} \label{eq:PPAP_final}
\begin{aligned}
\min_{ \bd, \bvarphi, \bm \Xi, \bZ, \bm \Gamma }
&  \max_{t=1,\dots,T} \! \| \bH^\dag ( \bvarphi \circ \!(\! \bd \diamond \!( \bs_{t}  +4L \bm \gamma_t )\! +\bm \mu_t )) \!+ \! \bB \bz_t\|_\infty^2 \\
{\rm s.t.}
& ~ -\bd + \balpha_c \leq_c  \bm \mu_t  \leq_c \bd - \balpha_c, \quad t=1,\dots,T, \\
& ~ \bm \gamma_t \in \mathbb{Z}_C^{K}, \  t =1,\dots,T,  ~ \bd \ge_c \balpha_c,\  |\bvarphi| = \bm 1,
\end{aligned}
\end{equation}
\else
\begin{equation} \label{eq:PPAP_final}
\begin{aligned}
\min_{ \bd, \bvarphi, \bU, \bZ, \bm \Gamma }
&  \max_{t=1,\dots,T} \! \| \bH^\dag ( \bvarphi \circ ( \bd \diamond( \bs_{t}  +4L \bm \gamma_t ) +\bm \mu_t )) +  \bB \bz_t\|_\infty^2 \\
{\rm s.t.}
& ~ -\bd + \balpha_c \leq_c  \bm \mu_t  \leq_c \bd - \balpha_c, \quad t=1,\dots,T, \\
& ~ \bm \gamma_t \in \mathbb{Z}_C^{K}, \quad t =1,\dots,T, \\
& ~ \bd \ge_c \balpha_c,\quad  |\bvarphi| = \bm 1,
\end{aligned}
\end{equation}
\fi
where $\boldsymbol{\Xi} = [\bm\mu_1,\ldots, \bm\mu_T]$, $\bZ = [\bz_1,\ldots,\bz_T]$, and $\bm \Gamma = [\bm \gamma_1,\ldots, \bm \gamma_T]$.
Note that,
as a direct extension of Proposition~\ref{prop:slp_zf}, we have $\bZ =\bm 0$ for TTP minimization.

We apply AM between $\bvarphi$, $(\bd,\bm \Xi)$ (respectively $(\bd,\bm \Xi,\bZ)$), and $\bm \Gamma$ for TTP minimization (respectively PPAP minimization).
The procedures for handling the $\bvarphi$ update, the $(\bd,\bm \Xi)$ update and the $(\bd,\bm \Xi,\bZ)$ update are the same as in the preceding development.
The $\bm \Gamma$ update is done by  sphere decoding~\cite{damen2003maximum} in the TTP minimization design, and by  $p$-sphere encoding~\cite{boccardi2006p} in the PPAP minimization design.
We can also consider the semi-ZF scheme wherein we pre-fix $\bd = \balpha_c$, as well as the nullspace-assisted ZF scheme (for PPAP minimization) wherein we pre-fix $\bd = \balpha_c$, $\bvarphi = \bone$.

The VP extension is numerically found effective in performance improvement, while the downside lies in its higher computational complexity of calling the sphere decoding (or the $p$-sphere encoding) algorithms.

\section{Simulation Results}
\label{sec:sim}

In this section, we provide numerical results to show the performance of the developed SLP schemes.
We aim to shed light onto how  different components, such as symbol perturbations and nullspace components, have their respective impacts on the system performance.

The simulation settings are as follows.
In each simulation trial, the channel matrix $\bH$ is randomly generated and follows an element-wise i.i.d. complex circular Gaussian distribution with zero mean and unit variance.
The symbols $s_{i,t}$'s are uniformly drawn from the QAM constellation.
The power of noise is set to $\sigma_v^2 = 1$.
The users share the same SEP requirement, i.e., $\eps_1 = \dots = \eps_K = \eps$.
Unless specified, the transmission block length is $T = 200$.
All the results to be reported are results averaged over  $1000$ Monte Carlo simulation trials.
The simulations were conducted by MATLAB on a small server with an Intel Core i7-6700K CPU and 16GB RAM.

To provide benchmarking, we consider the ZF scheme~\eqref{eq:ZF} and the SINR-constrained optimal linear beamforming (OLB) scheme~\cite{Bengtsson2001,bjornson2014optimal,Yu2007Transmitter}.
The implementation of OLB can be found in the supplemental material of this paper.
We also consider two representative SLP designs for TTP minimization:
1) the CIPM design~\cite{alodeh2017symbol} solved by CVX,
 and
2) the symbol-level optimization for conventional precoding (SLOCP) design~\cite{krivochiza2017low} solved by the non-negative least squares algorithm \cite{bro1997fast}.
As discussed in Section \ref{sec:rel_exsit},
these two SLP designs are SNR constrained.
To facilitate comparison, we repurpose these two SLP designs to the SEP-constrained designs.\footnote{\small
Following the spirit of the SEP characterization in Appendix~\ref{sec:proof_fact1}, one can show that if CIPM and SLOCP have their target SNRs chosen as $\zeta_i = \textstyle \frac{\rho}{2}[Q^{-1}(\frac{1-\sqrt{1-\eps_i}}{2})]^2$, then they will achieve the SEP quality guarantees ${\sf CSEP}_{i,t}  \leq \eps_i$.}

For clarity, we summarize all the tested precoding schemes in Table \ref{tb:schemes}.
We will refer to ``SLP'' as the SLP design that optimizes all the variables (e.g., Section~\ref{sec:TTP_AM} for TTP minimization), ``SLP-VP''   as the VP extension of ``SLP'', ``Null-ZF'' as the nullspace-assisted ZF scheme, and ``Null-VP'' as the nullspace-assisted VP scheme.

\ifconfver
\begin{table*}[t!]
	\else
	\begin{table}[ht!]
		\fi
		\centering
		\captionsetup{justification=centering}
		\caption{Summary of the tested precoding schemes}\label{tb:schemes}
		\renewcommand{\arraystretch}{1.05}
		\resizebox{\linewidth}{!}{%
			\begin{tabular}{ M{20mm}|M{16mm}| M{30mm} | M{38mm} |M{75mm} }
				\hline
				Name &  Scenario  &  Parameters to optimize & Fixed parameters & Formulations and methods\\ \hline\hline
                \multirow{2}{*}{ZF} & TTP& none & $(\bd,\bvarphi,\bU) = (\balpha_c,\bone,\bzero)$ & \multirow{2}{*}{\eqref{eq:ZF}, closed form}  \\ \cline{2-4}
			     & PPAP& none & $(\bd,\bvarphi,\bU,\bZ) = (\balpha_c,\bone,\bzero,\bzero)$ & \\
				\hline \hline
				\multirow{2}{*}{OLB \cite{Bengtsson2001,Yu2007Transmitter}} & TTP &  $(\bd,\bvarphi,\bU)$ &  none &
(81) in supplemental material, CVX  \\ \cline{2-5}
				& PPAP &  $(\bd,\bvarphi,\bU,\bZ)$ & none &
(82) in supplemental material, CVX \\
\hline \hline
{CIPM  \cite{alodeh2017symbol}} & { TTP}  & { $\bX$} &{ none} & { (12) in \cite{alodeh2017symbol}, CVX }\\
\hline \hline
{	SLOCP  \cite{krivochiza2017low}} & { TTP}  & { $\bU$} &{ $(\bd,\bvarphi)=(\balpha_c,\bone)$} & { (16) in \cite{krivochiza2017low}, the algorithm in~\cite{bro1997fast}}\\
				\hline \hline
				\multirow{2}{*}{Semi-ZF SLP} & TTP  &  $(\bvarphi,\bU)$ & $\bd = \balpha_c$ & \eqref{eq:TTP_ref}, AM with APG and PG  \\ \cline{2-5}
				& PPAP  &  $(\bvarphi,\bU,\bZ)$ & $\bd = \balpha_c$ & \eqref{eq:PPAP_pro}, LSE approximation, AM with APG and PG  \\
				\hline  \hline
				Null-ZF & PPAP  &  $\bZ$ & $(\bd,\bvarphi,\bU) = (\balpha_c,\bone,\bzero)$ & \eqref{eq:PPAP_pro}, LSE approximation, APG \\
				\hline \hline
				\multirow{2}{*}{SLP} & TTP  &   $(\bd,\bvarphi,\bU)$ & none & \eqref{eq:TTP_ref}, AM with APG and PG  \\ \cline{2-5}
				& PPAP  &  $(\bd,\bvarphi,\bU,\bZ)$ & none & \eqref{eq:PPAP_pro}, LSE approximation, AM with APG and PG  \\
				\hline  \hline
                \multirow{2}{*}{VP \cite{hochwald2005vector}} & TTP & $\bm \Gamma$ &  $(\bd,\bvarphi,\bm \Xi) = (\balpha_c,\bone,\bzero)$ &  \eqref{eq:TTP_slpvp_ref}, sphere decoding \\ \cline{2-5}
			     & PPAP  & $\bm \Gamma$ & $(\bd,\bvarphi,\bm \Xi,\bZ) = (\balpha_c,\bone,\bzero,\bzero)$ & \eqref{eq:PPAP_final}, $p$-sphere encoding \\
				\hline \hline
                Null-VP & PPAP  & $(\bZ, \bm \Gamma)$ & $(\bd,\bvarphi,\bm \Xi) = (\balpha_c,\bone,\bzero)$ & \eqref{eq:PPAP_final}, AM with APG (LSE approximation) and $p$-sphere encoding \\
				\hline \hline
                \multirow{2}{*}{SLP-VP} & TTP &  $(\bd,\bvarphi,\bm \Xi,\bm \Gamma)$ & none & \eqref{eq:TTP_slpvp_ref}, AM with APG, PG and sphere decoding \\ \cline{2-5}
			    & PPAP  &  $(\bd,\bvarphi,\bm \Xi,\bZ,\bm \Gamma)$ & none & \eqref{eq:PPAP_final}, AM with APG (LSE approximation), PG (LSE approximation) and $p$-sphere encoding \\
				\hline \hline
			\end{tabular}
		}
		\ifconfver
	\end{table*}
	\else
\end{table}
\fi

The implementation details of the SLP algorithms are as follows.
For the LSE approximation, we set the smoothing parameter as $\delta = L^2 / 25$.
The AM algorithm terminates when the relative change of the objective values of successive iterations is smaller than $10^{-3}$ or when the iteration number exceeds $10$.
The APG method stops when the difference of solutions between successive iterations is smaller than $10^{-3}$, or when the iteration number exceeds $300$.
The PG method is implemented under the same stopping criterion as that of APG.
SLP, Semi-ZF SLP and Null-ZF are initialized with the ZF solution.
Null-VP and SLP-VP are initialized by the solutions of VP and Null-VP, respectively.

The remaining parts of this section is organized as follows: Section~\ref{sec:sim1} and Section~\ref{sec:sim_PPAP} show the simulation results of the SLP schemes for TTP minimization and PPAP minimization, respectively.
Their VP extensions are considered in Section~\ref{sec:sim_SLP_VP}.

\subsection{SLP for TTP Minimization}
\label{sec:sim1}

First of all, we show the performance of the SLP schemes in the context of TTP minimization.
Figure~\ref{TP_fig1} shows the TTP performance versus the SEP requirement $\varepsilon$ for $(N,K) = (32,30)$ and for various QAM constellation sizes.
It is seen that the SLP schemes (SLP, Semi-ZF SLP, CIPM, SLOCP) outperform OLB and ZF;
CIPM and SLOCP are more than 1dB worse than SLP and Semi-ZF SLP.
Also, the performance gap decreases as the constellation size increases.
This trend is in agreement with the result in Theorem~\ref{Them1}.
We should pay attention to Semi-ZF SLP.
For $16$-QAM, Semi-ZF SLP outperforms ZF by $4.8$dB, which indicates that the designs of the symbol perturbations for OCPs and constellation phase can play a significant role in TTP reduction.
Also, it is interesting to see that Semi-ZF SLP exhibits nearly the same performance as SLP, which suggests that the choice of the constellation range $\bd = \balpha_c$ is a good heuristic.
Note that compared with SLP, Semi-ZF SLP is simpler in structures and much easier to optimize.
Thus, Semi-ZF SLP achieves a good balance between high performance and low computational complexity.

\ifconfver
\begin{figure}[htb!]
	\centering
	\begin{subfigure}[b]{\linewidth}
		\centering \includegraphics[width=0.8\linewidth]{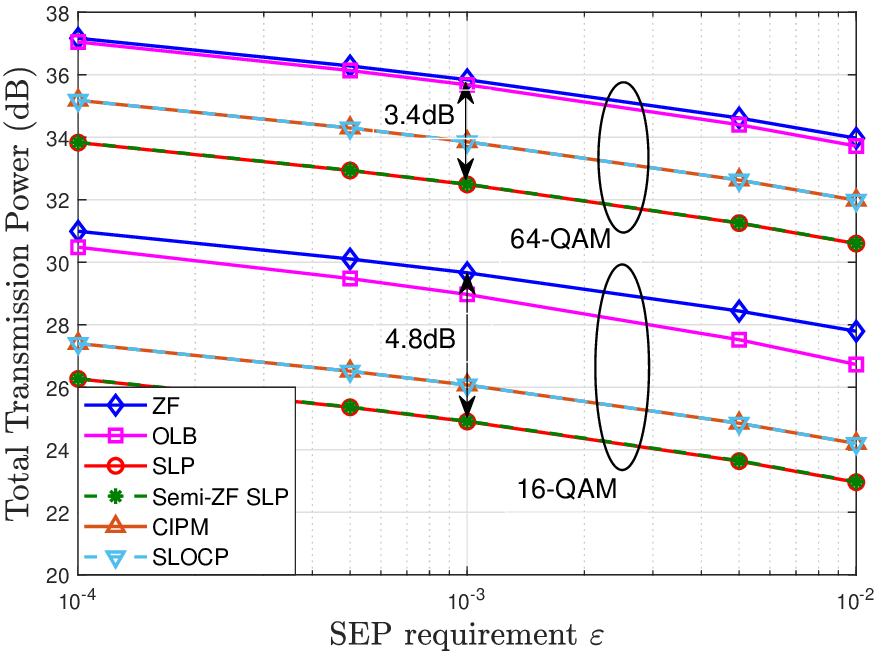}
		\caption{Performance for 16-QAM and 64-QAM.}
	\end{subfigure}
	~
	\begin{subfigure}[b]{\linewidth}
		\centering \includegraphics[width=0.8\linewidth]{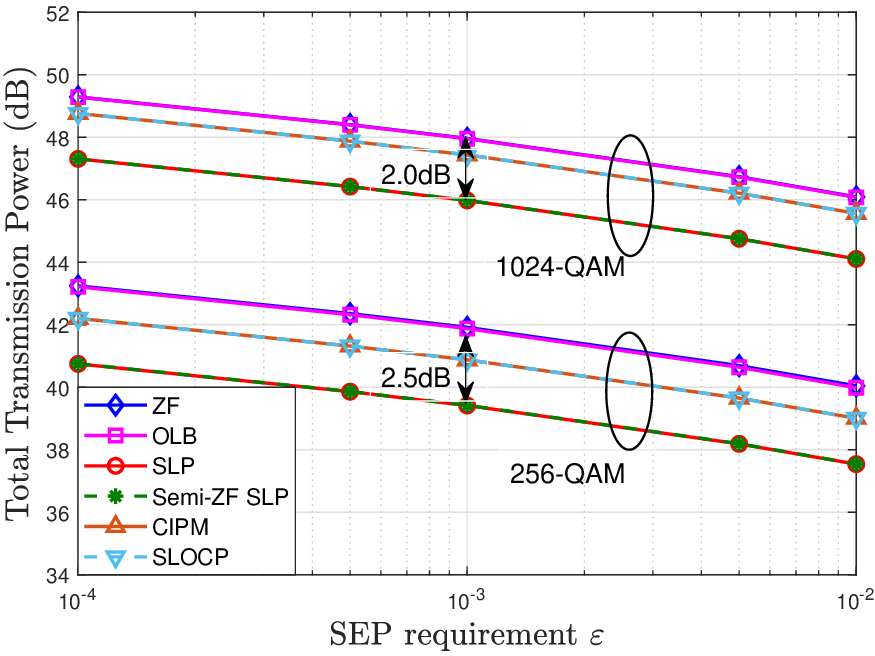}
		\caption{Performance for 256-QAM and 1024-QAM.}
	\end{subfigure}
	\caption{TTP versus the SEP requirements $\eps$. $(N,K)\! = \!(32,30)$.}\label{TP_fig1}
\end{figure}
\else
\begin{figure}[ht!]
	\centering
	\begin{subfigure}[b]{0.48\linewidth}
		\includegraphics[width=1\linewidth]{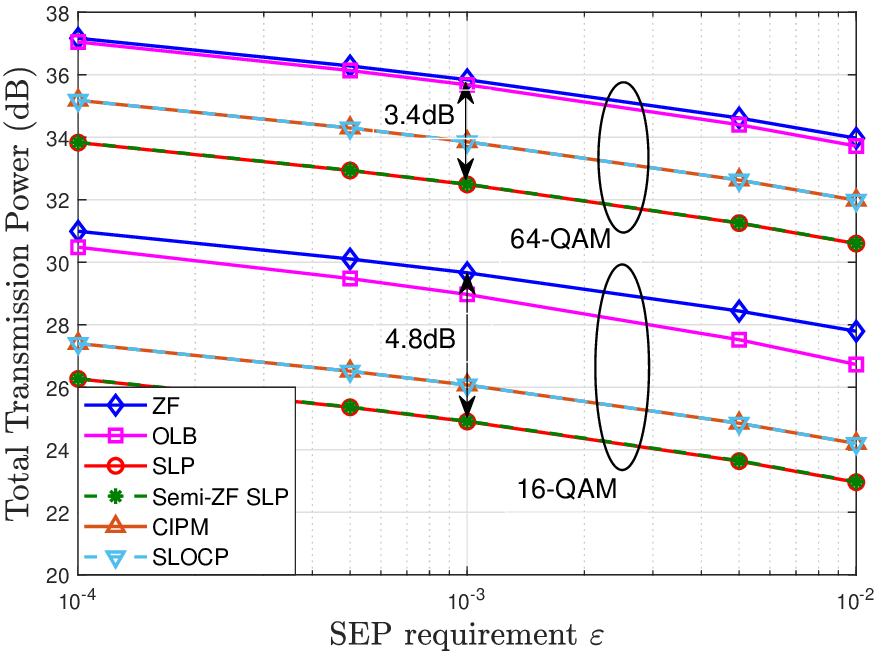}
		\caption{Performance for 16-QAM and 64-QAM.}
	\end{subfigure}
	\begin{subfigure}[b]{0.48\linewidth}
		\includegraphics[width=1\linewidth]{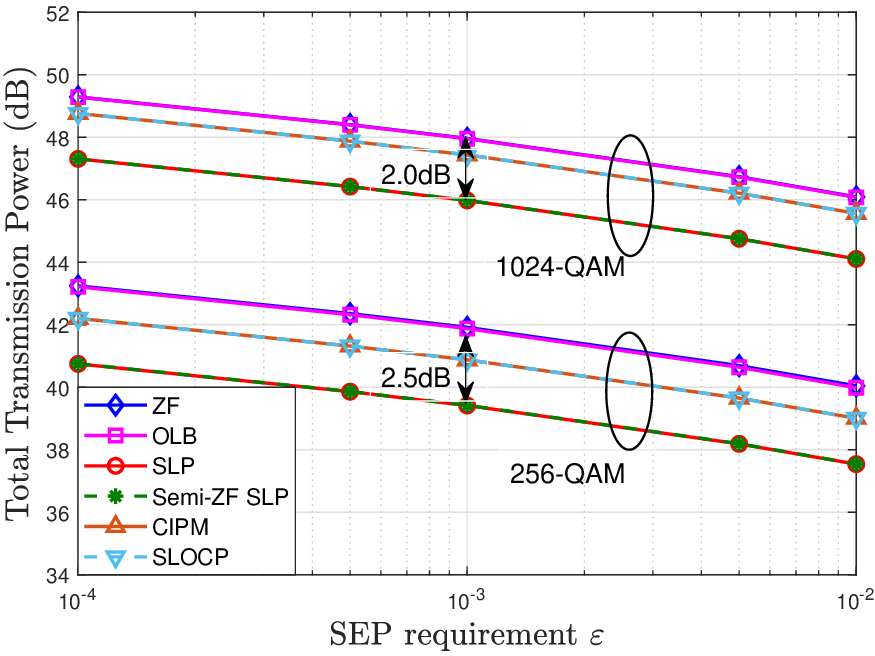}
		\caption{Performance for 256-QAM and 1024-QAM.}
	\end{subfigure}
	\caption{TTP versus the SEP requirements $\eps$. $(N,K) = (32,30)$.}\label{TP_fig1}
\end{figure}
\fi

\ifconfver
\begin{figure}[htb!]
    \centering
    \includegraphics[width=0.8\linewidth]{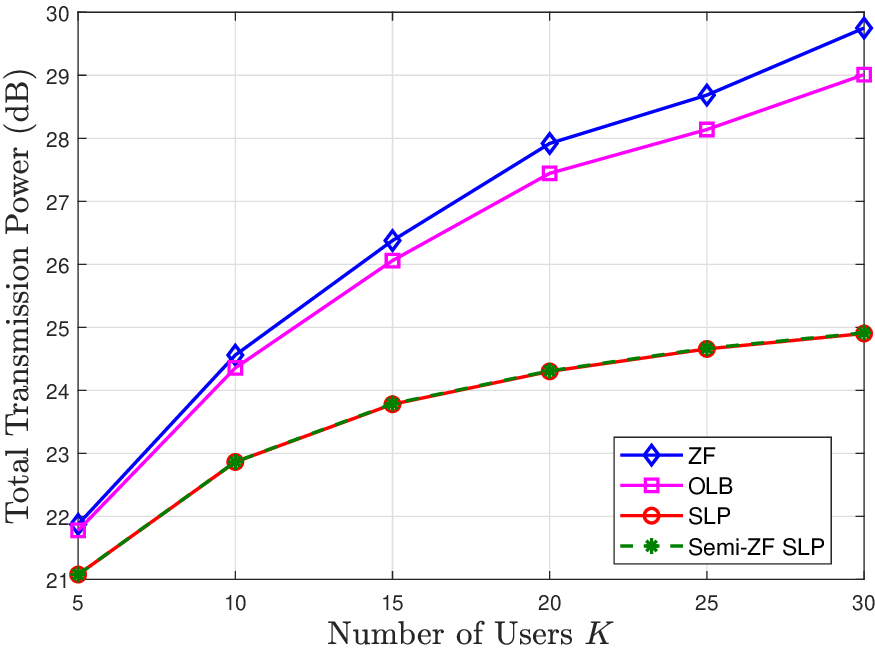}
    \caption{TTP versus the number of users $K$. $N = K+2$, $\eps = 10^{-3}$, 16-QAM.}
    \label{TP_fig2}
\end{figure}
\else
\begin{figure}[ht!]
    \centering
    \includegraphics[width=0.5\linewidth]{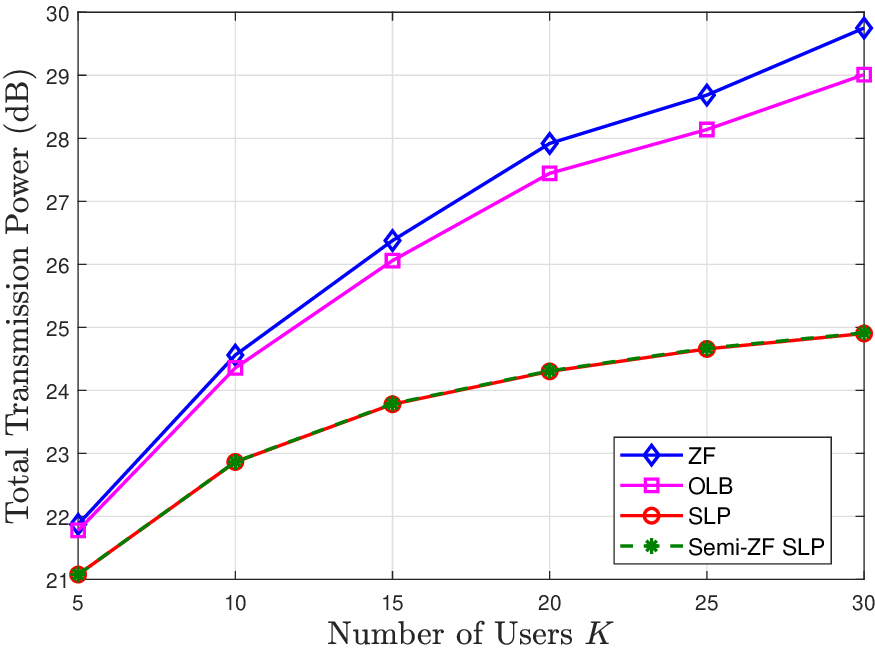}
    \caption{TTP versus the number of users $K$. $N = K+2$, $\eps = 10^{-3}$, 16-QAM.}
    \label{TP_fig2}
\end{figure}
\fi

Figure~\ref{TP_fig2} shows the TTP performance versus the problem size $K$.
We set $N=K+2$,  $\eps = 10^{-3}$ and use $16$-QAM constellation.
It is seen that the TTPs of the SLP schemes increase with $K$ at slower rates than those of OLB and ZF.
Again, we see that Semi-ZF SLP works well.

\begin{table}
	\centering
	\begin{tabular}{ c|c| c | c |c| c }
		\hline
		$T$ & $200$ & $400$  & $600$ & $800$ & $1000$ \\
\hline \hline
SLP & $6.44$ & $6.71$ & $7.94$ & $10.29$ & $12.84$ \\
\hline
Semi-ZF SLP & $0.12$ & $0.22$ & $0.33$ & $0.44$ & $0.54$ \\
\hline
CIPM & $62.48$ & $189.67$ & $384.76$ & $645.39$ & $969.70$ \\
\hline
SLOCP & ${\bf 0.09}$ & ${\bf 0.16}$ & ${\bf 0.25}$ & ${\bf 0.33}$ & ${\bf 0.40}$ \\
\hline
	\end{tabular}
\caption{Average runtime (in seconds) for each block. $(N,K) = (32,30)$, $\eps = 10^{-3}$, 16-QAM.}\label{TP_tab1}
\end{table}

In Table~\ref{TP_tab1}, we show the runtime performance of the SLP schemes w.r.t. the transmission block length $T$, including SLP, Semi-ZF SLP, CIPM and SLOCP.
It is seen that both Semi-ZF SLP and SLOCP are fast; SLOCP is slightly faster than Semi-ZF SLP.

Table~\ref{TP_ser} shows the actual average SEPs achieved by the various precoding schemes, where
we consider $(N,K) = (32,30)$ and 64-QAM constellation.
We see that the actual average SEPs are better than the required, although the differences are insignificant.
\begin{table}
	\centering
	\begin{tabular}{ c|c| c | c  }
		\hline
		$\varepsilon$ & $10^{-4}$ &  $10^{-3}$ & $10^{-2}$\\
\hline \hline
OLB & $8.0   \times   10^{-5}$ & $8.3   \times  10^{-4}$ & $8.9   \times  10^{-3}$\\
\hline
SLP & $8.3  \times  10^{-5}$ &  $9.1 \times  10^{-4}$ &  $9.3   \times  10^{-3}$\\
\hline
Semi-ZF SLP & $8.2   \times  10^{-5}$ & $9.1   \times  10^{-4}$ &  $9.3    \times   10^{-3}$\\
\hline
CIPM & $8.1  \times  10^{-5}$ &  $8.6  \times   10^{-4}$ &  $9.1  \times    10^{-3}$\\
\hline
SLOCP & $8.1   \times   10^{-5}$  & $8.5 \times   10^{-4}$ &  $9.0    \times   10^{-3}$\\
\hline
	\end{tabular}
\caption{Average SEPs for different SEP requirements $\eps$. $(N,K) = (32,30)$; 64-QAM.}\label{TP_ser}
\end{table}

\subsection{SLP for PPAP Minimization}
\label{sec:sim_PPAP}

Next, we test the SLP designs for PPAP minimization.
We use the complementary cumulative distribution function (CCDF) to measure the PPAP distribution, i.e.,
\[
{\rm CCDF}(x) = \Pr(\text{PPAP} \geq x).
\]
Note that given the same CCDF level, a smaller PPAP threshold $x$ means better performance.

\ifconfver
\begin{figure}[ht!]
	\centering
    \includegraphics[width=0.8\linewidth]{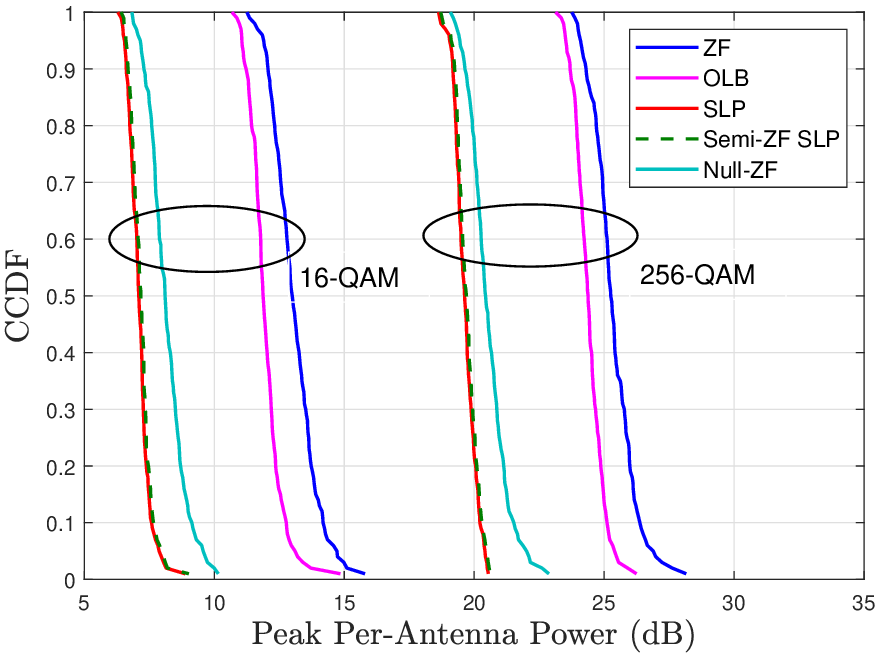}
	\caption{CCDF of PPAP. $(N,K) = (32,16)$, $\eps = 10^{-3}$.}\label{PAP_fig1}
\end{figure}
\else
\begin{figure}[ht!]
	\centering
    \includegraphics[width=0.5\linewidth]{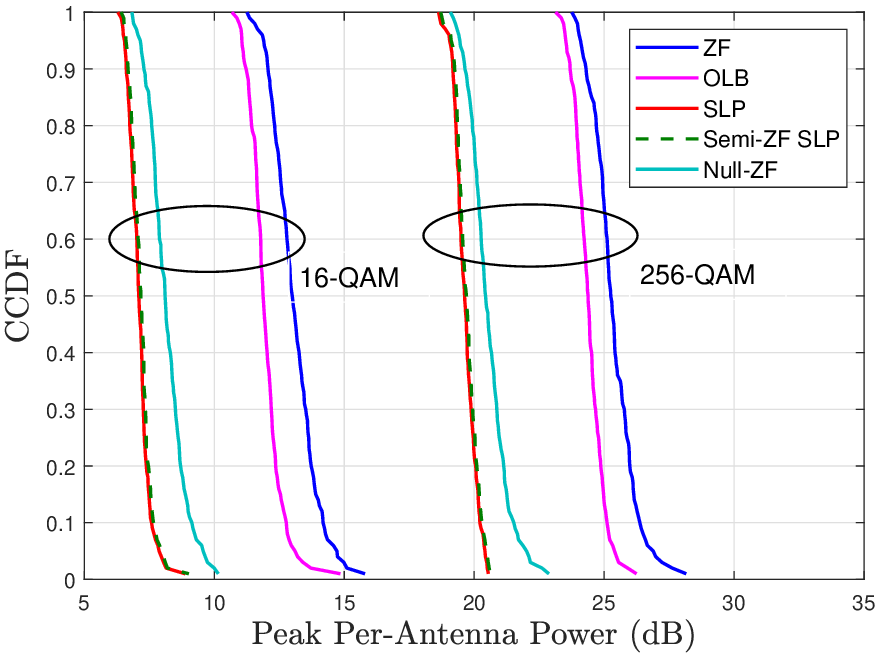}
	\caption{CCDF of PPAP. $(N,K) = (32,16)$, $\eps = 10^{-3}$.}\label{PAP_fig1}
\end{figure}
\fi

Figure~\ref{PAP_fig1} presents the CCDF of PPAP for $(N,K) = (32,16)$ and $\eps = 10^{-3}$. 
Our observations are as follows.
First, all the SLP schemes perform better than the OLB and ZF for 16-QAM and 256-QAM.
Different from the TTP minimization case in Figure~\ref{TP_fig1}, in this PPAP minimization case the benefits of SLP over ZF do not vanish as the QAM size increases.
Second, Semi-ZF SLP, SLP and  Null-ZF provide comparable performance, with Null-ZF  performing slightly worse.
Comparing  Null-ZF with ZF, we see that the incorporation of nullspace components contributes a lot to PPAP reduction.
Comparing  Null-ZF with  Semi-ZF SLP, we see that optimizing the symbol perturbations for OCPs is helpful, though the performance gain is not substantial.
Comparing Semi-ZF SLP with SLP, the nearly identical performance of the two again suggests that fixing the constellation range as $\bd = \balpha_c$ is a good heuristic.
Both Null-ZF and Semi-ZF SLP are computationally light and show promising performance.

Besides the PPAP, we also test the peak-to-average power ratio (PAPR) performance.
Specifically, we evaluate the worst PAPR among all the transmit antennas, defined as
$
\max_{n=1,\dots,N} \text{PAPR}_n,
$
where
\[
\text{PAPR}_n  \triangleq \frac{\max_{t=1,\dots,T}|x_{n,t}|^2}{\sum_{t=1}^T | x_{n,t} |^2 / T}
\]
is the PAPR of the $n$th transmit antenna.

In Figure~\ref{PAP_fig3}, we show the CCDF of the worst PAPR for 256-QAM, where $(N,K) = (32,16)$ and $\eps = 10^{-3}$.
We observe similar performance behaviours as the PPAP performance in Figure~\ref{PAP_fig1}.
Interestingly, although the SLP designs do not minimize the PAPR, the results indicate that minimizing the PPAP is helpful in reducing the PAPR.

\ifconfver
\begin{figure}[htb!]
    \centering
    \includegraphics[width=0.8\linewidth]{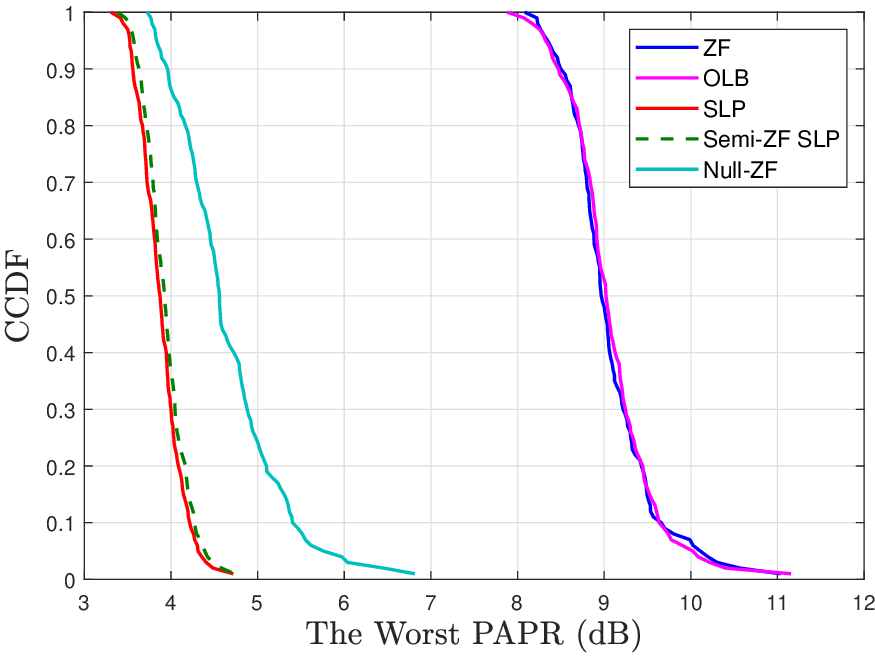}
    \caption{CCDF of the worst PAPR. $(N,K) = (32,16)$, $\eps = 10^{-3}$, 256-QAM.}
    \label{PAP_fig3}
\end{figure}
\else
\begin{figure}
    \centering
    \includegraphics[width=0.5\linewidth]{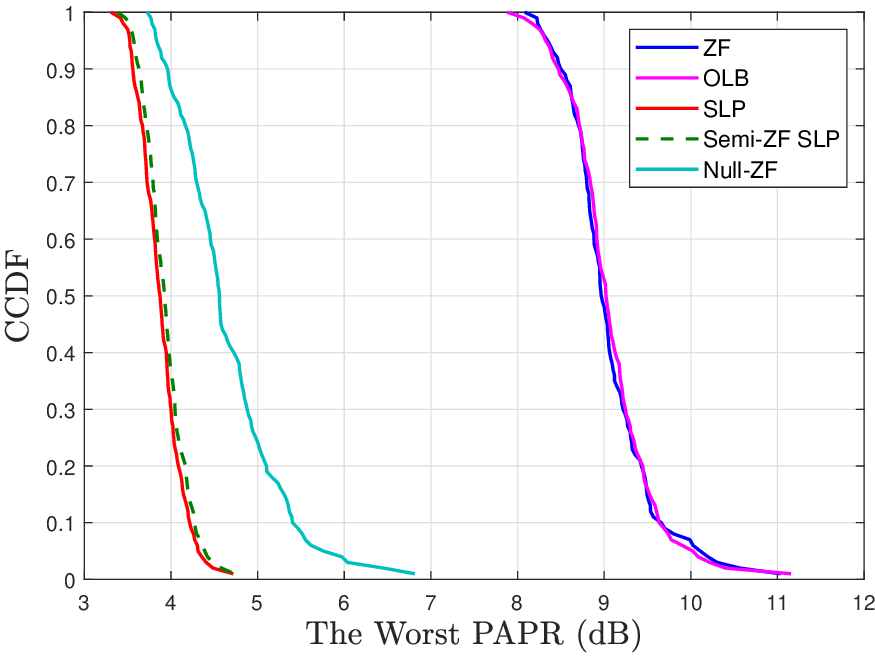}
    \caption{CCDF of the worst PAPR. $(N,K) = (32,16)$, $\eps = 10^{-3}$, 256-QAM.}
    \label{PAP_fig3}
\end{figure}
\fi

Let us test the runtime performance of SLP, Semi-ZF SLP and Null-ZF.
Table~\ref{PAP_tab1} shows the result. It is seen that Null-ZF is the most computationally efficient, Semi-ZF SLP is the second, and SLP is the slowest.

\begin{table}
	\centering
	\begin{tabular}{ c|c| c | c |c| c }
		\hline
		$T$ & $200$ & $400$  & $600$ & $800$ & $1000$\\
\hline \hline
SLP & $4.33$ & $8.79$ & $13.36$ & $17.72$ & $22.52$ \\
\hline
Semi-ZF SLP & $1.63$ & $3.00$ & $4.33$ & $5.57$ & $6.86$ \\
\hline
Null-ZF & ${\bf 0.40}$ & ${\bf 0.79}$ & ${\bf 1.19}$ & ${\bf 1.58}$ & ${\bf 1.98}$ \\
\hline
	\end{tabular}
\caption{Average runtime (in seconds) for each block. $(N,K) = (32,16)$, $\eps = 10^{-3}$, 16-QAM.}\label{PAP_tab1}
\end{table}

The above numerical results suggest that Semi-ZF SLP and Null-ZF are good candidates for the PPAP minimization design, offering a good balance in performance and complexity.

\ifconfver
\begin{figure}[htb!]
	\centering
	\includegraphics[width=0.8\linewidth]{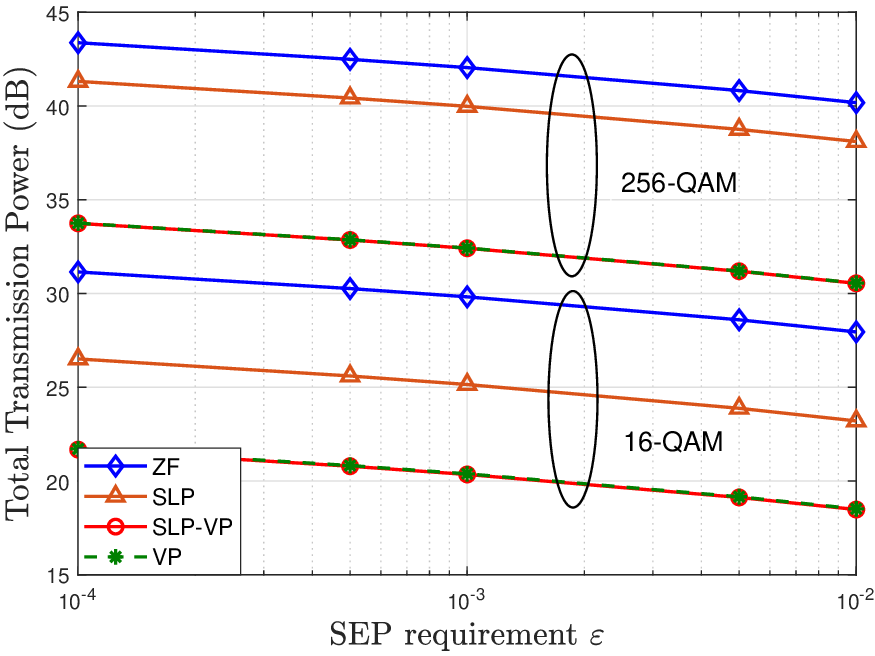}
	\caption{TTP versus the SEP requirements $\eps$ for TTP minimization. $(N,K) = (16,15)$.}
	\label{SLPVP_fig1}
\end{figure}
\else
\begin{figure}[ht!]
	\centering
	\includegraphics[width=0.5\linewidth]{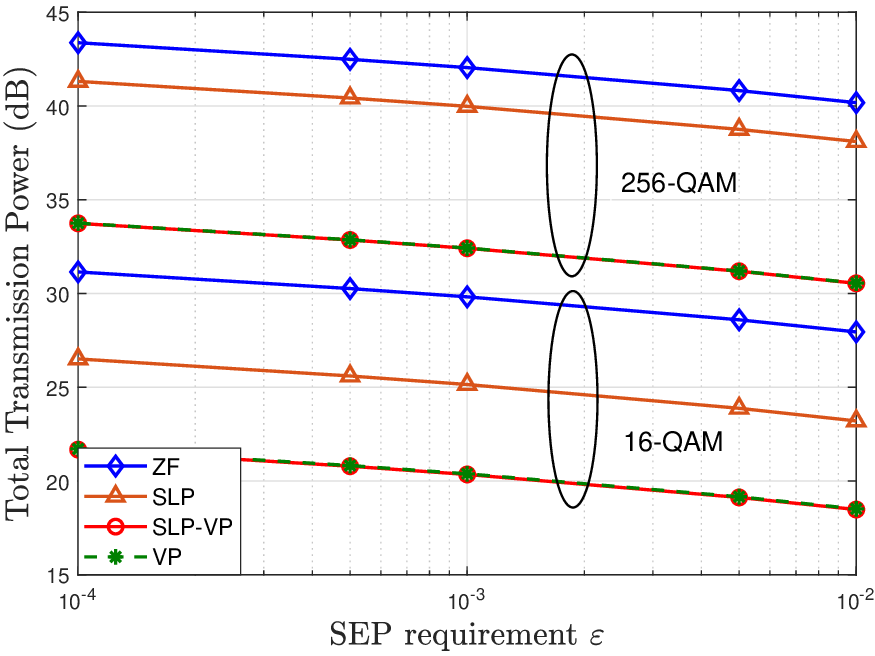}
	\caption{TTP versus the SEP requirements $\eps$ for TTP min. $(N,K) = (16,15)$.}
	\label{SLPVP_fig1}
\end{figure}
\fi

\ifconfver
\begin{figure}[htb!]
	\centering
	\includegraphics[width=0.8\linewidth]{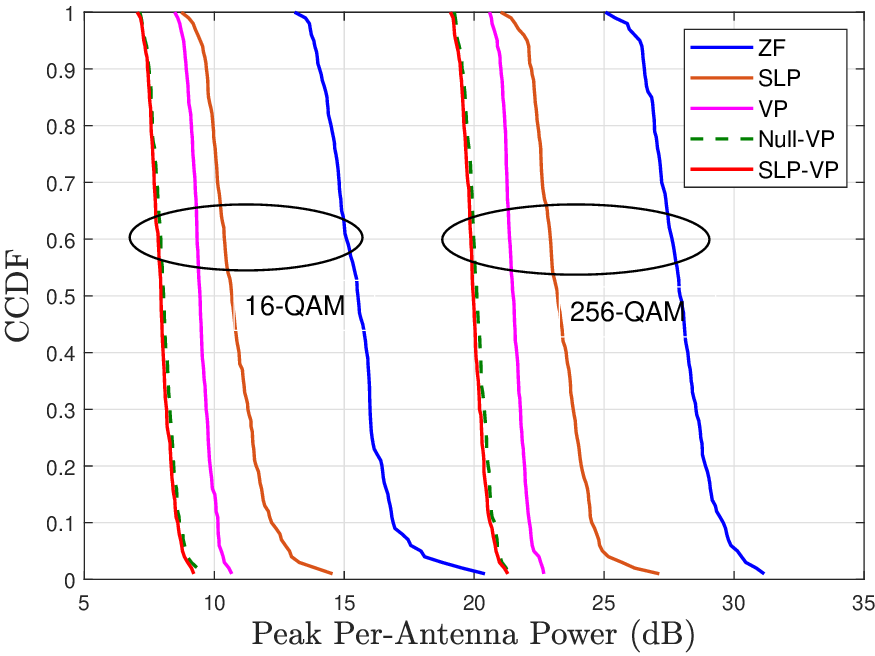}
	\caption{CCDF of the PPAP for PPAP minimization. $(N,K) = (16,8)$, $\eps = 10^{-3}$.}
	\label{SLPVP_fig2}
\end{figure}
\else
\begin{figure}[ht!]
	\centering
	\includegraphics[width=0.5\linewidth]{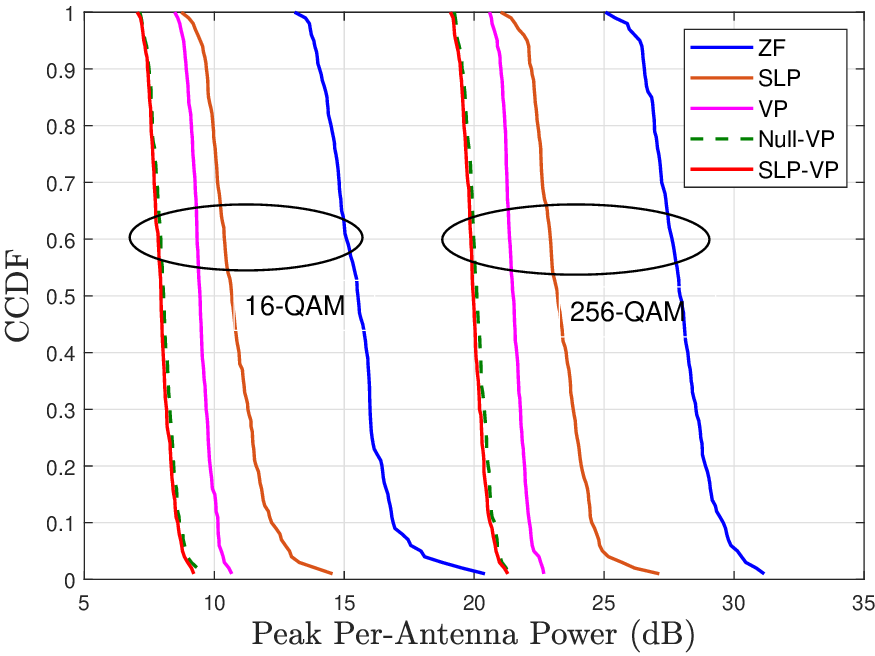}
	\caption{CCDF of the PPAP for PPAP min. $(N,K) = (16,8)$, $\eps = 10^{-3}$.}
	\label{SLPVP_fig2}
\end{figure}
\fi

\subsection{VP Extension of the SLP Schemes}
\label{sec:sim_SLP_VP}

Finally, we show the performance of the VP extensions of the SLP schemes.
We first consider the TTP minimization scenario.
The results are shown in Figure~\ref{SLPVP_fig1}, where we evaluate the TTP versus the SEP requirements $\eps$ for $(N,K) = (16,15)$.
It is seen that the VP extensions of both SLP and ZF provide much better performance than their no-VP counterparts.
We observe that SLP-VP and VP yield nearly identical performance.
A possible explanation is as follows.
The effect of modulo operation in the detection may be regarded as periodically and infinitely extending the QAM constellation with period $4L$ \cite{Maurer2011Vector}.
Therefore, there is no concept of OCPs for this extended QAM constellation.
On the other hand, the numerical results in Section~\ref{sec:sim1} suggest that optimizing the symbol perturbations for OCPs is  key to improving the performance of the SLP schemes for TTP minimization.

Next, we consider the PPAP minimization scenario.
In Figure~\ref{SLPVP_fig2}, we present the CCDF of the PPAP.
We choose $(N,K) = (16,8)$ and $\eps = 10^{-3}$.
Again, it is seen that the VP extensions bring significant performance improvement.
Moreover, we observe that Null-VP and SLP-VP achieve comparable performance, which again indicates that optimizing the nullspace components plays an important role in PPAP reduction.

\section{Conclusion}
\label{sec:conclusion}

Through the lens of ZF and VP precoding,
we studied SLP under SEP-constrained formulations and under QAM constellations.
The connections between SLP, linear precoding and VP precoding were shown by interpreting SLP as a ZF scheme with symbol perturbations, nullspace perturbations, and integer perturbations for the VP extension.
Taking insights from these connections, we developed a collection of SLP designs---from a more general design that gives the best performance in principle, to suboptimal but computationally more efficient designs; and from total transmission power minimization to peak per-antenna power minimization.
Simulation results were provided to examine the impacts of different design elements on the SLP performance.
A summary with our numerical examination is as follows.
\begin{enumerate}[1.]
\item
Symbol perturbations give rise to marked improvement with TTP reduction for lower QAM orders (this is also noted in the literature), but offer little gain once we consider the VP extension.
\item
Nullspace perturbations are useless in TTP reduction (this is known analytically), but are useful in PPAP reduction.
\item
The semi-ZF scheme, which employs a heuristic choice of the constellation range for simplifying the optimization, offers nearly identical performance as the more fully developed SLP designs, which optimizes the constellation range.
The same phenomena were observed for the VP extension.
It is worth noting that the semi-ZF SLP scheme resembles some existing SLP solutions~\cite{krivochiza2017low}.
\item
The VP-extended SLP designs yield significantly improved performance, although one should note that they also demand higher computational costs because of the need to optimize the integer perturbations.
\end{enumerate}

\appendix

\section*{Appendix}

\section{Proof of Fact~\ref{fac:SEP}}
\label{sec:proof_fact1}
Let
\begin{equation*}
\begin{split}
{\sf CSEP}^R_{i,t} &= {\rm Pr}( \Re(\hat{s}_{i,t}) \neq \Re(s_{i,t}) \mid s_{i,t} ), \\
{\sf CSEP}^I_{i,t} &= {\rm Pr}( \Im(\hat{s}_{i,t}) \neq \Im(s_{i,t}) \mid s_{i,t} ),
\end{split}
\end{equation*}
which are the conditional SEPs of the real and imaginary components of $\hat{s}_{i,t}$, respectively.
It is easy to verify that
\begin{equation}\label{eq:imp1}
{\sf CSEP}^R_{i,t} \! \leq 1 \! - \sqrt{1 \!-\! \eps_i},~
{\sf CSEP}^I_{i,t} \! \leq 1 \! - \sqrt{1 \!-\! \eps_i} \Rightarrow
{\sf CSEP}_{i,t}  \leq \eps_i.
\end{equation}
Following the same spirit in the real-valued conditional SEP analysis in Example~\ref{exa:SEP}, we have
\begin{equation*} \label{eq:SEP_eqs}
\begin{aligned}
{\sf CSEP}^R_{i,t} \begin{cases}
                  \leq  2 Q \Big( \frac{\sqrt{2}}{\sigma_v} ( d_i^R - |b_{i,t}^R| ) \Big)  , & \mbox{if } | \Re(s_{i,t}) | < 2L - 1, \\
                  = Q \Big( \frac{\sqrt{2}}{\sigma_v} ( d_i^R + b_{i,t}^R ) \Big)  , & \mbox{if } \Re(s_{i,t}) =  2L - 1\\
                 =Q \Big( \frac{\sqrt{2}}{\sigma_v} ( d_i^R -b_{i,t}^R ) \Big)   , & \mbox{if }  \Re(s_{i,t}) =  -2L + 1,
                  \end{cases}
\end{aligned}
\end{equation*}
where
$b_{i,t}^R = \Re(\varphi_i^* \bh_i^H\bx_t) - d_i^R \Re(s_{i,t}).$
Consequently,
\begin{equation}\label{eq:sep_suff_re}
\begin{aligned}
-d_i^R + a_{i,t}^R \leq \Re(\varphi_i^* \bh_i^H\bx_t )- d_i^R \Re(s_{i,t})  \leq d_i^R - c_{i,t}^R 
~\Rightarrow~ {\sf CSEP}^R_{i,t}  \leq 1 - \sqrt{1 - \eps_i},
\end{aligned}
\end{equation}
Similarly, the result in~\eqref{eq:sep_suff_re} also holds for ${\sf CSEP}^I_{i,t}$ by replacing ``$R$'' with ``$I$'' and ``$\Re$'' with ``$\Im$''.
Reorganizing~\eqref{eq:sep_suff_re} in a vector form yields~\eqref{eq:SEP_const1}.
The proof is done.

\section{Proof of Theorem \ref{Them1}}
\label{sec:proof_thm1}

We first prove $f_{\sf SLP} \le  f_{{\sf ZF}}$.
As the ZF scheme \eqref{eq:ZF} is feasible to Problem~\eqref{eq:TTP_thm}, we have
\begin{equation*}
	\begin{aligned}
	f_{\sf SLP} &\le \mathbb{E}_{\bs_t}\Big[\big\|\bx_t^{\sf ZF}\big\|_2^2 \Big]
= \mathbb{E}_{\bs_t} \left[ \|  \balpha_c \diamond \bs_{t}  \|_\bR^2 \right]
= \alpha^2 \rho\text{Tr}(\bR).
	\end{aligned}
\end{equation*}
Note that the last equation above is due to
$\mathbb{E}[s_{i,t}^* s_{j,t}] = 0$ for $i \ne j$ and
$\mathbb{E}[|s_{i,t}|^2] = \rho$.
Next, we prove $\kappa f_{{\sf ZF}} \le f_{\sf SLP}$.
The following two lemmas will be required, and their proofs are shown in the Appendices \ref{sec:proof_lemma1}-\ref{sec:proof_lemma2}.
\begin{Lemma}\label{Lemma1}
Consider
\begin{equation}\label{eq:lemma1}
\begin{aligned}
p^\star =
\min_{ \bx \in \Cbb^K }
\| \bx + \bb \|_{\bA}^2 \quad
{\rm s.t.} ~ -\bc \le_c \bx \le_c \bc,
\end{aligned}
\end{equation}
where $\bA$ is Hermitian positive definite.
Then, for any $\beta \ge 0$, we have
$$p^\star \ge \bb^H( \bA - \bA(\bA + \beta \bI)^{-1}\bA )\bb - \beta \|\bc\|_2^2.$$
\end{Lemma}

\begin{Lemma}\label{Lemma2}
Suppose $\bR$ is Hermitian positive definite.
\begin{enumerate}[(a)]
  \item For any $|\bvarphi| = \bone$, the matrices $\bR_{\bvarphi} \triangleq \text{Diag}(\bvarphi)^H \bR \text{Diag}(\bvarphi)$ and $\bR$ share the same eigenvalues.
  \item Let $\tilde{\bR}_\bvarphi = \bR_\bvarphi -  \bR_\bvarphi (\bR_\bvarphi+\beta \bI)^{-1}\bR_\bvarphi$, where $\beta\geq 0$.
We have
\begin{equation}\label{eq:eigen_phi}
\lambda_i(\tilde{\bR}_{\bm \varphi}) =  \frac{\lambda_i(\bR)  \beta}{\lambda_i(\bR)  + \beta},~ \forall i,
\end{equation}
where $\lambda_i(\bX)$ denotes the $i$th largest eigenvalue of $\bX$.
\end{enumerate}
\end{Lemma}

Firstly, we derive a lower bound for $g(\bd,\bvarphi)$.
Denote $\setI$ as the set of all ICPs of $\setS$, and let $\setD_{\setI}$ be the uniform distribution on $\setI$.
It holds that, for any $\beta \ge 0$,
\begin{equation}\label{eq:thm_proof1}
\begin{aligned}
&  g(\bd,\bvarphi)
=\frac{1}{(2L)^{2K}} \sum_{\bs_t \in \setS^K} \Big[ \min_{ \bu_t \in \setU(\bs_t,\bd) }\| \bd \diamond \bs_{t}  + \bu_t  \|_{\bR_{\bvarphi}}^2 \Big] \\
&\ge\frac{1}{(2L)^{2K}} \sum_{\bs_t \in \setI^K} \Big[ \min_{ -\bd + \balpha_c \le_c \bu_t \le_c \bd - \balpha_c }\|  \bd \diamond \bs_{t}  + \bu_t  \|_{\bR_{\bvarphi}}^2 \Big] \\
&=\Big( 1-\frac{1}{L} \Big)^{2K}\! \mathbb{E}_{{\bs}_t \sim \setD_{\setI}^K}\! \Big[ \! \min_{ -\bd + \balpha_c \le_c \bu_t \le_c \bd - \balpha_c } \!\! \| \bd \! \diamond \! {\bs}_{t} \!  + \! \bu_t  \|_{\bR_{\bvarphi}}^2 \Big]\\
&\ge \Big( 1-\frac{1}{L} \Big)^{2K}\! \mathbb{E}_{{\bs}_t \sim \setD_{\setI}^K} \Big[ (\bd \diamond {\bs}_t)^H \tilde{\bR}_\bvarphi (\bd \diamond {\bs}_t) \Big] - \beta \|\bd - \balpha_c\|_2^2  \\
&= \Big( 1-\frac{1}{L} \Big)^{2K}\! \sum_{i=1}^K \Big( \frac{\bar{\rho}}{2}\tilde{r}_{\bvarphi,i} |d_i|^2 - \beta |d_i - \alpha_c|^2 \Big),
\end{aligned}
\end{equation}
where
$\tilde{\bR}_\bvarphi \! = \! \bR_\bvarphi -  \bR_\bvarphi (\bR_\bvarphi+\beta \bI)^{-1}\bR_\bvarphi$;
$\bar{\rho} = \Exp_{s_{i,t} \sim \setD_{\setI}} [|s_{i,t}|^2]$;
$\tilde{r}_{\bvarphi,i}$ is the $(i,i)$th entry of $\tilde{\bR}_\bvarphi$;
the second equation is due to $\setI\subset \setS$;
the fourth equation follows from Lemma~\ref{Lemma1}.

Secondly, we specify the choice of $\beta\geq 0$ to obtain the desired result.
Plugging \eqref{eq:thm_proof1} into \eqref{eq:TTP_thm} gives
\begin{equation}\label{eq:thm_proof2}
\begin{aligned}
\!\!\!f_{\sf SLP} \!\ge\!
\Big( \! 1-\frac{1}{L} \Big)^{2K}\!\!\!\! \min_{ \substack{ \bd \ge_c \balpha_c,\\ |\bvarphi| = \bone} } \sum_{i=1}^K \Big( \frac{\bar{\rho}}{2}\tilde{r}_{\bvarphi,i} |d_i|^2 \! - \! \beta |d_i - \alpha_c|^2 \Big).
\end{aligned}
\end{equation}
Observe from \eqref{eq:thm_proof2} that, given any $|\bvarphi|=\bm 1$, the optimization over $\bd$ is decoupled for each $d_i$, i.e.,
\begin{equation}\label{eq:d_i}
\min_{d_{i} \ge_c \alpha_{c} } \Big( \frac{\bar{\rho}}{2}\tilde{r}_{\bvarphi,i} |d_i|^2 - \beta |d_i - \alpha_c|^2 \Big), 
\end{equation}
which is a one-dimensional quadratic program.
We choose $\beta = (\bar{\rho}/{2} - 1)\lambda_{\min}(\bR)$.
It can be shown that
the optimal solution to Problem \eqref{eq:d_i} is $d_i =\alpha_c$.
As a result,
\begin{equation}\label{eq:thm_proof3}
	\begin{aligned}
&	\min_{ \substack{ \bd \ge_c \balpha_c,\\ |\bvarphi| = \bone} } \sum_{i=1}^K  \Big( \frac{\bar{\rho}}{2}\tilde{r}_{\bvarphi,i} |d_i|^2 - \beta |d_i - \alpha_c|^2 \Big) \\
=~& \frac{\bar{\rho}}{2} |\alpha_c|^2  \min_{|\bvarphi| = \bone } \sum_{i=1}^K \tilde{r}_{\bvarphi,i} = \frac{\bar{\rho}}{2} |\alpha_c|^2  \sum_{i=1}^K \lambda_{i}(\tilde{\bR}_{\bm \varphi}) \\
\ge~ & \frac{\bar{\rho}}{2} |\alpha_c|^2  \frac{(\frac{\bar{\rho}}{2} - 1)\lambda_{\min}(\bR)}{\lambda_{\max}(\bR) + (\frac{\bar{\rho}}{2} - 1)\lambda_{\min}(\bR)}\sum_{i=1}^K\lambda_i(\bR),
	\end{aligned}
\end{equation}
where the second equality is due to Lemma~\ref{Lemma2}(a); the last inequality is due to Lemma~\ref{Lemma2}(b) and $ \lambda_i (\bR)\leq \lambda_{\max} (\bR)$ for all $i$.
By invoking
\begin{equation*}
\begin{aligned}
 \rho \!=\! 2(2L+1)(2L-1)/3,~
\bar{\rho} \!=\!2(2L-1)(2L-3)/3,
\end{aligned}
\end{equation*}
$\sum_{i=1}^K\lambda_i(\bR) = {\rm Tr}(\bR)$, and by plugging \eqref{eq:thm_proof3} into \eqref{eq:thm_proof2}, we get $\kappa f_{{\sf ZF}} \le f_{\sf SLP}$.
This completes the proof.

\section{Proof of Lemma \ref{Lemma1}}
\label{sec:proof_lemma1}
Problem~\eqref{eq:lemma1} can be equivalently transformed to
\begin{equation*}
\begin{aligned}
p^\star =
\min_{ \bx } & ~
\| \bx + \bb \|_{\bA}^2 \\
{\rm s.t.}
& ~ \Re(x_i)^2 \le \Re(c_i)^2, \quad i=1,\dots,K, \\
& ~ \Im(x_i)^2 \le \Im(c_i)^2, \quad i=1,\dots,K.
\end{aligned}
\end{equation*}
The Lagrangian associated with the above problem is
\begin{equation*}
\begin{aligned}
L (\bx, \boldsymbol{\nu}^R, \boldsymbol{\nu}^I) =
\| \bx + \bb \|_{\bA}^2 + \sum_{i=1}^K \nu_i^R (\Re(x_i)^2 - \Re(c_i)^2)  + \sum_{i=1}^K \nu_i^I (\Im(x_i)^2 - \Im(c_i)^2),
\end{aligned}
\end{equation*}
where $\boldsymbol{\nu}^R  \ge  \bzero$ and $\boldsymbol{\nu}^I \ge \bzero$ are the dual variables.
By the Lagrangian duality theory, it holds that $p^\star \ge \inf_{\bx} L (\bx, \boldsymbol{\nu}^R, \boldsymbol{\nu}^I)$ for any $\boldsymbol{\nu}^R \ge \bzero$ and $\boldsymbol{\nu}^I \ge \bzero$.
By choosing $\boldsymbol{\nu}^R = \boldsymbol{\nu}^I = \beta \bone$ with $\beta \ge 0$, we have
\begin{equation*}
	\begin{aligned}
	p^\star
	&\ge \inf_{\bx} L (\bx, \beta \bone, \beta \bone) \\
    &= \inf_{\bx} \{ \| \bx + \bb \|_{\bA}^2 + \beta \|\bx\|_2^2 - \beta \|\bc\|_2^2 \} \\
    &= \bb^H( \bA - \bA(\bA + \beta \bI)^{-1}\bA )\bb - \beta \|\bc\|_2^2,
	\end{aligned}
\end{equation*}
where the last equation is due to the fact that the optimization problem in the second equation 
has $\bx = -(\bA + \beta \bI)^{-1}\bA\bb$ as its optimal solution.
The proof is complete.

\section{Proof of Lemma \ref{Lemma2}}
\label{sec:proof_lemma2}
Denote the eigendecomposition of $\bR$ as $\bV \boldsymbol{\Lambda} \bV^H$, where $\bV \! \in \! \Cbb^{K\times K}$
is unitary, and $\boldsymbol{\Lambda}\! \in \! \Cbb^{K\times K}$ is diagonal whose diagonal elements
are the eigenvalues of $\bR$.
We have
$$\bR_{\bvarphi} = \text{Diag}(\bvarphi)^H \bV \boldsymbol{\Lambda} \bV^H \text{Diag}(\bvarphi) = \hat{\bV}\boldsymbol{\Lambda}\hat{\bV}^H,$$
where $\hat{\bV} = \text{Diag}(\bvarphi)^H \bV$.
It is seen that $\hat{\bV}$ is also unitary.
This means that the diagonal elements of $\boldsymbol{\Lambda}$ are also the eigenvalues of $\bR_{\bvarphi}$.
Therefore, $\bR_{\bvarphi}$ and $\bR$ share the same eigenvalues.

From the definition of $\tilde{\bR}_{\bm \varphi}$, we have
\begin{equation*}
\begin{aligned}
\tilde{\bR}_\bvarphi &= \bR_\bvarphi -  \bR_\bvarphi (\bR_\bvarphi+\beta \bI)^{-1}\bR_\bvarphi \\
&= \hat{\bV} (\boldsymbol{\Lambda} - \boldsymbol{\Lambda}(\boldsymbol{\Lambda} + \beta \bI)^{-1}\boldsymbol{\Lambda} ) \hat{\bV}^H.
\end{aligned}
\end{equation*}
It follows that the eigenvalues of $\tilde{\bR}_\bvarphi$ are
\begin{equation*}
\lambda_i(\tilde{\bR}_{\bm \varphi}) =  \lambda_i(\bR) - \frac{\lambda_i^2(\bR)}{\lambda_i(\bR)+\beta}
=\frac{\lambda_i(\bR)  \beta}{\lambda_i(\bR)  + \beta},~ \forall i.
\end{equation*}
The proof is complete.

\section{Derivation of Algorithm~\ref{AL_pro}}
\label{app:alg1_proof}
Observe from Problem~\eqref{eq:proj2} that given $d \ge \alpha$, the optimal $u_{t}$'s can be explicitly expressed as
\begin{equation}\label{ud}
u_{t} =
\begin{cases}
d - c_{t},& \quad \text{if} \  \tilde{u}_{t} \ge d -  c_{t} , \\
- d + a_{t},& \quad \text{if} \ \tilde{u}_{t}  \le - d + a_{t} , \\
\tilde{u}_{t},& \quad \text{otherwise},
\end{cases}
\end{equation}
for $t = 1,\dots,T$.
Therefore, by plugging \eqref{ud} into Problem~\eqref{eq:proj2}, the variable to optimize is only ${d}$, which leads to a simplified problem.
However, different intervals of $d$ will result in different forms of the ${u}_{t}$'s, and thus different forms of the summation term $\sum_{t=1}^T (u_{t} - \tilde{u}_{t})^2$ in the objective function.
We next show the formulations for $d$ lying in different intervals.
Define the  set that includes all the possible boundary points of the intervals of $d$ as
$$ \tilde{\setD}  \triangleq  \{{\alpha}\} \cup \setD_1 \cup \setD_2 \cup \{+\infty\},$$
where $ \setD_1 \! \triangleq \! \{ {c}_{t} + \tilde{u}_{t}, \forall t ~\big|~ {c}_{t} + \tilde{u}_{t} \! \ge \! {\alpha} \}$ and $\setD_2 \! \triangleq \! \{ {a}_{t} - \tilde{u}_{t} , ~\forall t ~\big|~ {a}_{t} - \tilde{u}_{t} \ge {\alpha} \}$.
Sort all the elements in $\tilde{\setD}$ in ascending order, which results in $\setD \! \triangleq \! \{\omega_1,\dots,\omega_{{\rm card}(\setD)}\}$ with $\omega_1 \! \leq \! \cdots \leq \omega_{{\rm card}(\setD)}$.
Then, the feasible region of ${d}$ can be divided into ${\rm card}(\setD)-1$ intervals, i.e.,
\begin{equation*}
\omega_{p} \leq  {d} \leq \omega_{p+1},\quad p = 1,\dots,{\rm card}(\setD)-1.
\end{equation*}
By \eqref{eq:proj2} and \eqref{ud}, the optimal ${d}$ restricted on the $p$th interval is obtained by solving the following quadratic program:
\begin{equation*}
\begin{aligned}
\min_{{d}} & ~ \sum_{t\in {\cal T}_p} ( {d}- {c}_{t} - \tilde{u}_{t})^2 + \sum_{t\in {\cal L}_p} (- {d} + {a}_{t} - \tilde{u}_{t})^2 + ( {d}-\tilde{d})^2  \\
{\rm s.t.} & ~ ~ \omega_{p} \leq  {d}  \leq \omega_{p+1},
\end{aligned}
\end{equation*}
where
${\cal T}_p \triangleq \{ t ~\big|~ \omega_{p+1} \le {c}_{t} + \tilde{u}_{t} \} $ and
${\cal L}_p \triangleq \{ t ~\big|~ \omega_{p+1} \le {a}_{t} - \tilde{u}_{t} \} $.
The above problem has a closed-form solution given by ${d}^{p} = \max \{ \omega_{p}, \min\{ \omega_{p+1}, \hat{d}^{p}\} \}$, where
\begin{equation*}
\hat{d}^{p} = \frac{ \sum_{t\in {\cal T}_p} ({c}_{t} +\tilde{u}_{t}) + \sum_{t\in {\cal L}_p} ({a}_{t} - \tilde{u}_{t}) + \tilde{d}}{ 1 +  {\rm card}({\cal T}_p)  + {\rm card}({\cal L}_p) }.
\end{equation*}
The corresponding optimal value for the $p$th interval is
\begin{equation*}
\textstyle f^{p} =  \sum\limits_{t\in {\cal T}_p} ( {d}^{p}- {c}_{t} - \tilde{u}_{t})^2 + \! \sum\limits_{t\in {\cal L}_p} (- {d}^{p} + {a}_{t} - \tilde{u}_{t})^2 + ( {d}^{p}-\tilde{d})^2.
\end{equation*}
After computing the $f^{p}$'s for all $p$, the ${d}^{p}$ that leads to the minimum $f^{p}$ is the optimal solution to Problem~\eqref{eq:proj2}.

\section{Proof of Fact~\ref{fac:zf}}
\label{app:factzf_proof}
Under the assumptions of Fact~\ref{fac:zf}, Problem~\eqref{eq:TTP_recap} becomes
	\begin{align}
    &\min_{\bm d \ge_c \bm \alpha_c, |\bm \varphi| = \bm 1 } ~ \mathbb{E}_{\bm s_t} \Big[ \|  \bm d \diamond \bm s_{t}  \|_{\bm R_{\bm \varphi}}^2 \Big] \notag \\
    =~&\min_{\bm d \ge_c \bm \alpha_c, |\bm \varphi| = \bm 1 } ~ \mathbb{E}_{\bm s_t} \Big[ \sum_{i=1}^{K} \sum_{j=1}^{K} \varphi_i^* (d_i \diamond s_{i,t})^* r_{ij} \varphi_j (d_j \diamond s_{j,t}) \Big]\notag \\
	=~&\min_{\bm d \ge_c \bm \alpha_c, |\bm \varphi| = \bm 1 } ~ \frac{\rho}{2} \sum_{i=1}^{K}  |\varphi_i|^2 r_{ii} |d_i|^2, \label{eq:ZF_fact3_opt}
	\end{align}
where
the last equation is due to $\mathbb{E}[(d_i \diamond s_{i,t})^*(d_j \diamond s_{j,t})] = 0$ for all $i \ne j$, and $\mathbb{E}[|d_i \diamond s_{i,t}|^2] = |d_i|^2 \rho /2$ for all $i$.
We see that $(\bm d, \bvarphi) = (\bm \alpha_c, \bone)$ is  an optimal solution to \eqref{eq:ZF_fact3_opt}.


\bibliographystyle{IEEEtran}

\clearpage
\setcounter{page}{1}
\title{Supplemental Material of ``Symbol-Level Precoding Through the Lens of Zero Forcing and Vector Perturbation''}
\maketitle

\setcounter{section}{0}

\section{Convergence Analysis}
\label{sec:basic_sol_pg}

\subsection{Preliminaries}

We begin by introducing some notations and definitions.
Let $f: \Rbb^n \rightarrow \Rbb\cup \{ +\infty \}$ be an extended-real-valued function.
Let $\setX \subseteq \Rbb^n$, and let
\[
I_{\setX}(\bx) =\begin{cases}
	0 , & \mbox{if } \bx \in \setX \\
	+\infty , & \mbox{otherwise}
\end{cases}
\]
be the indicator function of ${\cal X}$.
As defined previously, $\langle \cdot, \cdot \rangle$ is the inner product; $\nabla f(\bx)$ is the gradient of a differentiable function $f$.
A differentiable function $f$ is said to have $L_f$-Lipschitz continuous gradient on $\setX$ if
\begin{equation}\label{eq:Lip_con}
	\|\nabla  f(\bx) -\nabla f(\by) \|_2 \leq L_f \| \bx -\by \|_2, \ \forall \bx,\by \in {\cal X}.
\end{equation}
Also, ${\rm dist}(\bx,{\cal X}) \triangleq \inf_{\by \in {\cal X}} \|\bx-\by\|_2$ is the distance between $\bx$ and $\setX$.

Consider the minimization problem
\begin{equation}\label{eq:opt_general}
  \min_{\bx } f(\bx).
\end{equation}
It is a mathematically subtle subject to define what is a stationary point of Problem~\eqref{eq:opt_general} when $f$ is nonconvex and nonsmooth \cite{li2020understanding}.
Here we adopt the notion of critical points.
A point $\hat{\bx}$ is said to be a critical point of Problem~\eqref{eq:opt_general} if
\begin{equation}\label{eq:FO_cond}
    \bm 0 \in \partial f(\hat{\bx}),
\end{equation}
where $\partial f( \bx )$ is the limiting subdifferential of $f$ at $\bx$;
see, e.g., \cite{rockafellar2009variational,li2020understanding} and the references therein, for details.
If $f$ is differentiable, then $\partial f( \bx ) = \nabla f(\bx)$.
If $f$ is a sum of two functions, $f(\bx) = f_1(\bx) + f_2(\bx)$,
it is generally {\em not} true that $\partial f(\bx) = \partial f_1(\bx) +  \partial f_2(\bx)$.
But if $f_1$ is differentiable and $f(\bx) = f_1(\bx) + I_{\setX}(\bx)$, then we do have $\partial f(\bx) = \nabla f_1 (\bx) + \partial I_\setX(\bx)$.

The above concepts apply straightforwardly to functions of complex inputs, i.e., $f:\Cbb^n \rightarrow \Rbb$; see, e.g., \cite[Section~I.B]{shao2019framework}.

\subsection{PG Method for Nonconvex Constrained Problems}
\label{sec:hhhh}

Consider the following problem
\begin{equation*}
	\begin{aligned}
		\min_{\bx \in \setX } & ~ f(\bx),
	\end{aligned}
\end{equation*}
where $f:\Rbb^n \to \Rbb$ is differentiable; $\setX \subseteq \Rbb^n$ can be nonconvex.
To put into context, we rewrite the problem as
\begin{equation}\label{eq:pg_pro}
\begin{aligned}
\min_{\bx} & ~F(\bx) \triangleq f(\bx) +I_{\cal X} (\bx).
\end{aligned}
\end{equation}
Consider the following PG method for finding an approximate solution to Problem~\eqref{eq:pg_pro}:
given a starting point $\bx^0 \in \setX$ and a parameter $0 < \alpha < 1$, solve
\begin{align}
\bx^{\ell+1}  & \in
 \arg \min_{\bx} \| \bx - ( \bx^{\ell} - \alpha L_\ell^{-1} \nabla f(\bx^\ell) ) \|_2^2 + I_\setX(\bx) \nonumber \\
& = \Pi_\setX( \bx^{\ell} - \alpha L_\ell^{-1} \nabla f(\bx^\ell) ),
  \label{eq:pg_pro_alg1}
\end{align}
for $\ell= 0,1,\cdots$, where $L_\ell$ is such that
\begin{align}
f(\bx^{\ell+1}) \leq f(\bx^\ell) + \langle \nabla f(\bx^\ell), \bx^{\ell+1} - \bx^\ell \rangle + \frac{L_\ell}{2} \| \bx^{\ell+1} - \bx^\ell \|_2^2.
\label{eq:pg_pro_alg2}
\end{align}
The above $L_\ell$'s can be obtained by standard methods;
see, e.g., \cite{beck2017first}, for details.
We are interested in the question of under what conditions the above PG method will lead to convergence to a critical point of Problem~\eqref{eq:pg_pro}.

The above convergence question is relevant to the phase optimization problem in the main paper;
specifically, in Problem \eqref{eq:phi} and in the AM of the PPAP minimization in Section \ref{sec:PPAP}.
The problems are instances of Problem~\eqref{eq:pg_pro}, while Algorithm~\ref{AL_PG} is identical to the above PG method.

In signal processing, convergence analyses of the PG methods are arguably well-known for the case of convex $\setX$; see, e.g., \cite{beck2017first} and the references therein.
Convergence analyses for nonconvex $\setX$ are, however, possibly less known.
In fact, the convergence question for nonconvex $\setX$ was already answered by mathematical optimization researchers \cite{attouch2013convergence,xu2017globally} as a special case of some general frameworks.
In particular,  Attouch {\em et al.}~\cite{attouch2013convergence} developed a powerful framework that shows critical-point convergence for a general class of problems, and they did so by using the  Kurdyka–Łojasiewicz property elegantly.

While the convergence question was solved, there is a much simpler convergence proof if we focus just on Problem~\eqref{eq:pg_pro}.
The proof, interestingly, resembles that for the more well-known case of convex $\setX$ (e.g., \cite{beck2017first}).
For the reader's interest, we show the proof.
Let us first describe the result.

\begin{Prop}\label{prop:conv_PG}
Consider the PG method \eqref{eq:pg_pro_alg1}--\eqref{eq:pg_pro_alg2} for Problem~\eqref{eq:pg_pro}.
Suppose that
\begin{enumerate}[i)]
  \item $f^{\star} \triangleq \inf_{ \bx \in {\cal X}}  f(\bx) > - \infty$;
  \item $f$ has $L_f$-Lipschitz continuous gradient on ${\cal X}$;
  \item every $L_\ell$ satisfies $L_{\ell} \in [c_1 L_f  , c_2 L_f  ]$ for some $0<c_1 \leq c_2<+\infty$ (true for a pertinent choice of $L_{\ell}$ \cite{beck2017first}).
\end{enumerate}
Then,
\begin{enumerate}[(a)]
\item the sequence $\{ \bx^\ell \}_{\ell \geq 0}$ generated by the PG method satisfies the descent property $f(\bx^\ell) \geq f(\bx^{\ell+1})$ for all $\ell \geq 0$;
\item $\{ \bx^\ell \}_{\ell \geq 0}$ exhibits a sublinear convergence rate property
\begin{equation}\label{eq:PG_conv}
	\begin{split}
		\min_{\ell=0,\ldots, J} {\rm dist}(\bm 0, \partial F(\bx^{\ell+1}))
		\leq  \sqrt{\frac{C}{J+1}},
	\end{split}
\end{equation}
where $$\textstyle C =  \frac{4\big(1+ \frac{c_2^2}{\alpha^2} \big)L_f  (f(\bx^{0}) -f^{\star})}{ \big( \frac{1}{\alpha} -1 \big) c_1 };$$

\item
any limit point of $\{\bx^\ell\}_{\ell \geq 0}$ is a critical point of Problem~\eqref{eq:pg_pro}.
\end{enumerate}
\end{Prop}
It is worth noting that the convergence rate result in {\it (b)} was not explicitly mentioned in the aforementioned literature, although the key ideas leading to {\it (b)} follow those in the literature.

\medskip
\noindent{\it Proof of Proposition~\ref{prop:conv_PG}}:
Firstly we show {\it (a)}.
Define
\[
h(\bx|\tilde{\bx},\beta) = f(\tilde{\bx}) + \langle \nabla f(\tilde{\bx}), \bx - \tilde{\bx} \rangle + \frac{\beta}{2} \| \bx - \tilde{\bx} \|_2^2,
\]
and rewrite \eqref{eq:pg_pro_alg1} and \eqref{eq:pg_pro_alg2} as
\begin{align}
	\bx^{\ell+1} & \in \arg \min_{\bx} h(\bx| \bx^\ell, L_\ell/\alpha) + I_\setX(\bx),
	\label{eq:proof_conv_t1} \\
	f(\bx^{\ell+1}) & \leq h(\bx^{\ell+1}|\bx^\ell,L_\ell), \label{eq:proof_conv_t2}
\end{align}
respectively.
We see from \eqref{eq:proof_conv_t1} that
\begin{equation} \label{eq:proof_conv_t3}
	h(\bx^{\ell+1}| \bx^\ell, L_\ell/\alpha) \leq h(\bx^{\ell}| \bx^\ell, L_\ell/\alpha) = f(\bx^\ell).
\end{equation}
Applying \eqref{eq:proof_conv_t3}  to \eqref{eq:proof_conv_t2} gives
\begin{equation} \label{eq:proof_conv_t4}
	f(\bx^\ell) - f(\bx^{\ell+1}) \geq \left( \tfrac{1}{\alpha} - 1 \right) \frac{L_\ell}{2} \| \bx^{\ell+1} - \bx^\ell \|_2^2,
\end{equation}
which leads to {\it (a)}.

Second we show {\it (b)}. From \eqref{eq:proof_conv_t4},
\begin{align}
	& f(\bx^0) -f^\star \geq f(\bx^0) - f(\bx^{J+1}) \nonumber \\
	\geq & \left( \tfrac{1}{\alpha} - 1 \right) \frac{c_1 L_f}{2}  \sum_{\ell=0}^J \| \bx^{\ell+1} - \bx^\ell \|_2^2 \nonumber \\
	\geq & \left( \tfrac{1}{\alpha} - 1 \right) \frac{c_1 L_f (J+1)}{2}  \min_{\ell =0,1,\ldots,J} \| \bx^{\ell+1} - \bx^\ell \|_2^2. \label{eq:proof_conv_t5}
\end{align}
Moreover, since $\bx^{\ell+1}$ is a critical point of the problem in \eqref{eq:proof_conv_t1},  $\bx^{\ell+1}$ satisfies
\begin{align}
	\bm 0  &
	\in \nabla h( \bx^{\ell+1} | \bx^\ell, L_\ell/\alpha ) +
	\partial I_\setX(\bx^{\ell+1})  \nonumber \\
	& = \nabla f(\bx^\ell) + \frac{L_\ell}{\alpha} ( \bx^{\ell+1} - \bx^\ell ) + \partial I_\setX(\bx^{\ell+1}). \label{eq:proof_conv_t6}
\end{align}
It follows that
\begin{subequations} \label{eq:proof_conv_t7}
	\begin{align}
		& {\rm dist}(\bm 0, \partial F(\bx^{\ell+1}))^2 = {\rm dist}(\bm 0, \nabla f(\bx^{\ell+1}) + \partial I_\setX(\bx^{\ell+1})  )^2   \nonumber \\
		& \quad = \min_{\bv \in \partial I_\setX(\bx^{\ell+1})} ~ \| \nabla f(\bx^{\ell+1}) + \bv \|_2^2 \nonumber \\
		& \quad \leq \left\| \nabla f(\bx^{\ell+1})  - \left[ \nabla f(\bx^\ell) + \tfrac{L_\ell}{\alpha} ( \bx^{\ell+1} - \bx^\ell ) \right] \right\|_2^2
		\label{eq:proof_conv_t7a} \\
		& \quad \leq 2 L_f^2 \| \bx^{\ell+1} - \bx^\ell \|_2^2 + \tfrac{2 c_2^2 L_f^2}{\alpha^2} \| \bx^{\ell+1} - \bx^\ell \|_2^2, \label{eq:proof_conv_t7b}
	\end{align}
\end{subequations}
where \eqref{eq:proof_conv_t7a} is due to \eqref{eq:proof_conv_t6};
\eqref{eq:proof_conv_t7b} is due to $\| \bx + \by \|_2^2 \leq 2\| \bx \|_2^2 + 2 \| \by \|_2^2$ and the assumptions in ii)--iii).
Applying \eqref{eq:proof_conv_t5} to \eqref{eq:proof_conv_t7} leads to the result in {\it (b)}.

Lastly we show {\it (c)}.
Suppose that there exists a convergent subsequence $\{ \bx^{\ell_i} \}_{i\geq 0}$ of $\{ \bx^\ell \}_{\ell \geq 0}$.
Let $\bar{\bx}$ be the limit of $\{ \bx^{\ell_i} \}_{i\geq 0}$.
Observe that, for all $\bx \in \setX$,
\begin{subequations} \label{eq:proof_conv_t8}
	\begin{align}
		h(\bx| \bx^{\ell_i}, c_2 L_f/\alpha) & \geq  h( \bx| \bx^{\ell_i}, L_{\ell_i}/\alpha)
		\nonumber \\
		&
		\geq h( \bx^{\ell_i+1} | \bx^{\ell_i}, L_{\ell_i}/\alpha) \label{eq:proof_conv_t8a} \\
		& \geq f(\bx^{\ell_i+1}) \geq f(\bx^{\ell_{i+1}}), \label{eq:proof_conv_t8b}
	\end{align}
\end{subequations}
where \eqref{eq:proof_conv_t8a} is due to \eqref{eq:proof_conv_t1};
\eqref{eq:proof_conv_t8b} is due to \eqref{eq:proof_conv_t2} and
the result $f(\bx^\ell) \geq f(\bx^{\ell+1})$ in {\it (a)}.
Taking limit $i\rightarrow \infty$ on both sides of \eqref{eq:proof_conv_t8} gives
\[
h(\bx| \bar{\bx}, c_2 L_f/\alpha) \geq f(\bar{\bx}), \quad \text{for all $\bx \in \setX$,}
\]
which implies
\begin{equation} \label{eq:proof_conv_t9}
\bar{\bx} \in \arg \min_{\bx} h(\bx| \bar{\bx}, c_2 L_f/\alpha) + I_\setX(\bx)
\end{equation}
Since $\bar{\bx}$ is a critical point of Problem \eqref{eq:proof_conv_t9}, we have
\[
\bm 0 \in \nabla h(\bar{\bx}| \bar{\bx}, c_2 L_f/\alpha) + \partial I_\setX(\bar{\bx}) = \nabla f(\bar{\bx})  +  \partial I_\setX(\bar{\bx})
\]
which shows that $\bar{\bx}$ is a critical point of Problem~\eqref{eq:pg_pro}, the result in ${\it (c)}$.
\hfill $\blacksquare$

\subsection{A Proximal AM Method and Its Convergence}

Consider the problem
\begin{equation}
  \min_{\bx_1 \in \setX_1, \bx_2 \in \setX_2 } f(\bx_1 ,\bx_2 ), \nonumber
\end{equation}
where $f: \Rbb^{n_1+ n_2}\rightarrow \Rbb$ is differentiable; $\setX_1 \subseteq \Rbb^{n_1}$ and $\setX_2 \subseteq \Rbb^{n_2}$ can be nonconvex.
Let us define $\bx = (\bx_1,\bx_2)$ and $\setX = \setX_1 \times \setX_2$, and rewrite the problem as
\begin{equation} \label{eq:opt_prob}
	\min_{\bx} F(\bx) \triangleq f(\bx ) + I_\setX(\bx);
\end{equation}
note that $I_\setX(\bx) = I_{\setX_1}(\bx_1) \times I_{\setX_2}(\bx_2)$.
We are interested in the following proximal method for finding an approximate solution to Problem~\eqref{eq:opt_prob}: given $\bx^0 \in \setX$, $\tau > 0$,
\begin{subequations}\label{eq:Prox_AM2}
	\begin{align}
		& \bx_1^{k+1} \approx \arg\min_{ \bx_1 }  f (\bx_1,\bx_2^{k}) + I_{\setX_1}(\bx_1) ,\label{eq:AM_xi2}
		\\
		&  \bx_2^{k+1} \approx \arg\min_{\bx_2} f (\bx_1^{k+1},\bx_2)+\frac{\tau}{2} \|  \bx_2 - \bx_2^k \|_2^2 +  I_{\setX_2}(\bx_2), \label{eq:AM_phi2}
	\end{align}
\end{subequations}
for $\ell=0,1,\ldots,$ where ``$\approx$'' means that we solve the problems approximately.
The above proximal AM method is a variant of the proximal AM in \cite{attouch2010proximal};
the notable difference is that the original proximal AM method requires the problems in \eqref{eq:AM_xi2}--\eqref{eq:AM_phi2} to be exactly solved.
As a variation of \cite[Lemma 3.1]{attouch2010proximal}, we have the following critical-point convergence result.

\begin{Prop}\label{prop:conv}
	Consider the proximal AM method \eqref{eq:Prox_AM2} for  Problem \eqref{eq:opt_prob}.
	Suppose that
	\begin{enumerate}[i)]
		\item $f^{\star} \triangleq \inf_{ \bx \in {\cal X}}  f(\bx) > - \infty$, and $\setX$ is closed;
		\item $ f$ has $L_f$-Lipschitz continuous gradient on $\setX$;
		\item $\bx^{k+1}_1$ is a critical point of the problem in \eqref{eq:AM_xi2}, and
		$\bx_2^{k+1}$ is a critical point of the problem in \eqref{eq:AM_phi2};
		\item the following coordinate descent property holds
			\begin{align*}
				f(\bx_1^k,\bx_2^{k}) & \geq f(\bx_1^{k+1},\bx_2^{k}), \\
				f(\bx_1^{k+1},\bx_2^{k}) & \geq f(\bx_1^{k+1},\bx_2^{k+1}) + \frac{\tau}{2} \|  \bx_2^{k+1} - \bx_2^k \|_2^2.
			\end{align*}

	\end{enumerate}
	Then,
	\begin{enumerate}[(a)]
		\item the sequence $\{ \bx^k \}_{k \geq 0}$ generated by the proximal AM method satisfies the descent property $f(\bx^k) \geq f(\bx^{k+1})$ for all $k \geq 0$;
		\item $\{ \bx^k \}_{k \geq 0}$ exhibits a sublinear convergence rate property
		\[
		\min_{k=0,1,\cdots,K} {\rm dist}(\bm 0, \partial F(\bx^{k+1})) \leq \sqrt{\frac{C}{K+1}},
		\]
		where $C= 2 ( L_f^2/\tau + \tau)( f(\bx^0) - f^\star)$;

\item  any limit point  of $\{ \bx^k \}_{k \geq 0}$ is a critical point of Problem~\eqref{eq:opt_prob}.
	\end{enumerate}

\end{Prop}

We will give the proof later.
There are applications for which assumption {\it ii)} in Proposition \ref{prop:conv}, the Lipschitz continuous gradient assumption with $f$, may not be satisfied.
For such cases we can consider the following alternative.

\begin{Corollary}\label{cor:conv_AM}
 	The same result in Proposition~\ref{prop:conv} holds if we replace assumption ii) by the following conditions:
	\begin{enumerate}
		\item[ii.a)] $f$ is twice differentiable on $\setX$;
		\item[ii.b)] $\{ \bx^k \}_{k \geq 0}$ is a bounded sequence.
	\end{enumerate}
\end{Corollary}

In the following we give the proof of Proposition~\ref{prop:conv} and Corollary~\ref{cor:conv_AM}.
The reader may jump to the next subsection for the application to the SLP designs in the main paper.

\medskip
\noindent{\it Proof of Proposition~\ref{prop:conv}}:
From assumption {\it iv)}, we see that
\begin{equation} \label{eq:tt1}
	\frac{\tau}{2} \| \bx_2^{k+1} - \bx_2^k \|_2^2 \leq f(\bx^k) - f(\bx^{k+1}).
\end{equation}
We thereby have {\it (a)}.
To show {\it (b)}, observe from \eqref{eq:tt1} that
\begin{equation} \label{eq:tt2}
	\frac{\tau}{2} \sum_{k=0}^{K} \| \bx_2^{k+1} - \bx_2^k \|_2^2 \leq f(\bx^0) - f^\star < + \infty,
\end{equation}
which implies
\begin{equation} \label{eq:tt2_5}
	 \min_{k=0,1,\ldots,K} \| \bx_2^{k+1} - \bx_2^k \|_2^2 \leq \frac{2}{\tau (K+1)} ( f(\bx^0) - f^\star ).
\end{equation}
Moreover, from assumption {\it iii)}, we have
\begin{align*}
\bzero & \in \nabla_{\bx_1} f (\bx_1^{k+1}, \bx_2^{k}) + \partial I_{\mathcal{X}_1}(\bx_1^{k+1}), \\
\bzero & \in \!  \nabla_{\bx_2} f (\bx_1^{k+1}, \bx_2^{k+1}) +  \tau (\bx_2^{k+1} - \bx_2^{k}) \! + \! \partial I_{\setX_2}(\bx_2^{k+1});
\end{align*}
($\nabla_{\bx_i} f$ denotes the gradient w.r.t. $\bx_i$).
The above equations can be rewritten as
\begin{equation} \label{eq:tt3}
	\bv^{k+1} \in \nabla f(\bx^{k+1}) + \partial I_\setX(\bx^{k+1}),
\end{equation}
where $\bv^{k+1}= (\bv_1^{k+1}, \bv_2^{k+1})$ has
\begin{align}
	\bv_1^{k+1} & = \nabla_{\bx_1} f (\bx_1^{k+1}, \bx_2^{k+1}) - \nabla_{\bx_1} f (\bx_1^{k+1}, \bx_2^{k})   \label{eq:tt4}  \\
	{\bv}_2^{k+1} = & - \tau (\bx_2^{k+1} - \bx_2^{k}). \label{eq:tt5}
\end{align}
By assumption {\it ii)}, we can bound ${\bv}_1^{k+1}$ as
\begin{align} \label{eq:tt6}
	\| {\bv}_1^{k+1} \|_2 \leq L_f \| \bx_2^{k+1} - \bx_2^{k} \|_2
\end{align}
It follows from \eqref{eq:tt3}--\eqref{eq:tt6} that
\begin{align}
	{\rm dist}(\bzero, \partial F(\bx^{k+1}))^2 & = {\rm dist}(\bzero, \nabla f(\bx^{k+1}) + \partial I_{\setX}(\bx^{k+1}))^2 \nonumber \\
	& \leq \|  \bv^{k+1}  \|_2^2 \nonumber \\
	& \leq ( L_f^2 + \tau^2 ) \| \bx_2^{k+1} - \bx_2^{k} \|_2^2. \label{eq:tt7}
\end{align}
Applying \eqref{eq:tt2_5} to \eqref{eq:tt7} leads to {\it (b)}.

To show {\it (c)}, suppose that $\{ \bx^k \}_{k\geq 0}$ has a convergent subsequence $\{ \bx^{k_i} \}_{i \geq 0}$.
Let $\bar{\bx}$ be the limit point of $ \{\bx^{k_i}\}_{i\geq 0}$.
Since $f$ is continuous on $\setX$ and we have $\bx^k, \bar{\bx} \in \setX$, we get $F(\bx^{k_i} )\rightarrow F(\bar{\bx})$.
From \eqref{eq:tt2} and \eqref{eq:tt7}, we observe that $\sum_{k=0}^{+\infty} \| \bv^{k+1} \|_2^2 \leq +\infty$, which means that $\bv^{k+1}\rightarrow \bm 0$ as $k\rightarrow \infty$.
Also, note from \eqref{eq:tt3} that $\bv^{k } \in \partial F(\bx^{k } )$.

As an elementary result, it is known that if $\by_i\rightarrow \bar{\by}$, $F(\by^i) \rightarrow F(\bar{\by})$, $\bu^i \in \partial F(\by^i)\rightarrow \bar{\bu}$, then $\bar{\bu}\in \partial F(\bar{\by})$; see, e.g., \cite[Remark 2.1(b)]{attouch2010proximal}.
Applying this result to our problem by $\by^i = \bx^{k_i}$, $\bu^i = \bv^{k_i}$, we get $\bm 0 \in \partial F(\bar{\bx})$. \hfill $\blacksquare$

\medskip
\noindent{\it Proof of Corollary~\ref{cor:conv_AM}}:
In the proof of Proposition~\ref{prop:conv},  we only used assumption {\it ii)} in \eqref{eq:tt6}.
Under the new assumption {\it ii.b)}, there exists a finite bound $M$ that bounds $\bx^k$; specifically, $\| \bx^k \|_2 \leq M$ for all $k$.
Since $\| \bx_1^k \|_2 \leq M$ and $\| \bx_2^k \|_2 \leq M$ are also true,
$(\bx_1^{k+1}, \bx_2^k)$ is also bounded by $M$.
Hence $\{ \bx^{k+1} \}_{k\geq 0}$ and $\{ (\bx_1^{k+1}, \bx_2^k) \}_{k \geq 0}$ lie in $\tilde{\setX} = \setX \cap \{ \bx \in \Rbb^{n_1+ n_2} \mid \| \bx \|_2 \leq M \}$, which is compact.
As an elementary fact, a twice differentiable function has Lipschitz continuous gradient on a compact set.
By setting $L_f$ in \eqref{eq:tt6} as the Lipschitz constant of $\nabla f$ on $\tilde{\setX}$, we complete the proof.
\hfill $\blacksquare$

\subsection{Application of Proximal AM to SLP Designs}

Now we study the application of the proximal AM framework in the last subsection to the SLP designs in the main paper, with the focus on critical-point convergence.

We start with the TTP-minimization SLP design \eqref{eq:TTP_recap}.
We treat the SLP design \eqref{eq:TTP_recap} as an instance of Problem \eqref{eq:opt_prob}, with
\[
f= f_{\sf TTP}, ~ \bx_1  = (\bd, \bU), ~ \setX_1 = \setW, ~ \bx_2  =\bvarphi, ~
\setX_2  = {\cal P}.
\]
The AM scheme \eqref{eq:AM} for the SLP design \eqref{eq:TTP_recap} in the main paper, upon adding a proximal term $\frac{\tau}{2} \| \bvarphi - \bvarphi^k \|_2^2$ in \eqref{eq:AM1}, is identical to the proximal AM method in \eqref{eq:Prox_AM2}.
In the main paper,
the AM scheme \eqref{eq:AM} solves Problem~\eqref{eq:AM_xi2} optimally via the APG method (Algorithm~\ref{AL_APG}); and it approximates Problem~\eqref{eq:AM_phi2} via the PG method (Algorithm~\ref{AL_PG}), which was studied in the last last subsection.
Let us add one more condition, namely, that we use $\bx_2^k$ to initialize the PG method for Problem~\eqref{eq:AM_phi2}.
Then we can verify that the assumptions   {\it iii)-iv)} in Proposition~\ref{prop:conv} are all satisfied; the PG results in Proposition~\ref{prop:conv_PG} are needed.
Hence, by Proposition~\ref{prop:conv}, we can conclude the following:
the AM scheme \eqref{eq:AM} for the SLP design \eqref{eq:TTP_recap} in the main paper, under the above described modification, guarantees convergence to a critical point if we assume that $f$ has Lipschitz continuous gradient on $\setX$.

However, there is a caveat: we are unable to show that $f$ has Lipschitz continuous gradient on $\setX$.
Fortunately we can use Corollary~\ref{cor:conv_AM}.
To describe, consider the following assumption.

\begin{Asm} \label{asm5}
For each user $i$, there exists a symbol $s_{i,t}$ such that $| \Re(s_{i,t}) | > 1$; and that, for each $i$, there exists a symbol $s_{i,t}$ such that $| \Im(s_{i,t}) | > 1$.	
\end{Asm}

\begin{Prop}\label{fact:Lip_cons}
	Consider the TTP-minimization SLP design \eqref{eq:TTP_recap}.
	Suppose that Assumption~\ref{asm5} holds.
	The AM scheme \eqref{eq:AM} under the above described modification generates a bounded sequence $\{ \bx^k \}_{k \geq 0}$.
	By Corollary~\ref{cor:conv_AM} and by the above discussion, the modified AM scheme guarantees convergence to a critical point of the SLP design \eqref{eq:TTP_recap}.
\end{Prop}

We will show the proof later.
The above result also applies to the PPAP-minimization SLP design \eqref{eq:PPAP_pro}.
Concisely we have
\begin{align*}
& f= \hat{f}_{\sf PPAP} \text{~(cf.,\eqref{eq:hat_ppap})}, ~ \bx_1  = (\bd, \bU, \bZ), \\
& \setX_1 = \setW \times \Cbb^{(N-K) \times T},
 \bx_2  =\bvarphi, ~
\setX_2  = {\cal P}.
\end{align*}
We consider the same proximal AM scheme as above.
As an extension of Proposition~\ref{fact:Lip_cons}, we have
\begin{Corollary}\label{fact:Lip_cons2}
	Consider the PPAP-minimization SLP design \eqref{eq:PPAP_pro} under the log-sum-exponential approximation \eqref{eq:hat_ppap}.
	All the results in Proposition~\ref{fact:Lip_cons} apply.
\end{Corollary}

\medskip

\noindent{\it Proof of Proposition~\ref{fact:Lip_cons} and Corollary~\ref{fact:Lip_cons2}}:
We first consider Proposition~\ref{fact:Lip_cons}.
The variable $\bvarphi^k$ is bounded, naturally, and the nontrivial part lies in the boundedness of $(\bd^k,\bU^k)$.
By the descent property in Proposition~\ref{prop:conv}.{\it (a)}, we have $f(\bx^0) \geq f(\bx^k)$ for all $k$.
For convenience, let $\bx = \bx^k$.
We get
\begin{align}
	f(\bx^0) & \geq	
	\frac{1}{T}  \|  \bd \diamond \bs_{t}  + \bu_t  \|_{\bR_{\bvarphi}}^2
	\geq \frac{1}{T} \lambda_{\rm min}(\bR) \| \bd \diamond \bs_{t}  + \bu_t  \|_2^2
	\nonumber \\
	& \geq \frac{1}{T} \lambda_{\rm min}(\bR) | d^R_{i,t} \Re(s_{i,t}) + \Re(u_{i,t})|^2
	\label{eq:ttt1}
\end{align}
for all $i,t$; note that $\lambda_{\rm min}(\bR_{\bvarphi}) = \lambda_{\rm min}(\bR) > 0$.
Suppose that $(i,t)$ is such that $\Re(s_{i,t}) > 1$.
By Fact~\ref{fac:SEP}, we have $\Re(u_{i,t}) \geq -d_i^R + c_i$, where $c_i = \alpha_i$ if $s_{i,t} < 2L-1$ and $c_i= \beta_i$ if $s_{i,t}= 2L-1$.
Applying this result to \eqref{eq:ttt1} gives
\[
\sqrt{\frac{f(\bx^0) T}{\lambda_{\rm min}(\bR)}} \geq d^R_{i,t} \Re(s_{i,t})+ \Re(u_{i,t}) \geq d^R_{i,t} ( \Re(s_{i,t}) - 1 ) + c_i.
\]
Since $\Re(s_{i,t}) > 1$, the above inequality suggests that $d_i^R$ is bounded above.
Since $d_i^R\geq \alpha_i\geq 0$, $d_i^R$ is bounded.
Under a bounded $d_i^R$, we see from \eqref{eq:ttt1} that $\Re(u_{i,t})$ is bounded for all $t$.
Similarly we can show the same bound result when $\Re(s_{i,t}) < -1$ and when we consider the imaginary counterparts.
We hence conclude that, under Assumption~\ref{asm5}, every $(\bd^k,\bU^k)$ is bounded.
This completes Proposition~\ref{fact:Lip_cons}.

The proof of Corollary~\ref{fact:Lip_cons2} is similar. We have
\begin{align}
	f(\bx^0) & \geq \hat{f}_{\sf PPAP}(\bd,\bU,\bZ, \bvarphi) \nonumber  \\
	& \geq \max_{t=1,\ldots,T} \| \bH^\dag ( \bvarphi \circ ( \bd \diamond \bs_{t}  + \bu_t ) ) +  \bB \bz_t \|_\infty^2 \nonumber \\
	& \geq \frac{1}{N} \| \bH^\dag ( \bvarphi \circ ( \bd \diamond \bs_{t}  + \bu_t ) ) +  \bB \bz_t \|_2^2  \nonumber \\
	& = \frac{1}{N} ( \|  \bd \diamond \bs_{t}  + \bu_t  \|_{\bR_{\bvarphi}}^2 + \|  \bz_t \|_2^2 ), \label{eq:ttt2}
\end{align}
where we have used $\delta \log (\sum_{i=1}^n e^{x_i/\delta}) \geq \| \bx \|_\infty$,
$\bB^H \bH^\dag = \bzero$, and $\bB^H \bB =\bI$.
Eq.~\eqref{eq:ttt2} shows that $\bz_t$ is bounded;
by    the proof of Proposition~\ref{fact:Lip_cons} shown above, we readily see from \eqref{eq:ttt2} that $(\bd,\bU)$ is bounded.
\hfill $\blacksquare$

\section{Optimal Linear Beamforming}

In this section, we briefly review the optimal linear beamforming (OLB) scheme in~\cite{Bengtsson2001,Yu2007Transmitter} and describe its implementation in our numerical simulations.

Under the linear precoding scheme $\bx_t =  \sum_{i=1}^K \bw_i s_{i,t}$,
the OLB scheme designs the beamforming vectors $\bw_1,\ldots, \bw_K$ by minimizing the average total transmission power (TTP) subject to signal-to-interference-and-noise ratio (SINR) constraints; specifically,
\begin{equation} \label{eq:olb_tp_pro}
\begin{aligned}
\min_{ \bw_1, \dots, \bw_K } & ~  \Exp [\|\bx_t\|_2^2] = \textstyle \sum_{i=1}^K \rho\|  \bw_i \|^2_2 \\
{\rm s.t.} & ~{\sf SINR}_i \triangleq \frac{ \rho | \bh_i^H \bw_i |^2}{ \sum_{j \neq i} \rho | \bh_i^H \bw_j |^2 + \sigma_v^2} \geq \zeta_i, ~ \forall i,
\end{aligned}
\end{equation}
where $\zeta_i$ is the SINR requirement of the $i$th user.
As a variation of \eqref{eq:olb_tp_pro}, we can also consider peak per-antenna average power minimization

\begin{equation}\label{eq:olb_pe_pro}
\begin{aligned}
\min_{ \bw_1, \dots, \bw_K  } & ~ \max_{n=1,\dots,N} ~ \Exp[|x_{n,t}|^2] = \textstyle \sum_{i=1}^K  \rho |w_{n,i}|^2 \\
{\rm s.t.} & ~ {\sf SINR}_i \geq \zeta_i, ~ \forall i.
\end{aligned}
\end{equation}

In the simulation, the SINR requirement $\zeta_i$ of both Problems \eqref{eq:olb_tp_pro} and \eqref{eq:olb_pe_pro} are chosen to satisfy the symbol error probability (SEP) requirement~\eqref{eq:SEP_const}, which can be achieved by the following fact.

\begin{Fact} \label{fac:linsep}
Consider the OLB design in \eqref{eq:olb_tp_pro} or \eqref{eq:olb_pe_pro}.
Suppose that the multiuser interferences (MUIs) are approximated as complex circular Gaussian random variables.
Then any feasible beamforming solution to \eqref{eq:olb_tp_pro} or \eqref{eq:olb_pe_pro} satisfies the SEP requirements in \eqref{eq:SEP_const}
if we choose
\[
\zeta_i = \frac{\rho}{2}\Big[Q^{-1}\Big(\frac{1-\sqrt{1-\eps_i}}{2}\Big)\Big]^2.
\]
\end{Fact}

\noindent{\it Proof}: Plugging the transmitted signals of the linear precoding scheme \eqref{eq:LB} into the system model~\eqref{eq:sig_mod}, we get
\[
\textstyle y_{i,t} = \bh_i^H \bw_i {s_{i,t}} + \sum_{j \neq i} \bh_i^H \bw_j {s_{j,t}} + v_{i,t},
\]
where $\eta_{i,t}\triangleq \sum_{j \neq i} \bh_i^H \bw_j {s_{j,t}}$ is the MUI.
By assuming that $\eta_{i,t}$ is a complex circular Gaussian random variable, we have $\eta_{i,t}  \sim  \CN(0, \sum_{j\neq i} \rho | \bh_i^H \bw_j |^2 )$.
Then, we  model
\[
\eta_{i,t} + v_{i,t} = \textstyle \Big( \sqrt{  \sum_{j\neq i} \rho | \bh_i^H \bw_j |^2  + \sigma_v^2 } \Big) \xi_{i,t},
~
\xi_{i,t} \sim \CN(0,1).
\]
By further assuming that $\bh_i^H \bw_i \ne 0$, we have
\[
  \frac{y_{i,t}}{\bh_i^H \bw_i} =  s_{i,t} + \sqrt{ \frac{ \rho }{{\sf SINR}_i} } \xi_{i,t}.
\]
By the basic SEP result in digital communications (e.g. \cite{proakis2001digital}), or by the SEP derivation in Section~\ref{sec:model}, we have
\[
\textstyle{\sf CSEP}_{i,t}^R \leq 2 Q\big( \sqrt{ \frac{2 {\sf SINR}_i}{\rho} } \big),~
{\sf CSEP}_{i,t}^I \leq 2 Q\big( \sqrt{ \frac{2 {\sf SINR}_i}{\rho} } \big).
\]
By the relation \eqref{eq:imp1} and the invertibility of the $Q$ function, the desired result is obtained.
\hfill $\blacksquare$

\medskip

Problems~\eqref{eq:olb_tp_pro} and \eqref{eq:olb_pe_pro} can be transformed to convex problems \cite{Bengtsson2001} and then solved by available convex optimization softwares, such as CVX \cite{grant2008cvx}.

\end{document}